\RequirePackage[2020-02-02]{latexrelease}
\documentclass[11pt,nofootinbib,aps,letterpaper,
preprintnumbers,superscriptaddress,onecolumn,showkeys]{revtex4}
\usepackage{amsmath}
\usepackage{eurosym}
\usepackage{amssymb}
\usepackage{graphicx}
\usepackage{color}
\usepackage{braket}
\usepackage{amsmath}
\usepackage{amssymb}
\usepackage{subfigure}
\usepackage{amsfonts}
\usepackage[left=1.5cm,right=1.5cm,top=1.5cm,bottom=1.5cm]{geometry}
\usepackage[usenames,dvipsnames,svgnames]{xcolor}  
\usepackage{hyperref}  
\definecolor{beamer@PRD}{RGB}{46,48,146}
\hypersetup{
breaklinks=true,
pdfstartview={FitH},    
colorlinks=true,       
linkcolor=magenta,          
citecolor=blue,        
filecolor=magenta,      
urlcolor=teal,           
anchorcolor=red,      
linktocpage=true
}
\begin{document}
\date{\today}
\newcommand\be{\begin{equation}}
\newcommand\ee{\end{equation}}
\newcommand\bea{\begin{eqnarray}}
\newcommand\eea{\end{eqnarray}}
\newcommand\bseq{\begin{subequations}} 
\newcommand\eseq{\end{subequations}}
\newcommand\bcas{\begin{cases}}
\newcommand\ecas{\end{cases}}
\newcommand{\p}{\partial}
\newcommand{\f}{\frac}

\title{Shadow of black hole surrounded by magnetized plasma: Axion-plasmon cloud}

\author {\textbf{Mohsen Khodadi}}
\email{m.khodadi@ipm.ir}
\affiliation{School of Astronomy, Institute for Research in Fundamental Sciences (IPM)
	P. O. Box 19395-5531, Tehran, Iran}
\affiliation{School of Physics, Institute for Research in Fundamental Sciences (IPM)
P. O. Box 19395-5531, Tehran, Iran}
\affiliation{Physics Department, College of Sciences, Shiraz University, Shiraz 71454, Iran}
\affiliation{Biruni Observatory, College of Sciences, Shiraz University, Shiraz 71454, Iran}

\begin{abstract}
By exploiting the extreme environment of the black hole (BH) as a potential place for axion-photon interaction, we use an axion-producing model of the magnetized plasma
to study the shadow of an asymptotically flat rotating BH immersed into an axion-plasmon cloud. By aiming to reveal footprints of axion in the dark shadow of BH, we in this paper explore the influence of the fixed axion-plasmon background on the motion of incident photons around the rotating BH. Under some free parameter settings, we find that axion-plasmon cloud around rotating BH affects the shape and size of the shadow in such a way that its role is distinguishable from non-magnetized plasma and standard vacuum solutions. By being limited to high rotation BH, we show that the size of the BH shadow increases as the axion-plasmon coupling gets strong.  Interestingly, our analysis indicates that as the mass of axion gets heavier, it can leave a subtle imprint of itself on the shadow.
Conversely, in the non-rotating limit (Schwarzschild), by recovering the spherical symmetry of the shadow shape of BH, its size decreases. 
In coordination with the trend of change in shadow size, the investigation of the energy emission from the BH surrounded by the magnetized plasma shows that the maximal energy emission rate from the rotating BH in the presence of axion-plasmon cloud increases compared to the non-magnetized plasma and the vacuum solutions. Subsequently, by relaxing the rotation, the axion-plasmon cloud causes a decrease in the energy emission rate from the BH.

\end{abstract}
\maketitle
\section{Introduction}

One of the most exciting challenges that we are facing in our comprehension of the Universe is the fact that most of the total matter density is not in the form of standard model (SM) baryons~\cite{Schindler:2001sd}. More precisely, the baryon content measured from the abundance of light elements~\cite{Kirkman:2003uv} does not match the matter content inferred from other probes such as the cosmic microwave background~\cite{WMAP:2008lyn, Planck:2013pxb, Planck:2018vyg}, gravitational lensing~\cite{Trimble:1987ee}, and the evolution of large-scale structures~\cite{Kolb:1990vq, Springel:2005nw}.
This discrepancy has been interpreted as an indication for the existence of a non-baryonic form of matter known as dark matter (DM)~\cite{Bertone:2004pz}. Among the possible candidates for explaining the missing DM abundance are primordial black holes (PBHs)~\cite{Carr:1974nx}, massive neutrinos, and hypothetical weakly interacting massive particles (WIMPs) that appear in various extensions of the SM once new, exotic elements such as extra dimensions~\cite{Kolb:1983fm, Servant:2002aq} or supersymmetry~\cite{Jungman:1995df} are invoked.

It has been also argued that an extremely light boson that populates the Universe in a coherent wave-like state might also be responsible for the observation of DM~\cite{Press:1989id, Sin:1992bg, Hu:2000ke, Riotto:2000kh}. In this regard, the privileged candidate is a hypothetical pseudo-Nambu-Goldstone boson called the axion. Historically, the term ``axion'' was first used by Wilczek~\cite{Wilczek:1977pj} and Weinberg~\cite{Weinberg:1977ma} in their seminal papers, in which they show that this new pseudoscalar particle should exist within the solution of the strong-CP problem in QCD proposed by Peccei and Quinn (PQ)~\cite{Peccei:1977hh,Peccei:1977ur}. In the theory, a new $U(1)$ symmetry is introduced which gets spontaneously broken below an energy scale $f_a$.\footnote{The couplings between the axion and other fields are suppressed by the inverse of the symmetry breaking scale, $\sim 1/f_a$. This occurs in ``invisible axion'' models, in which the spontaneous symmetry breaking occurs at energy scales much above the electroweak scale and the QCD axion turns out to be experimentally viable. See Refs.~\cite{Sikivie:2006ni, Marsh:2015xka, DiLuzio:2020wdo, Sikivie:2020zpn} for recent reviews of the QCD axion.}

The properties of the QCD axion are closely related with the detailed history of the early Universe. The relic QCD axions are produced in the early Universe both as hot and cold populations. While the cold and non-relativistic population can explain the missing DM since it was never in thermal equilibrium with the rest of the plasma~\cite{Abbott:1982af, Dine:1982ah, Preskill:1982cy, Stecker:1982ws} 
(see also \cite{Park:2012ru, Jeong:2013oza, Stern:2014wma}), hot QCD axions which are thermally produced in the early plasma are not a suitable DM candidate since they move too fast to gather into galactic halos~\cite{Hannestad:2007dd, Archidiacono:2013cha, Giare:2020vzo, Du:2016zcv}. The idea of the QCD axion as a CDM particle widely has been studied in the literature, see e.g.\ Refs.~\cite{Hertzberg:2008wr, Visinelli:2009kt, Visinelli:2014twa, Noh:2017sdj}. In this view, QCD axions would solve two major problems in modern physics and astrophysics at once. Besides, the discovery of the Higgs particle is potentially in favor of other scalars since it is in essence the proof for the existence of a spin-zero boson that emerges from a spontaneous symmetry breaking~\cite{Chadha-Day:2021szb}. Meantime, it is essential to make a theoretical framework suitable for low-energy phenomena related to CDM that can safely address the non-relativistic limit of relativistic quantum field theories \cite{Namjoo:2017nia,Firouzjaee:2018pkp,Friedrich:2018qjv,Braaten:2018lmj,Eby:2018ufi,Salehian:2020bon,Salehian:2021khb}.

Dropping the requirement that the axion is related to the QCD theory leads to a generalization of these pseudoscalar particles known as axion-like particles (ALPs), which are generally extremely light and have very weak coupling with quarks, leptons, and photons. There is growing theoretical realization that emphasize the appearing of ALPs in the context of string theory-based extensions of the SM~\cite{Svrcek:2006yi, Conlon:2006tq}, which reinvigorates the study of these hypothetical particles over a vast range of masses. ALPs can also be DM over a vast range of masses and couplings. As a result, one can still imagine ALPs as a popular candidate for the role of DM~\cite{Arias:2012az, Visinelli:2017imh}. People usually collectively refer to the QCD axion and ALPs simply as ``axions''.

Regardless of their origin, it is conceivable that some coupling of the axion with charged fermions exists, which inevitably leads to an effective axion-photon coupling $g$. The interaction of the axion with the photon is one of the main channels exploited for direct detection techniques~\cite{Sikivie:1983ip, Sikivie:1985yu, Duffy:2009ig, Kahn:2016aff, Millar:2016cjp} and gives rise to axion electrodynamics~\cite{Wilczek:1987mv}, see also Refs.~\cite{Visinelli:2013mzg, Beck:2014aqa}. If axions exist, they would also alter astrophysical environments through the interaction with charged plasma such as at the onset of supernova events~\cite{Mikheev:1998bg}, in the magnetosphere of a neutron star~\cite{Pshirkov:2007st, Huang:2018lxq, Hook:2018iia, Edwards:2020afl, Safdi:2018oeu}, or in the interior of stars and white dwarfs~\cite{Caputo:2020quz}.
One of the known applications of axion in the context of astrophysics comes from the axions-photons conversion (convert of axion and photon to each other) which in the presence of a MF acts as a coolant mechanism of stars, known as the Primakov mechanism~\cite{DiLuzio:2020wdo}.
In this regard, we consider yet another environment in which this conversion might lead to a possible signal, namely the accretion region around a rotating BH.

Despite that role of the environment through which the light rays pass is ignored in most frameworks related to compact objects as BH, by taking it into account, one obtain non-trivial physics. In other words,  be enclosing an astrophysical object by the plasma causes the trajectories of light rays to change compared to whose counterpart in a vacuum background. It is for that plasma indeed plays the role of a dispersive medium (with the refraction index opposite 1) so that the photons with various frequencies follow separate trajectories. Usually, for the description of the physics of such a medium, it is utilized a non-magnetized plasma and a linearized gravity theory~\cite{Er:2013efa,Rogers:2015dla,Rogers:2016xcc}.
For instance, various works done entitled the role of non-magnetic plasma on the BH shadow~\cite{Atamurotov:2015nra,Perlick:2015vta, Abdujabbarov:2015pqp, Huang:2018rfn, Chowdhuri:2020ipb, Wang:2021irh, Badia:2021kpk, Badia:2021kig, Ozel:2021ayr} and gravitational lensing~\cite{Bisnovatyi-Kogan:2010flt, Tsupko:2013cqa,Bisnovatyi-Kogan:2015dxa,Liu:2016eju,Bisnovatyi-Kogan:2017kii,Hensh:2021nsv, Atamurotov:2021hoq}, as two famous classical tests of GR.
Nevertheless, in the light of the pattern recorded of the magnetic field (MF) and electron temperature within the plasma close to the supermassive M87* BH by the Event Horizon Telescope (EHT) \cite{EventHorizonTelescope:2021srq, EventHorizonTelescope:2021btj}, it expects that the plasma around BH is magnetized. In other words, these worth observational achievements include this message is that the physics of supermassive BHs, have close relation with the plasma and MF their around. 

Given that the axion mass can cover a wide range of scales, $m_a \gtrsim \mathcal{O}(10^{-22}{\rm\,eV})$, with macroscopic Compton wavelengths from the order of magnitude of galaxies to table-top sizes, thereby the axion has been attracted the attention of many laboratory searches, see e.g.\ Refs.~\cite{Adler:2008gk, ADMX:2009iij, Irastorza:2011gs,  CAST:2017uph, MADMAX:2019pub, Zarei:2019sva, Shakeri:2020sin}. Recently proposed one experiment~\cite{Tercas:2018gxv,Mendonca:2019eke} in which the magnetized plasma medium is prone to produce and detection of axion.
In this setup, an electron beam passes via a magnetized plasma yield a beam-plasma instability called plasmon which after interacting with the applied strong MF converts to axions \cite{Tercas:2018gxv,Mendonca:2019eke}. The axions produced after passing an intervening wall, convert to photons, and finally will be recorded by a single-photon microwave detector. Theoretically, the main finding of this model is that a magnetized plasma can be an active source of axions and ALPs.\footnote{More exact, in the framework of the scheme proposed here, the axions track via a new quasi-particle called axion-plasmon (AP) polariton that, in essence, is a dynamical response of the plasma to the presence of axions~\cite{Tercas:2018gxv, Visinelli:2018zif}.} In short, the presence of a MF provides a bed for the interaction of axions with the plasma. It means that the magnetized plasmas formed around the compact stars and BHs can be the potential astrophysical places to detect axions. With this assumption that the experimental model~\cite{Tercas:2018gxv, Mendonca:2019eke} can provide an approximate description of the magnetized plasma around BH, thereby one can by taking into account its parametrization of the refraction index of the magnetized plasma probe the imprint of axion on Kerr BH shadow.
From the view of optical phenomena, this is well-motivated since recently in Ref.~\cite{McDonald:2019wou} studied different properties of light propagation within axion-plasma background and has been shown that in the bed of plasma increases the sensitivity to axion-induced optical phenomena. In this manner, one can propose a scenario to follow the imprint of axions using confront with the astronomical observations obtained of the supermassive M87* BH shadow. More than two years after the announcement of the relativistic images\footnote{The terminology of relativistic images was coined by Virbhadra and Ellis~\cite{Virbhadra:1999nm}.} of BH in the center of the M87* galaxy \cite{EventHorizonTelescope:2019dse,EventHorizonTelescope:2019ggy} and very recently in the center of SgrA* \cite{EventHorizonTelescope:2022xnr,EventHorizonTelescope:2022wok} by the EHT, we are witnessing the publication of numerous papers on the topic of BH shadow in the facing of EHT. Although it is not possible to introduce all of these papers, here we provide a very summarized list of them~\cite{Davoudiasl:2019nlo, Vagnozzi:2019apd, Jusufi:2019nrn, Allahyari:2019jqz, Li:2019lsm, Rummel:2019ads, Narang:2020bgo, Khodadi:2020jij, Neves:2020doc, Kumar:2020yem, Khodadi:2020gns, Lee:2021sws, Li:2021mzq, Khodadi:2021gbc, Heydari-Fard:2021pjc, Jusufi:2021fek, Badia:2021yrh, Jha:2021bue, Zeng:2021mok, Vagnozzi:2022moj, Jusufi:2022loj, Uniyal:2022vdu, Heumann:2022cyi, Ghosh:2022kit, Khodadi:2022pqh, KumarWalia:2022aop, Pantig:2022qak}. In the meantime, some interesting works are trying to show that BH shadow at interplay with superradiant instability phenomenon, potentially can be used to detect and or constraints light bosons candidates to the axion-CDM, see Refs. \cite{Roy:2019esk,Cunha:2019ikd,Creci:2020mfg,Roy:2021uye} for more details. 

Overall, probing the axions via astrophysics effects arising from axion-photon conversion in magnetized sources is a tempting idea.
With this idea in mind that the magnetized plasma around BH may be prone to provide a bed to produce axions, in this paper, we consider a Kerr BH enclosed into an axion-plasmon cloud which its parameterization of refraction index comes from Refs. \cite{Tercas:2018gxv,Mendonca:2019eke}. Generally speaking, it is well-motivated since depending on properties of the surrounding plasma, the strength of MF can change commonly in the range $10–10^8$ G for each particular BH candidate \cite{Tursunov:2019oiq}. For example, 
thanks to the findings published by the team of EHT, we now know that the plasma medium enclosing the supermassive M87* BH enriched by MF of order $1-30$G \cite{EventHorizonTelescope:2021srq, EventHorizonTelescope:2021btj}. Also already has shown that that supermassive BH located in the center of the milk way galaxy, i.e., Sgr A$^*$ should be surrounded by a magnetized plasma of order $10-100$G \cite{Eatough:2013nva}.
However, we may be faced with the query that, whether one can safely utilize the refraction index of a laboratory-based plasma configuration such as \cite{Tercas:2018gxv,Mendonca:2019eke} for the environment of BH.
The essential condition to calculate the plasmon-axion conversion probability in the experimental configuration \cite{Tercas:2018gxv,Mendonca:2019eke}
is that $\omega_{e} \gg gB_0$, i.e., the plasma frequency $\omega_{e}$ be very bigger than the product of axion-photon coupling and the MF. We will show that in the light of the first estimation released by EHT of the MF strength and average plasma density around supermassive M87* BH \cite{EventHorizonTelescope:2021srq, EventHorizonTelescope:2021btj}, the condition $\omega_{e} \gg gB_0$ satisfies. Note that in practice, the present estimation obtained of MF is not so strong that one expects to leave some notable phenomenological imprints of the axion-plasmon cloud on the BH shadow. However, one can optimistically interprets the current measurements of EHT as encouraging estimations motivating us to take the first steps towards making a speculative scenario for probing the imprint of the axion-photon interaction around BH. 
In this way, under some free parameter settings, one can find an intuition of the influence of axion-plasmon cloud on the shadow geometry by following the contribution of axions to the shape and size of BH shadow. Theoretically, such a speculative scenario motivated from the first observational hints of EHT may open up a potential prospect of exploiting BHs and their environments as extreme places for tracing the most prone candidate for CDM, i.e., the axion.  In Ref. \cite{Chen:2021lvo} authors by having the data of polarization properties of the radiation near the M87* BH \cite{EventHorizonTelescope:2021srq,EventHorizonTelescope:2021btj}, have constrained the axion-photon coupling and axion mass arising from Kerr BH superradiant instability. Recently, by employing the axion-producing model proposed in \cite{Tercas:2018gxv,Mendonca:2019eke} has been studied the contribution of axions to the deflection of the angle of light lensed by the Schwarzschild BH \cite{Atamurotov:2021cgh}. 
It would be interesting to mention that in \cite{Junior:2021dyw} and \cite{Wang:2021ara, Zhong:2021mty} respectively have been computed shadow of Schwarzschild and Kerr Melvin electro-vacuum solutions enclosed in a strong and uniform MF without the presence of plasma.

The outline of the paper is as follows. In Sec. \ref{secs.3}, by serving two different distribution functions for the plasma, we study the effect of the axion-plasmon cloud on the Kerr BH shadow in details. In this Sec, we also calculate the shadow radius and distortion parameter.
We do analysis of Sec. \ref{secs.3} for case of non rotation in Sec. \ref{secs.4}. The influence of axion-plasmon cloud on the energy emission rate of the both Kerr and non-Kerr BHs, is studied in Sec. \ref{emission}. Finally, we present a summary of results in Sec. \ref{con}.
Throughout this paper, for simplicity, we work in natural units with $c=G_N=\hbar=1$. 

\section{Rotating BH shadow in the presence of an axion-plasmon background}
\label{secs.3}

The space-time metric for a rotating BH of mass $M$ and angular momentum ${\bf J}$ in the Boyer-Lindquist coordinates is
\begin{eqnarray}
	\label{metric}
	\mathrm{d}s^2 &=& -\mathrm{d}t^2 + \frac{2Mr}{\Sigma}\left(\mathrm{d}t - a\sin^2\theta\mathrm{d}\phi\right)^2 + \frac{\Sigma}{\Delta}\mathrm{d}r^2 + \Sigma\mathrm{d}\theta^2 + \left(r^2+a^2\right)\sin^2\theta\mathrm{d}\phi^2\,,\\
	\Delta &=& r^2 -2 M r + a^2\,,\\
	\Sigma &=& r^2+a^2 \cos^2\theta\,,
\end{eqnarray}
where $t$ is the cosmic time and the spherical coordinates $(r, \theta,\phi)$ refer to a sphere centered around the BH and of radius $r$. In the units chosen, both the mass $M$ and the spin parameter $a\equiv |{\bf J}|/M$ have dimensions of length. For future use, note that the time-time component of the metric is $g_{tt} = -(1-2 M r/\Sigma)$. In this metric, the event horizon is obtained by setting the denominator of the ${\rm d}r^2$ term to zero, $\Delta = 0$, with solutions
\begin{equation}
	r_{\pm}=M \pm \sqrt{M^2-a^2}\,,
\end{equation}
which has real solutions whenever the {\it Kerr bound} is assured, $a \leq M$. Follow the standard procedure, here we assume that the Kerr bound is assured at all times, so that the BH singularity is always protected by a horizon \footnote{This, in essence, is the weak cosmic censorship hypothesis conceived by Roger Penrose in 1969 \cite{Penrose:1969pc}.}.
Kerr BHs are characterized by the presence of a region called the ``ergosphere'' which is situated between the outer event horizon $r_H\equiv r_+$ and the static limit horizon $r_t\equiv M+\sqrt{M^2-a^2\cos^2\theta}$ at which the argument of {\rm d}t vanishes. 

Given our interested in a scenario in which the axially symmetric BH is surrounded by a magnetized plasma, a quantity that separates it from standard Kerr vacuum solution is the refraction index of the plasma $n=n(x^{i}, \omega)$ in which $\omega$ is the photon frequency measured by an observer with velocity $u^\alpha$. For a non-magnetized cold plasma and in spherical symmetry with radial coordinate $r$, the refractive index is~\cite{synge1960relativity}
\begin{equation}
	\label{eq:coldplasma}
	n(r, \omega) = \sqrt{1-\frac{\omega_e^2(r)}{\omega(r)}}\,,
\end{equation}
where the plasma frequency depends on the electron mass $m_e$, the charge of the electron $e$, and the number density of electrons $N(r)$ as \footnote{However, there may also be ions in the plasma, in addition to electrons, which are not accounted for throughout the paper by assuming that they are immobile.}
\begin{equation}
	\label{eq:freqelectron}
	\omega_e^2(r) = \frac{4\pi e^2}{m_e}\,N(r)\,.
\end{equation}
For example, for the electron densities within range measured by EHT i.e., $N = 10^{4}-10^{7}{\rm\,cm^{-3}}$, the plasma frequency is $\omega_e\approx 10^{-9}-10^{-6}\,$eV ( in SI units the
plasma frequency reads as $\omega_e^2(r)=\frac{e^2}{m_e \epsilon_0}\,N(r)$, which $\epsilon_0=8.8\times10^{-12}Kg^{-1}.m^{-3}.e^2.s^2$, is the permittivity of free space). To explain the dependency of the frequency $\omega(r)$ on $r$, we should utilize the Hamiltonian addressing the photon around a BH surrounded by plasma. It takes the following form \cite{synge1960relativity}
\begin{equation}
	\label{generalHamiltonian}
	H(x^\alpha, p_\alpha)=\frac{1}{2}\left( g^{\alpha \beta} p_\alpha p_\beta + (n^2-1)\left( p_\beta u^\beta \right)^2 \right)\,,
\end{equation}
where $x^\alpha$ and $p_\alpha$ are the spacetime coordinates and the photon four-momentum, respectively. For a stationary spacetime, one can introduce a timelike Killing vector $\xi^\alpha$ obeying the Killing equations $\xi_{\alpha ;\beta}+\xi_{\beta;\alpha}=0$, see e.g. Ref.~\cite{Rezzolla:2004hy}, so that the frequency of a photon for the timelike Killing vector $\omega_0 \equiv -k^\alpha\xi_{\alpha}$ is constant. 
An observer moving with four-velocity $u^{\alpha}$ would measure a different frequency,
\begin{equation}
	\label{rs}
	\omega(r)= \frac{\omega_0}{\sqrt{-g_{tt}}} = \frac{\omega_0}{\left(1-\frac{2 M r}{\Sigma}\right)^{1/2}}\,.    
\end{equation}
It is easy to see, $\omega(r\longrightarrow\infty) = \omega_0 = -p_t$, meaning that 
$\omega_0$ indeed denotes the energy of the photon from the perspective of a distant observer.
This explains why the frequency in Eq.~\eqref{eq:coldplasma} depends on the radial coordinate $r$. The expression in Eq.~\eqref{rs} can be alternatively derived from the equation of motion related to the Hamiltonian Eq.~\eqref{generalHamiltonian}.

In this paper we are interested in the magnetized plasma which is prone to production of axion. In this regard, we take  a generalized electromagnetic theory in which regarded the axion-photon coupling~\cite{Sikivie:1983ip, Sikivie:1985yu, Wilczek:1987mv}, see also Refs.~\cite{Visinelli:2013mzg, Beck:2014aqa}
\begin{align}
&	\mathcal{L} = R-\frac{1}{4}F_{\mu\nu}F^{\mu\nu}-A_\mu J_e^\mu+\mathcal{L}_\varphi+\mathcal{L}_{\text{int}}\,, \label{Lag1}\\ 
&	\mathcal{L}_\varphi = \nabla_\mu\varphi^*\nabla^\mu\varphi-m_\varphi^2|\varphi|^2\,,\label{Lag2}\\ 
&	\mathcal{L}_{\text{int}} =- \frac{g}{4} F_{\mu\nu}\tilde F^{\mu\nu}\,,
\label{Lag3}
\end{align}
where $R$ is the Ricci scalar, $F_{\mu\nu}:=\partial^\mu A^\nu-\partial^\nu A^\mu$ and $\tilde F^{\mu\nu}:=\epsilon^{\mu\nu\sigma\rho}F_{\sigma\rho}$ respectively the electromagnetic field tensor with the four-vector potential $A^\mu$ and its dual which is an antisymmetric pseudo-tensor. Here, source $J_e^\mu=(\rho_e,\boldsymbol{j_e})$ refers to the four-vector current of electrons which $\rho_e$ and $\boldsymbol{j_e}$ are charge density and the current density, respectively. Moreover, $\mathcal{L}_\varphi$ and $\mathcal{L}_{\text{int}}$ are the Lagrangian terms for the axion energy density and the axion-photon interaction with the coupling $g$, respectively.

In the presence of the axion-plasmon contribution which comes from a magnetized plasma with the homogeneous MF $B_0$ in the $z$-direction, the refractive index is expressed as~\cite{Mendonca:2019eke}
\begin{equation}
	\label{eq:n1}
	n^2(r,\omega) = 1- \frac{\omega_e^2}{\omega^2}-\frac{\Omega^4}{\omega^2(\omega^2-\omega_{\varphi}^2)}-\frac{f_0}{\gamma_{0}}\left[\frac{\omega_e^2}{(\omega - k u_0)^2}+\frac{\Omega^4}{(\omega-k u_0)^2(\omega^2-\omega_{\varphi}^2)}\right]\,,
\end{equation}
where $\omega=\omega(r)$ is the photon frequency, the electron plasma frequency $\omega_{e} = \omega_{e}(r)$ is given in Eq.~\eqref{eq:freqelectron}, $\omega_{\varphi}$ is the axion frequency in the plasma, and the Rabi frequency $\Omega=\sqrt{gB_0\,\omega_{e}}$ describes the strength of the axion-plasmon coupling. The parameter $f_0$ denotes the fraction of the electrons in the beam propagating inside the plasma with velocity $u_0$, and the corresponding Lorentz factor $\gamma_0$.

As earlier mentioned, in the derivation of the refractive index above, the presence of ions in plasma in addition to electrons was not taken into account, assuming that ions are immobile. A notable point concerning the role of the axion scalar field in the refractive index at hand is that in the absence of it, Eq. (\ref{eq:n1}), addresses beam-plasma instability \cite{Plasma}. Indeed, the dispersion relation of such a plasma (free of axion scalar field) in case of satisfying propagation vector (or wavevector) $k\leq\frac{\omega_e}{u_0}\big(1+(f_0/\gamma_0)^{1/3}\big)^{3/2}$ can become complex i.e., $\omega \simeq \omega_r+i\gamma_p$ which its imaginary part denotes the instability growth rate \cite{Mendonca:2019eke}. Recall that beams of electrons (also ions) in plasma environments are a source of free energy, which can be transferred to waves through instability. Since the value of Rabi frequency, $\Omega$ is sufficiently small, and does not result in a considerable change in the instability of plasma, so the statement above still can be right for axion-plasmon described by refractive index  (\ref{eq:n1}).
Note, that $gB_0$ is generally extremely small, with $gB_0 \sim 10^{-22}\,$eV for a coupling strength $g \sim 10^{-12}{\rm\,GeV^{-1}}$, which experimentally \footnote{Recently, in the light of radio data of the Bullet  Cluster \cite{Chan:2021gjl}, by considering some benchmark axion mass ranges around $\sim10^{-5}$ eV, has been extracted a constraint of order of magnitude $\sim(10^{-12}-10^{-11})$ GeV$^{-1}$ for axion-photon coupling $g$.}
($g<0.66\times10^{-10}~ GeV^{-1}$ \cite{CAST:2017uph}) is allowed and for a MF of order of magnitude measured by EHT, i.e., $B_0 \sim 10\,$G. Now it is easy to see that $\omega_{e} \gg gB_0$, in agreement with the essential condition to calculate the plasmon-axion conversion probability in the experimental configuration
\cite{Mendonca:2019eke}. As a relevant comment, if in future measurements of EHT, the estimations of MF strength and the average plasma density increase the condition $\omega_{e} \gg gB_0$ holds yet.

The axion frequency $\omega_{\varphi}$ in (\ref{eq:n1}) is related to the axion mass $m_\varphi$ and the MF $B_0$ as~\cite{Mendonca:2019eke}
\begin{equation}\label{mass}
	\omega_{\varphi} = \sqrt{m_\varphi^2 + g^2B_0^2 + k^2}\,.
\end{equation}
 Given the point that the role of the electron beam scenario near the BH still is not very clear so we prefer to ignore the contribution of the $f_0$ terms in (\ref{eq:n1}). As a result, the refractive index written in Eq.~\eqref{eq:n1}, re-express as
\begin{equation}
	\label{eq:n2}
	n^2(r)=1- \frac{\omega_{\text{e}}^2(r)}{\omega^2(r)}\left(1+\frac{ g^2B_0^2}{\omega^2(r)-\omega_{\varphi}^2}\right)\,,
\end{equation}
where $\omega(r)$ is determined from Eq.~(\ref{rs}). In the absence of the MF $B_0$ or the axion-photon coupling $g$, the refractive index reduces to the case for a non-magnetized electron plasma in Eq.~\eqref{eq:coldplasma}. It means that to have a magnetized plasma, the presence of both strong MF and axion-photon coupling is essential and inevitable. In other words, the origin of magnetized plasma comes from an axion-plasmon background.

Throughout this paper, for plasma number density $N(r)$ we will employ two types distribution functions: the radial power-law density $N(r)=N_0 r^{-h} (h\geq0)$ \cite{Rogers:2015dla} and the exponential density $N(r)=N_0 e^{-r/l_0}$ \cite{Er:2013efa}. Here $N(r)$ and $n^2(r)$ are related to each other through Eq. (\ref{eq:freqelectron}).
Note that for the BH shadow immersed by the plasma be separable from its vacuum counterpart ($n^2=1$), it is essential that $\omega_{e}^2\ll \omega^2$. As showed above for a distant observer $\omega(r)=\omega_0$. By fixing $\omega_0\sim 10^{-4}$ eV equal with the same frequency that EHT observed the M87*, i.e.,  $230$GHz equivalence to a wavelength of $1.3$mm, the condition $\omega_{e}^2\ll \omega^2$ satisfies, meaning that the current resolution of EHT is sufficient to separating the non-magnetized plasma from its vacuum counterpart. In what follows, to find an intuition of the influence of axion-plasmon cloud on the shadow geometry by following the contribution of axions to the shape and size of BH shadow, we merely set a favorable choice of the free parameters. Before going into details, it is significant to mention that in this paper, the axion-plasmon cloud around a Kerr-spacetime modeled via a massive complex scalar field $\varphi$ minimally coupled to gravity i.e., $\int d^4x \sqrt{-g} ~\mathcal{L}_{total}~~~(\mathcal{L}_{total}=\mathcal{L}+\mathcal{L}_{\varphi}+\mathcal{L}_{int}$) (Eqs. (\ref{Lag1})--(\ref{Lag3})). Besides, the underlying spacetime can be described by the Kerr BH as long as the axion field is regarded as a perturbation to spacetime. More exactly, this occurs just in the linear level of approximation which is known as test field approximation i.e., the scalar field is not very dense and still can be studied in fixed spacetime geometry. Interestingly, this approximation works well for most situations of interest such axion producing model at hand. In other words, the produced axions feedback to the plasma through the modified Maxwell equations such that mathematically its effect is formulated by assuming perturbations around the electron density as $n_e=n_0+\tilde{n}$ in the quasi-neutrality plasma and $\varphi=\tilde{\varphi}$. Actually, the refractive index (\ref{eq:n1}) is a result of taking the mentioned decomposition into the Fourier modes $(\tilde{n},\tilde{\varphi})\sim \exp[i kz-i\omega t]$, and keeping the linear terms \cite{Tercas:2018gxv,Mendonca:2019eke}. It means that the axion-producing model at hand is not capable of generating the high-dense complex scalar field to break the test field approximation and subsequently affect the Kerr geometry. Consequently, without worry, one can be treated with the axion scalar field as a perturbation, leaving the Kerr spacetime unchanged.

\subsection{Photon motion and BH Shadow}
Here, we consider the shadow cast by BH surrounded by the plasma described above.
By placing the BH between the light source and the infinity observer due to trapping the light beam by BH, the observer will see a black spot on the bright background and interpret it as the shadow of BH. The shaded area that appeared in the sky for the observer indeed is similar to a spot dark enclosed into a bright ring. 
To put it another, the BH is surrounded by a region known as \textit{`` photon sphere''} in which exist closed, circular photon orbits
While in the outer edge of the shadow is a closed curve, including unstable photons that can escape bound orbits and reach a faraway observer. Indeed, the edge of the shadow acts as a separator of the captured and scattered orbits so that the infinite observer sees a dark central silhouette enclosed in a bright ring.
In order to determine the boundary of the shadow (locus of ray that can escape bound orbits), the equation of motion of photons are essential \cite{chandra98}. 

To do so usually is used the following Hamilton-Jacobi equation \footnote{A standard method to find photon paths through the refractive plasma medium around any given spacetime metric is the use of the Hamiltonian in the geometric optics limit \cite{synge1960relativity}. More exactly, through the Hamilton-Jacobi approach, one can define the equation of motion of the photons for exterior spacetime geometry of BH, provided that it for lightlike geodesics is separatable. In other words, this approach lets us do an analytical study of issues such as BH shadow and lensing within the framework of geometrical optics. This has applications in astrophysics since the real environment surrounding BHs, in essence, is plasma. Indeed the worth of the Hamilton-Jacobi approach is to disclose the duality between trajectories and wavefronts. This duality means that light could be either investigated by geometric optics- a branch of optics where light is described by rays- or by wavefront. It is well-known that this duality can be extended to any system described by a Lagrangian formalism, including systems with gravity. Given that the refraction index of the plasma is indeed a function of the photon four-momentum $p^{\alpha}$, so Eq. (\ref{eq:n1}) via the Hamilton-Jacobi equation (\ref{p3}) can safely apply to the underlying gravity system i.e., Kerr BH.} \cite{synge1960relativity,Bisnovatyi-Kogan:2010flt}
\begin{equation}
\frac{\partial S}{\partial
\sigma}=-\frac{1}{2}\Big(g^{\alpha\beta}p_{\alpha}p_{\beta}-(n^2-1)(p_{t}
\sqrt{-g^{tt}})^{2}\Big),~~~~p_{\alpha}=\partial S/\partial x^\alpha \label{p3}
\end{equation} with this assumption that the plasma is a static medium with four-velocity $u^{\alpha}=(\sqrt{-g^{tt}},0,0,0)$. Concerning the massless particles such as photon, the Jacobi action $S$ using method of separation of variables, be written as 
\begin{equation}
S=- {\cal E} t + {\cal L} \phi +
S_{r}(r)+S_{\theta}(\theta).\label{p4}
\end{equation} 
Here, $\cal L$, $\cal E$ respectively are interpreted as angular momentum and energy of the photons that are conservative quantities. 
Through the Hamilton-Jacobi equation, one can for trajectories of the photons obtain the following set of equations
\begin{eqnarray}
\Sigma\frac{dt}{d\sigma}&=&a ({\cal L} - n^2 {\cal E} a
\sin^2\theta)+ \frac{r^2+a^2}{\Delta}\left((r^2+a^2)n^2 {\cal
E} -a {\cal L} \right), \label{teqn}
\\
\Sigma\frac{d\phi}{d\sigma}&=&\left(\frac{{\cal L}}{\sin^2\theta}
-a  {\cal E}\right)+\frac{a}{\Delta}\left((r^2+a^2) {\cal E}
-a {\cal L} \right), \label{pheqn}
\\
\Sigma\frac{dr}{d\sigma}&=&\sqrt{\mathcal{R}}, \label{reqn}
\\
\Sigma\frac{d\theta}{d\sigma}&=&\sqrt{\Theta}, \label{theteqn}
\end{eqnarray} where the functions $\mathcal{R}(r)$ and $\Theta(\theta)$ have the following form
\begin{eqnarray}
\mathcal{R}&=&\left((r^2+a^2) {\cal E} -a {\cal L}
\right)^2+(r^2+a^2)^2(n^2-1){\cal E}^2 	-\Delta\left(\mathcal{K}+({\cal L} -a {\cal E})^2\right)\ , \label{9}
\\
\Theta&=&\mathcal{K}+\cos^2\theta\left(a^2  {{\cal
E}^2}-\frac{{\cal L}^2}{\sin^2\theta}\right) -(n^2-1) a^2 {\cal E}^2 \sin^2\theta\ , \label{10}
\end{eqnarray} with the Carter constant as $ \mathcal{K} $. The equation of motion of photons given by expressions (\ref{teqn})--(\ref{theteqn}), address the boundary the BH surrounded by plasma. To derive the apparent shape of the the BH surrounded by plasma we have to consider the closed orbits around it. Due to dependency of the equations of motion on conserved quantities ${\cal E}$, ${\cal L}$ and also the Carter constant ${\cal K}$, it is convenient to parametrize them using the normalised parameters $\xi={\cal L/E}$ and $\eta= {\cal K/E}^2$. Note that the origin of these two carter constants comes from the separability of the Hamilton-Jacobi equation \cite{Perlick:2017fio,Bezdekova:2022gib}. The silhouette of the BH shadow in the presence of the plasma can be found using the conditions
\[{\cal R}(r)=0=\partial {\cal R}(r)/\partial r . \]
Through these equations, the relevant expressions for the parameters $\xi$ and $\eta$ are written as
\begin{eqnarray}
\xi = - \frac{A_2}{A_1} - \sqrt{\big(\frac{A_2}{A_1}\big)^2 -\frac{A_3}{A_1}},~~~ \label{xiexp}
\eta= \frac{(r^2+a^2-a\xi)^2 +(r^2+a^2)^2 (n^2-1)}{\Delta} -(\xi-a)^2, \label{etaexp}
\end{eqnarray}
where
\begin{eqnarray}
A_1= \frac{a^2}{\Delta},~~~	A_2= \frac{r^2-a^2}{M-r}\frac{Ma }{\Delta},~~~ A_3= n^2 \frac{(r^2+a^2)^2}{\Delta}+\frac{2r (r^2+a^2)n^2 +(r^2+a^2)^2 n n'}{M-r}.
\end{eqnarray}
Note that the prime here refers to the differentiation of refractive index with respect to radial coordinate $r$. As a consistency check, one can easily show that by relaxing the plasma parameters, i.e., setting $n=1$ and $n'=0$, acquires the following standard expressions for vacuum case \cite{chandra98}
\begin{eqnarray}\label{IP}
\xi = \frac{r^2(r - 3M) + a^2 (r + M)}{a (M - r)}~,~~~~~~~~
\eta = \frac{r^3\big(4Ma^2-r(r-3M)^2\big)}{a^2(M - r)^2}~.
\end{eqnarray}

The boundary of the BH's shadow can be fully specified via the expressions $\xi$ and $\eta$ in (\ref{xiexp}). However, the shadow will be observed at \textit{''observer's sky''}, by introducing the following celestial coordinates 
\begin{eqnarray}  \label{alpha1}
\alpha=\lim_{r_{0}\rightarrow \infty}\left(
-r_{0}^{2}\sin\theta_{0}\frac{d\phi}{dr}\right),~~~~
\label{beta1}
\beta=\lim_{r_{0}\rightarrow \infty}r_{0}^{2}\frac{d\theta}{dr}~,
\end{eqnarray}  related to the real astronomical measurements. Here,
$r_0$ and $\theta_0$ denote the distance between the observer and the BH, and the inclination
angle between the line of sight of the observer and the rotational axis of the BH, respectively. By putting the equations of motion (\ref{teqn})-(\ref{theteqn}), into  (\ref{alpha1}), the relations related to the celestial coordinates $(\alpha,\beta)$ for the case BH surrounded by plasma acquire as follow
\begin{eqnarray}
\alpha= -\frac{\xi}{n\sin\theta_0}\, \label{alpha},~~~~
\beta=\frac{\sqrt{\eta+a^2(1-n^2\sin^2\theta_0)-\xi^2\cot^2\theta_0 }}{n}. \label{beta}
\end{eqnarray} 
By setting $n=1$, the above celestial coordinates $(\alpha,\beta)$ recovers the same thing expected for the standard vacuum case. Now, we are in a suitable position to display the shadow of rotating BH surrounded by a magnetized plasma enriched by the presence of axion. In this way, one can probe the role of environmental parameters around the BH such as electron plasma, magnetic field, and axion on the rotating BH shadow. Due to support of the high rotation in the BHs by EHT observation \cite{EventHorizonTelescope:2019pgp} and data analysis \cite{Tamburini:2019vrf}, preferably until the end of this paper, the value of spin parameter $a$ takes from within range $0.8<a<1$. 

\begin{figure}[ht]
\includegraphics[scale=0.33]{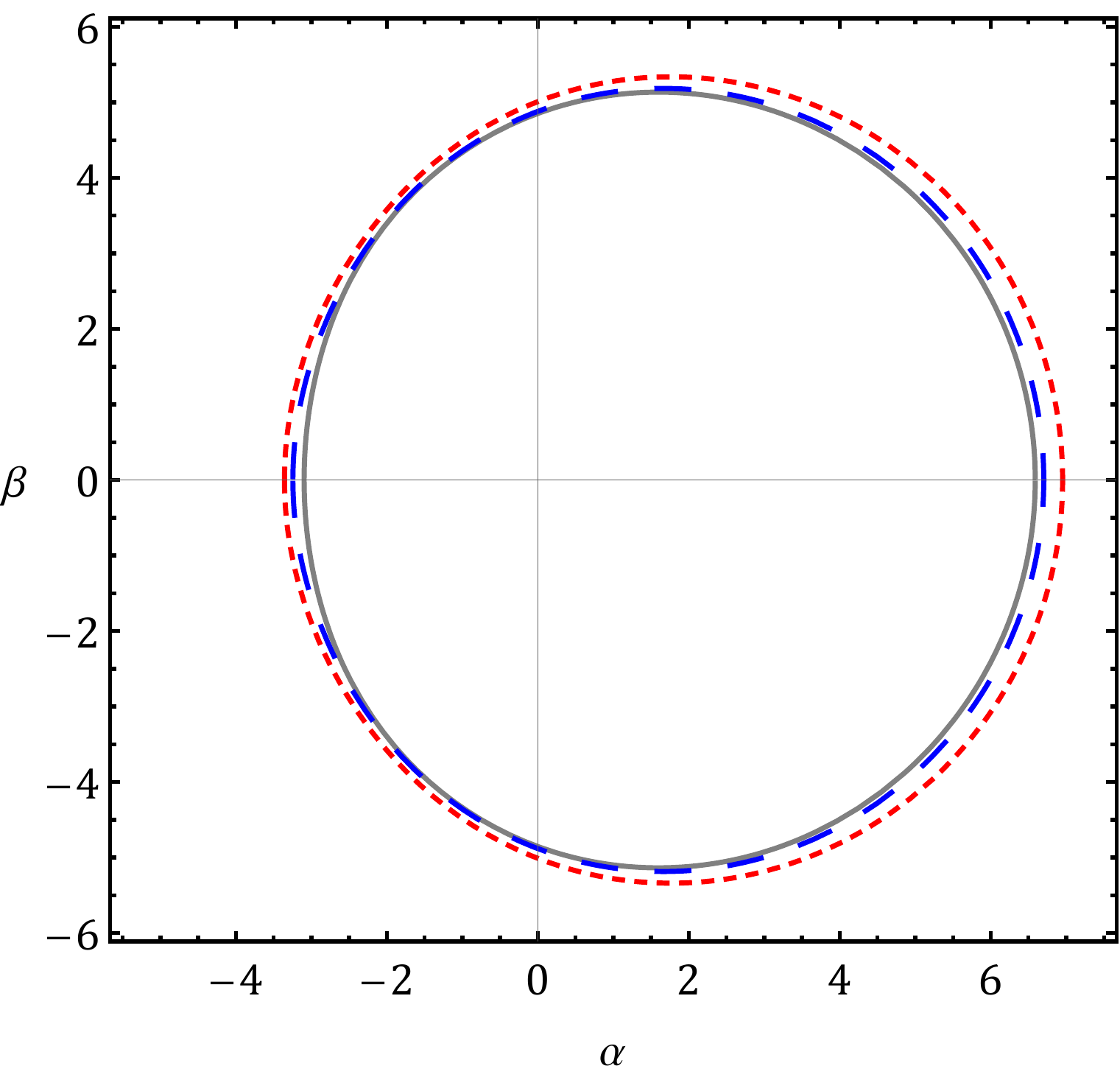}~~~
\includegraphics[scale=0.33]{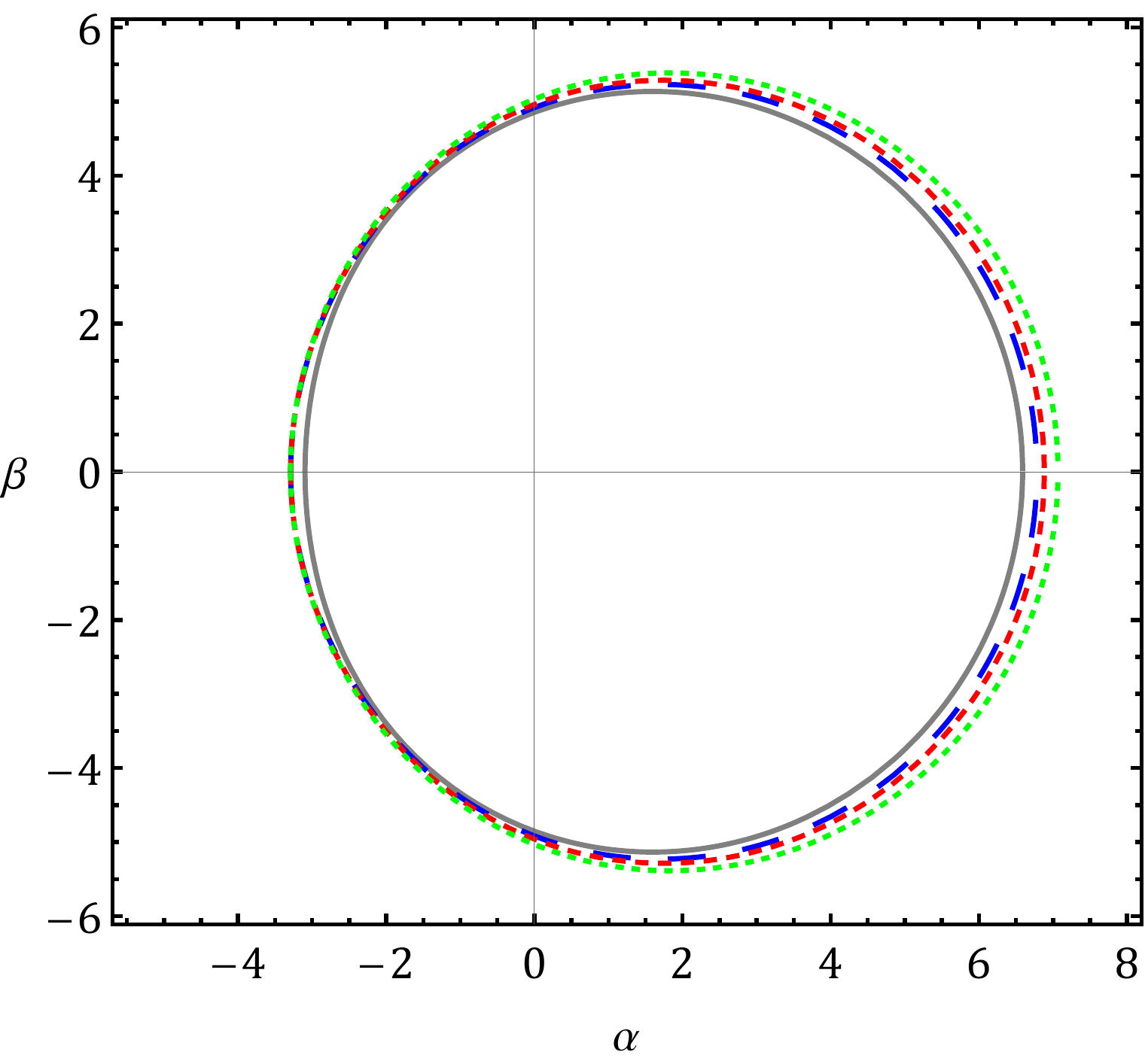}~~~
\includegraphics[scale=0.33]{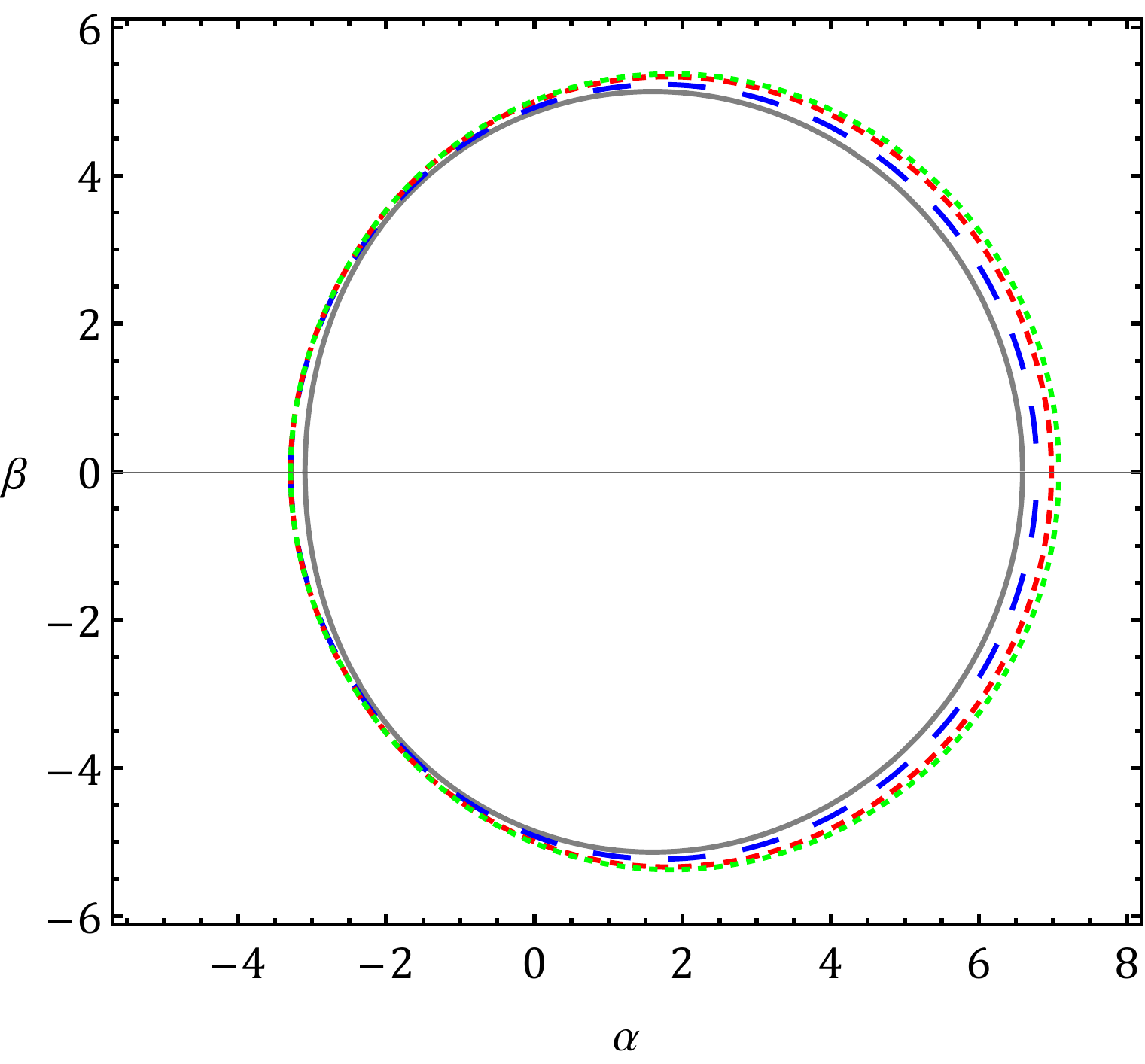}
\caption{\textit{The shadow of rotating BH $(a=0.9)$ surrounded by a plasma with homogeneous power-law density distribution  ($h=0$) for different sets of  $\{\chi_k,\chi_{ap},\chi_\varphi\}$. \textbf{Left:} $\{0,0,0\}$ (gray-solid), $\{0.35,0,0\}$ (blue-long dashed ), $\{0.7,0,0\}$ (red-dashed). \textbf{Middle:} $\{0,0,0\}$ (gray-solid), $\{0.5,0,0\}$ (blue-long dashed), $\{0.5,0.35,0.1\}$ (red-dashed), $\{0.5,0.7,0.1\}$ (green-dotted). \textbf{Right:} $\{0,0,0\}$ (gray-solid), $\{0.5,0,0\}$ (blue-long dashed), $\{0.5,0.5,0.25\}$ (red-dashed), $\{0.5,0.5,0.5\}$ (green-dotted). We have set $\theta_0=\pi/3$ and $M=1$.}}
\label{Shadow1}
\end{figure}

\begin{figure}[ht]
\includegraphics[scale=0.33]{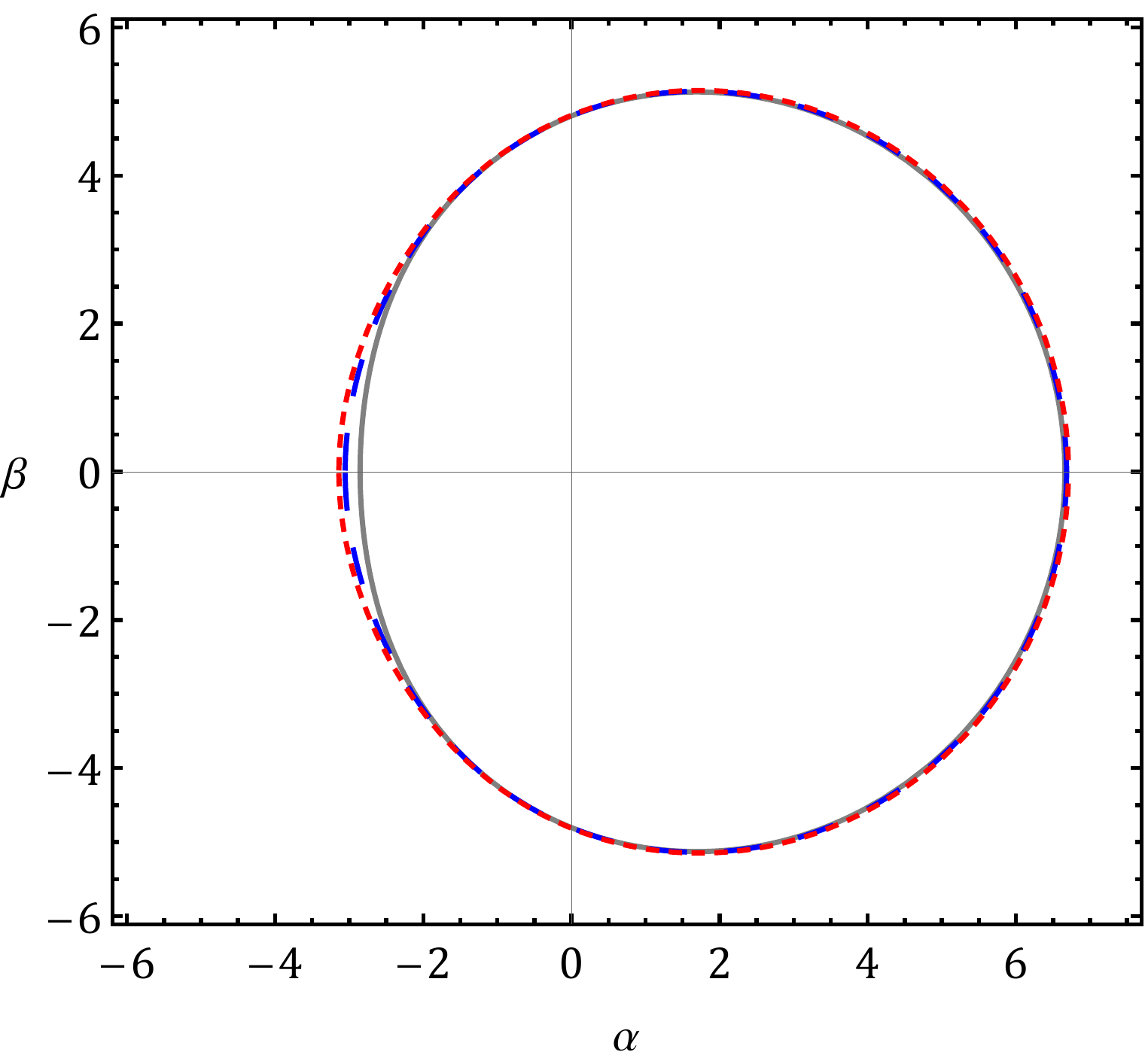}~~~
\includegraphics[scale=0.33]{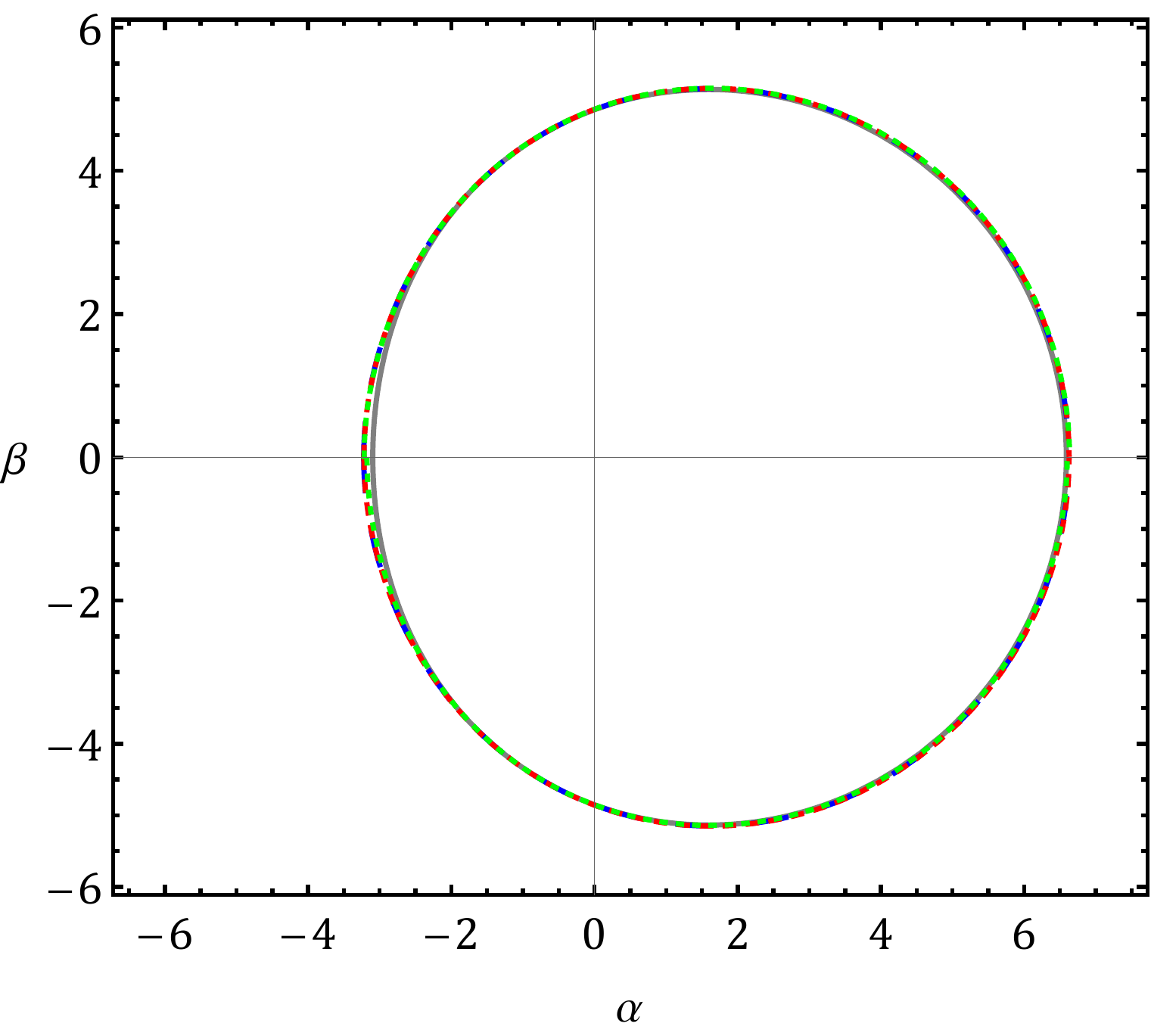}~~~
\includegraphics[scale=0.33]{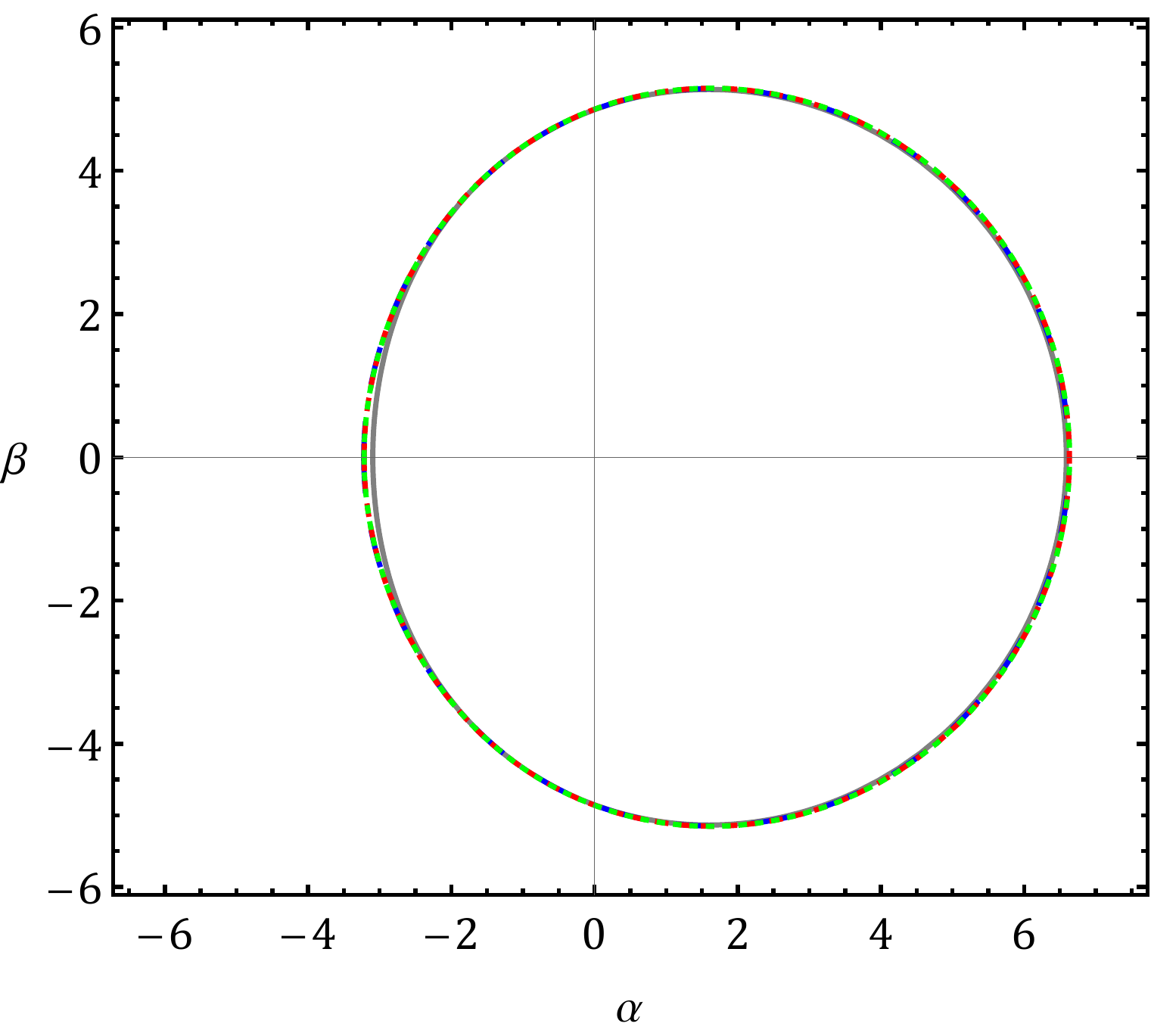}
\caption{\textit{The shadow of rotating BH $(a=0.9)$ surrounded by a plasma with inhomogeneous power-law density distribution ($h=1$) for different sets of $\{\chi_k,\chi_{ap},\chi_\varphi\}$. \textbf{Left:} $\{0,0,0\}$ (gray-solid), $\{0.35,0,0\}$ (blue-long dashed), $\{0.7,0,0\}$ (red-dashed). \textbf{Middle :} $\{0,0,0\}$ (gray-solid), $\{0.5,0,0\}$ (blue-long dashed), $\{0.5,0.3,0.2\}$ (red-dashed), $\{0.5,0.6,0.2\}$ (green-dotted). \textbf{Right:} $\{0,0,0\}$ (gray-solid), $\{0.5,0,0\}$ (blue-long dashed), $\{0.5,0.5,0.3\}$ (red-dashed), $(0.5,0.5,0.6)$ (green-dotted). We have set $\theta_0=\pi/3$ and $M=1$. }}
\label{Shadow2}
\end{figure}

Let us begin with a plasma which its number density obey of a radial power-law function as $N(r)=N_0 r^{-h} (h\geq0)$. For cases  $h=0$ and $h\neq0$, it address  a homogeneous and in-homogeneous distribution for the number density of plasma, respectively. By taking it into the refractive index (\ref{eq:n2}), we have 
\begin{eqnarray}
n^2(r)=1+\chi_k g_{tt}r^{-h}\bigg(1-\frac{\chi_{ap} g_{tt}}{1+\chi_\varphi g_{tt}}\bigg)~,	\label{nn}
\end{eqnarray} where $\chi_k=\frac{k}{\omega_0^2}~~(k=\frac{4 \pi e^2N_0}{m_e})$, $\chi_{ap}=\frac{g^2B_0^2}{\omega_0^2}$ and $\chi_\varphi=\frac{\omega_\varphi^2}{\omega_0^2}$ are three dimensionless parameters that characterize the plasma at hand.
Now, by regarding the refractive index (\ref{nn}) into the celestial coordinates $(\alpha,\beta)$, we can release the relevant shadows for different sets of  $\{\chi_k,\chi_{ap},\chi_\varphi\}$, as one can seen in Figs~\ref{Shadow1} and  \ref{Shadow2} for the homogeneous ($h=0$) and in-homogeneous power-law number density ($h=1$), respectively. 
Before going to details of these curves should be mentioned that, in general, their trend is independent of the value of inclination angle $\theta_0$ between the observer and the axis of the rotation. A common point of these figures is that the presence of plasma with power-law number density around a rotating BH results in an increase in the size of the shadow. Of course with this difference that in the case of in-homogeneous distribution this effect is not very sensible, just unlike the homogeneous case. As another interesting point, unlike in-homogeneous power-law number density, the magnetized plasma in the case of homogeneous leaves an effect separable from electron plasma on the shadow.  More exactly, for the case of homogeneous, axion-plasmon background leaves some separable imprints relative to non-magnetic plasma on the shadow. However, such effects on the shadow for the case of in-homogeneous power-law number density are hardly separable.

\begin{figure}[ht]
\includegraphics[scale=0.33]{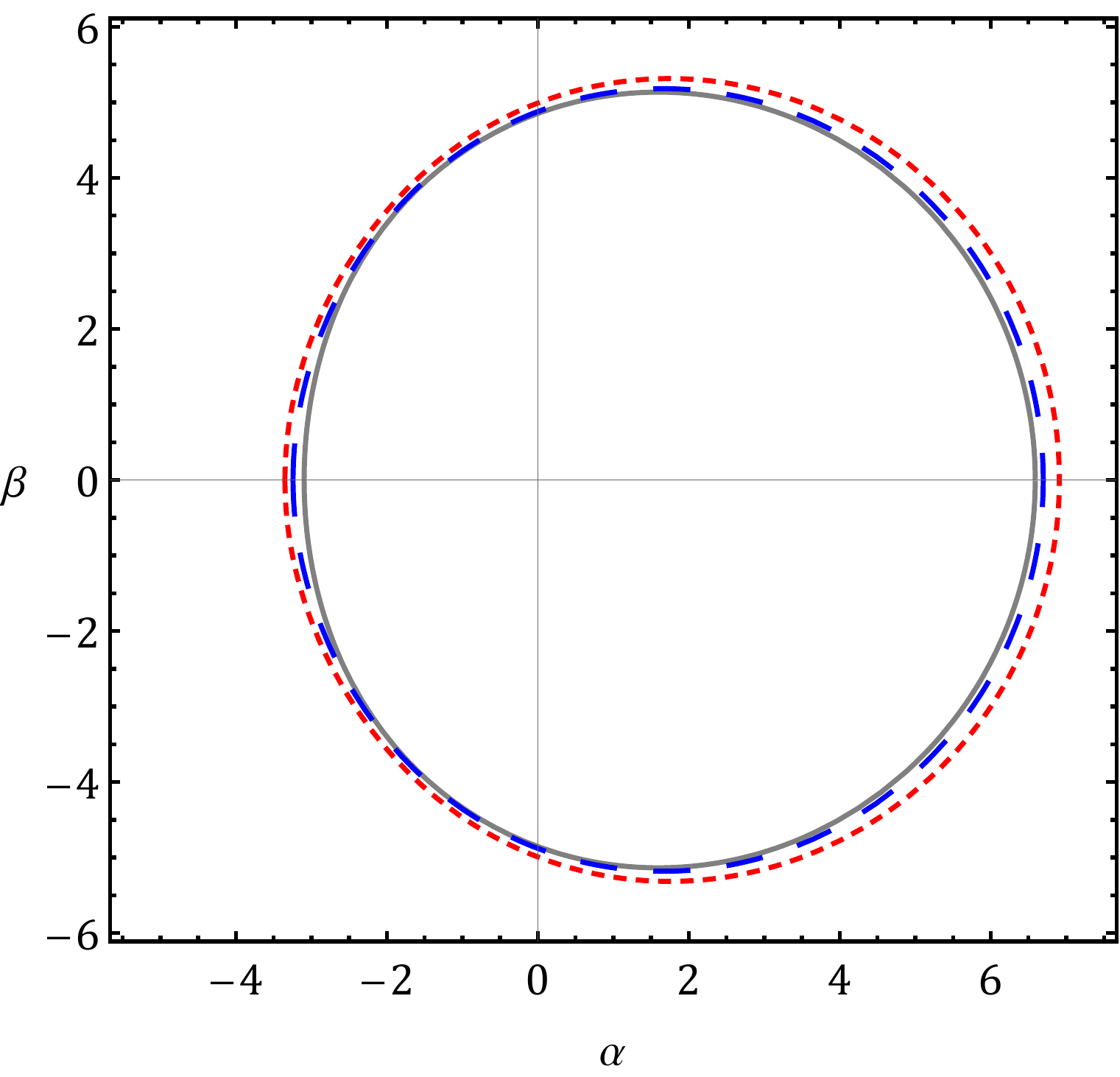}~~~
\includegraphics[scale=0.33]{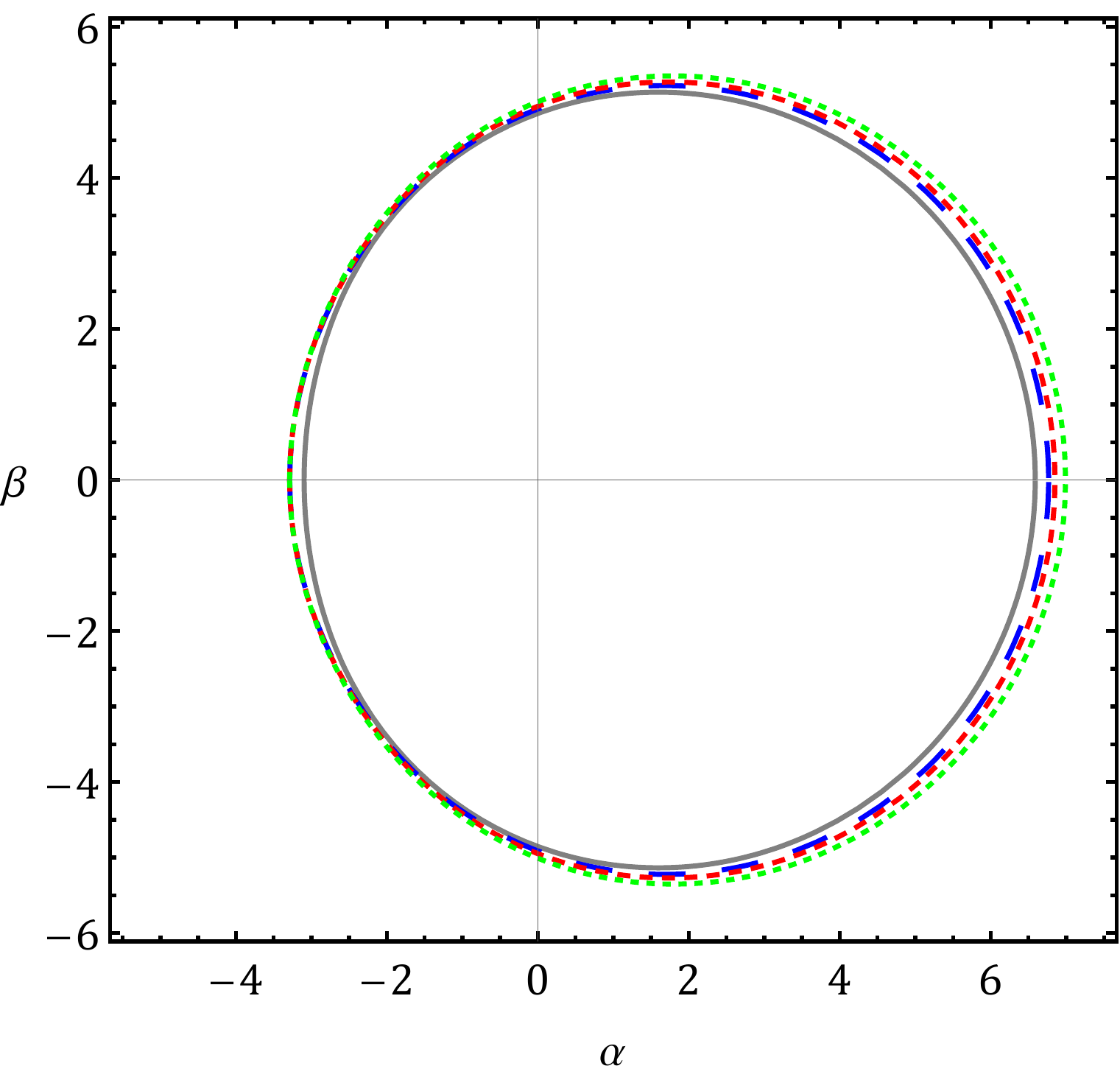}~~~
\includegraphics[scale=0.33]{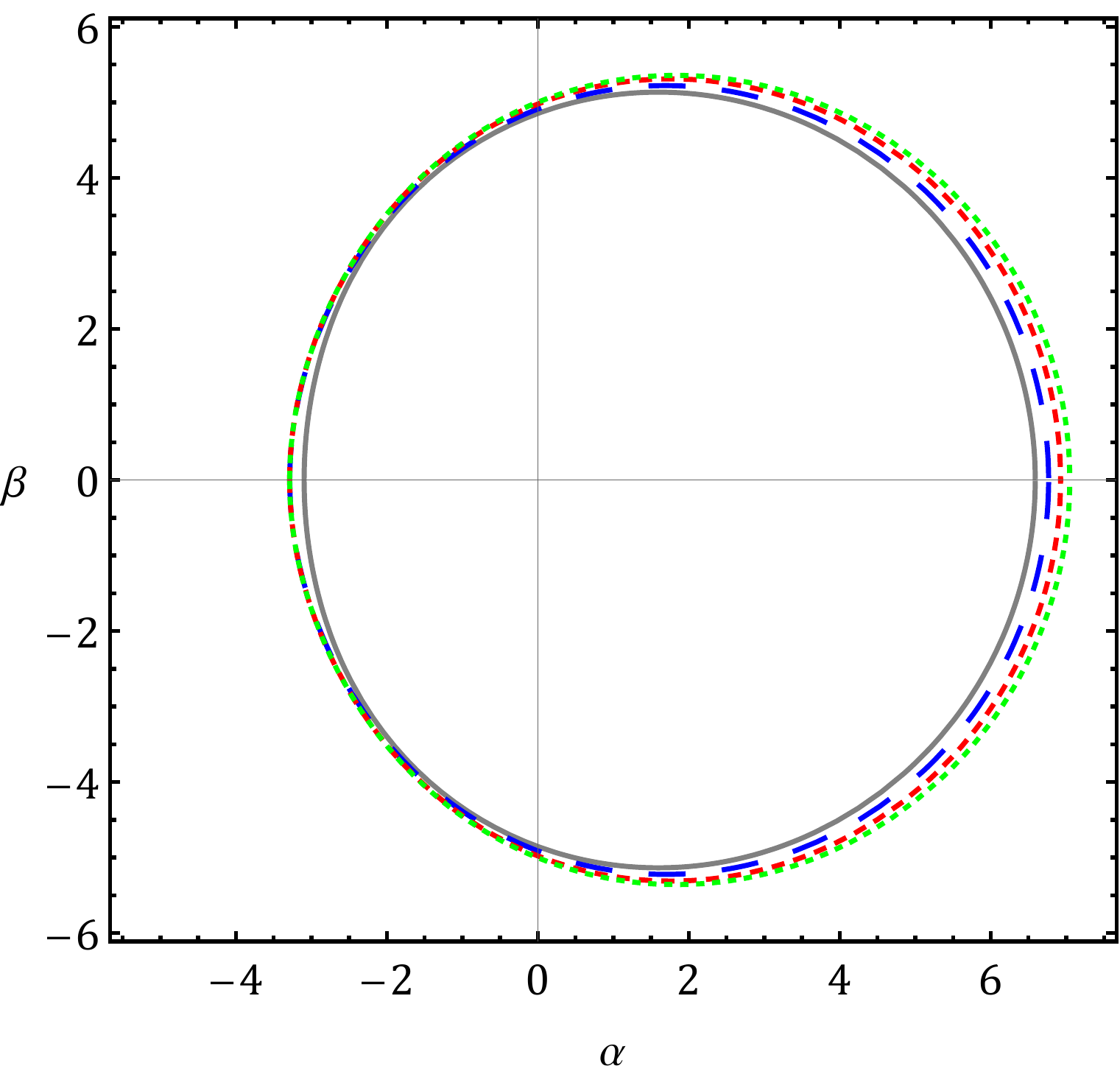}
\caption{\textit{The shadow of rotating BH $(a=0.9)$ surrounded by a plasma with inhomogeneous exponentially density distribution for different sets of $(\chi_\kappa,\chi_{ap},\chi_\varphi)$. \textbf{Left:} $\{0,0,0\}$ (gray-solid), $\{0.35,0,0\}$ (blue-long dashed), $\{0.7,0,0\}$ (red-dashed). \textbf{Middle:} $\{0,0,0\}$ (gray-solid), $\{0.5,0,0\}$ (blue-long dashed), $\{0.5,0.35,0.1\}$ (red-dashed), $\{0.5,0.7,0.1\}$ (green-dotted). \textbf{Right:} $\{0,0,0\}$ (gray-solid), $\{0.5,0,0\}$ (blue-long dashed), $\{0.5,0.5,0.25\}$ (red-dashed), $\{0.5,0.5,0.5\}$ (green-dotted). We have set $\theta_0=\pi/3$, $M=1$ and $l_0=50$.} }
	\label{Shadow3}
\end{figure}

Now, let us follow our analysis with a plasma which its number density obey of a exponentially function as $N(r)=n_0 e^{-r/l_0}$ in which $n_0$ and $l_0$ are the central density of plasma and the scale radius
of the central region of plasma, respectively. 
In this case, the refractive index (\ref{eq:n2}) reads as 
\begin{eqnarray}
n^2(r)=1+\chi_\kappa g_{tt}e^{-r/l_0}\bigg(1-\frac{\chi_B g_{tt}}{1+\chi_\varphi g_{tt}}\bigg),	\label{nnn}
\end{eqnarray} where $\chi_\kappa=\frac{\kappa}{\omega_0^2}~~(\kappa=\frac{4 \pi e^2n_0}{m_e})$, and $\chi_{ap}$, $\chi_\varphi$ with the same definition in 
(\ref{nn}). It is clear that for the case of homogeneous distribution, i.e., $N(r)=n_0$, we are not expected to get a different result from what is shown in Fig~\ref{Shadow1}. However, for the case of in-homogeneous distribution, we see in Fig~\ref{Shadow3} the curves are separable, just unlike their counterparts in power-laws one.
In agreement with the case of power-law number density, here also the presence of the plasma, whether magnetized or non-magnetized, results in increases in the shadow size and shape change of rotating BH. The point that is phenomenological valuable here is that plasma with in-homogeneous exponentially number density under some parameter selections lets us probe the imprints of axion-plasmon cloud via BH shadow since they leave some additional effects separable from the electron plasma and vacuum solution. It is important to note that the differentiation of the non-magnetized plasma from its magnetized counterpart has close relation with the value of $l_0$ such that as it gets smaller, the possibility of differentiation disappears.

\begin{figure}[htbp]
\centering
\includegraphics[scale=0.4]{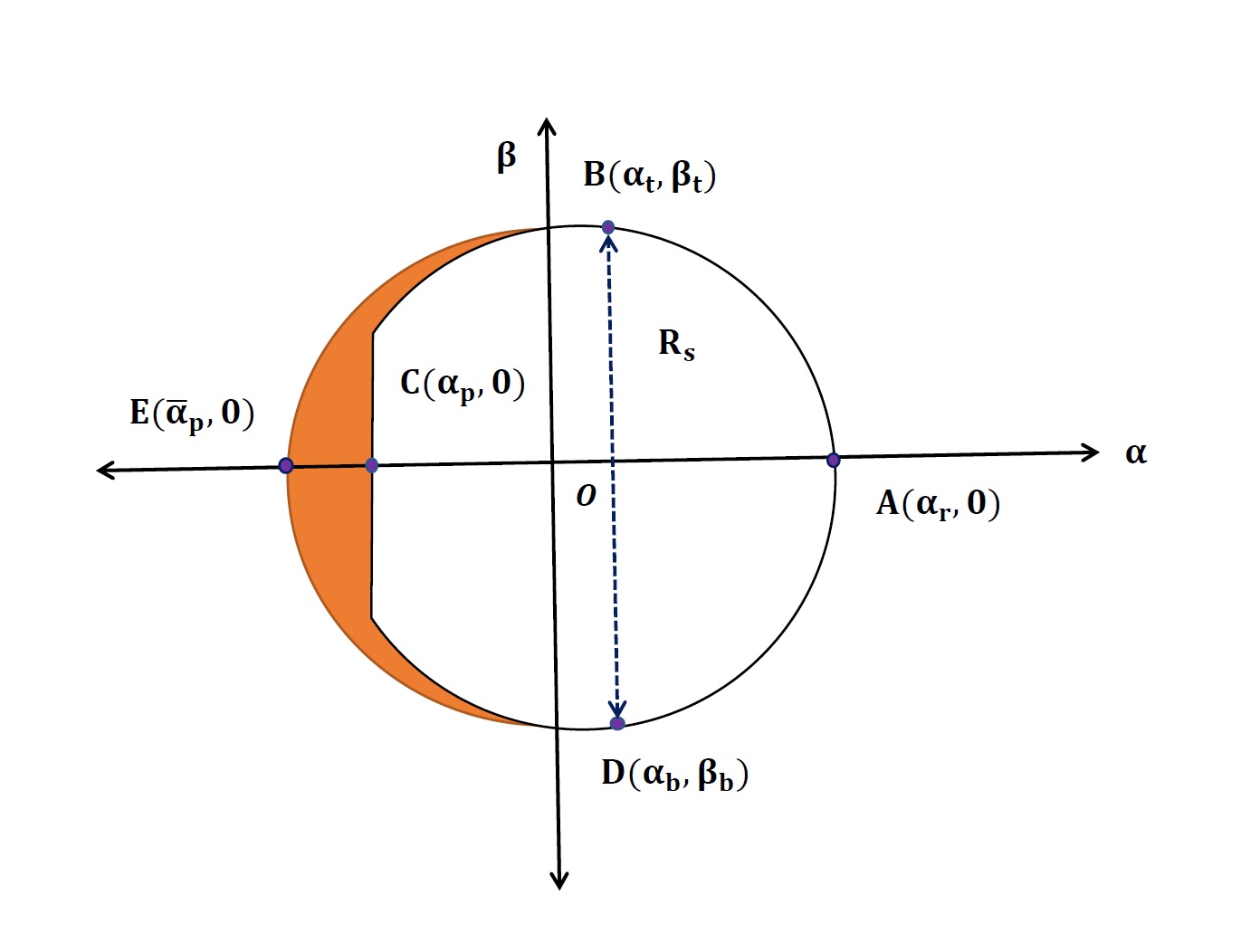}
\caption{\textit{Schematic representation of the BH shadow (whit region) and the reference (or perfect) circle (colored region).}}	\label{shadow_ref}
\end{figure}
\begin{figure}[ht]
	\includegraphics[scale=0.3]{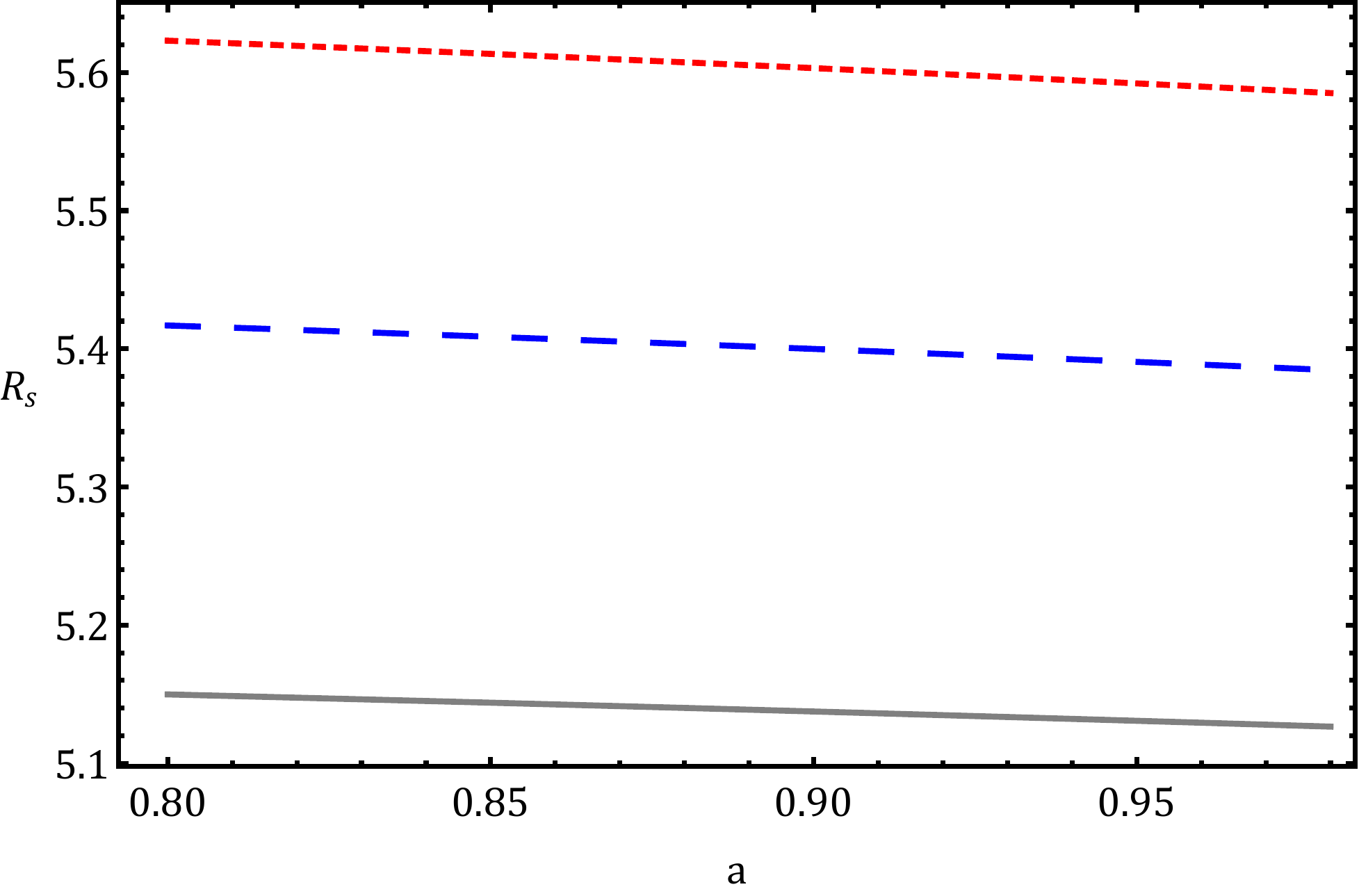}~~~
	\includegraphics[scale=0.3]{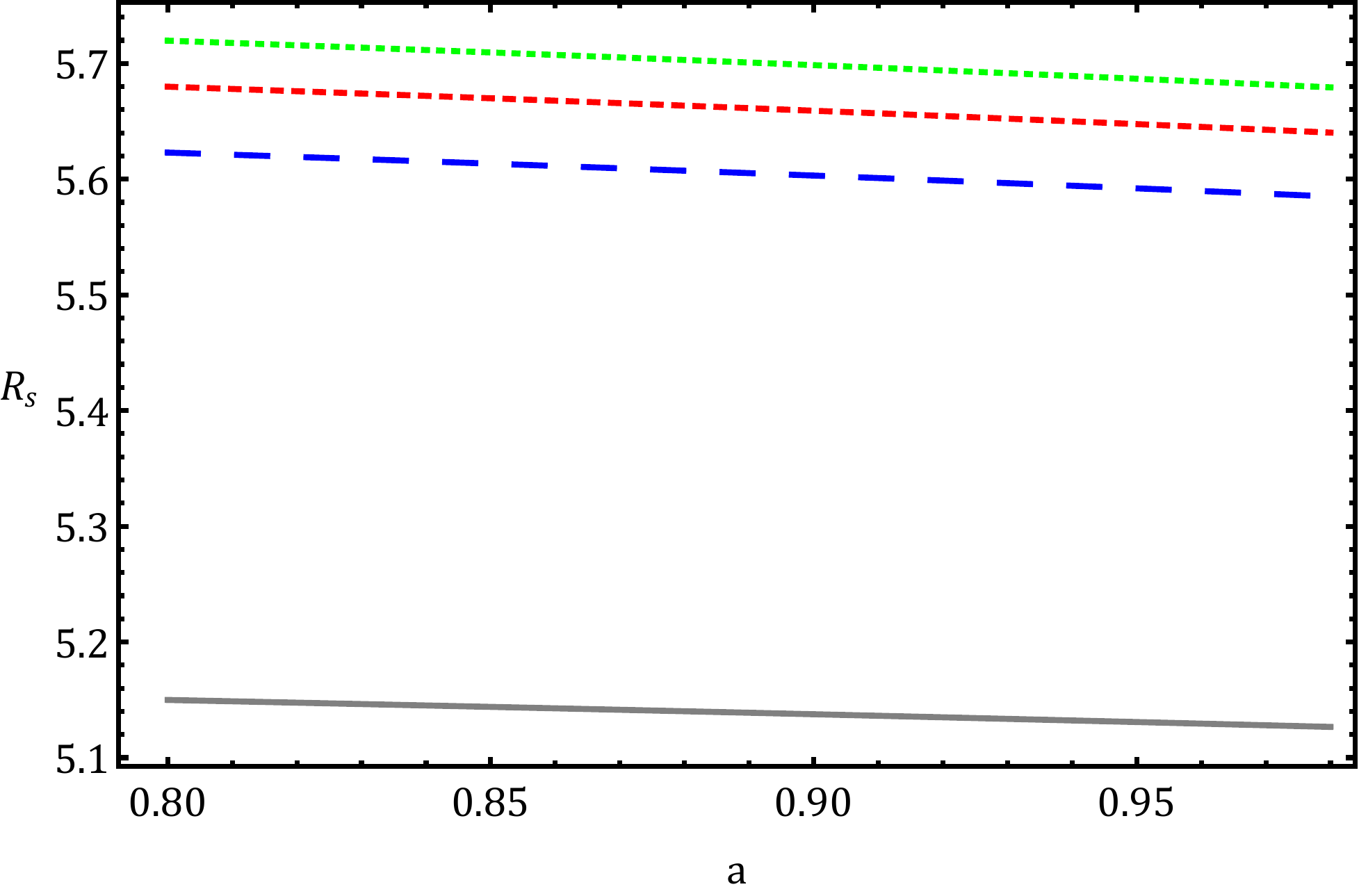}~~~
	\includegraphics[scale=0.3]{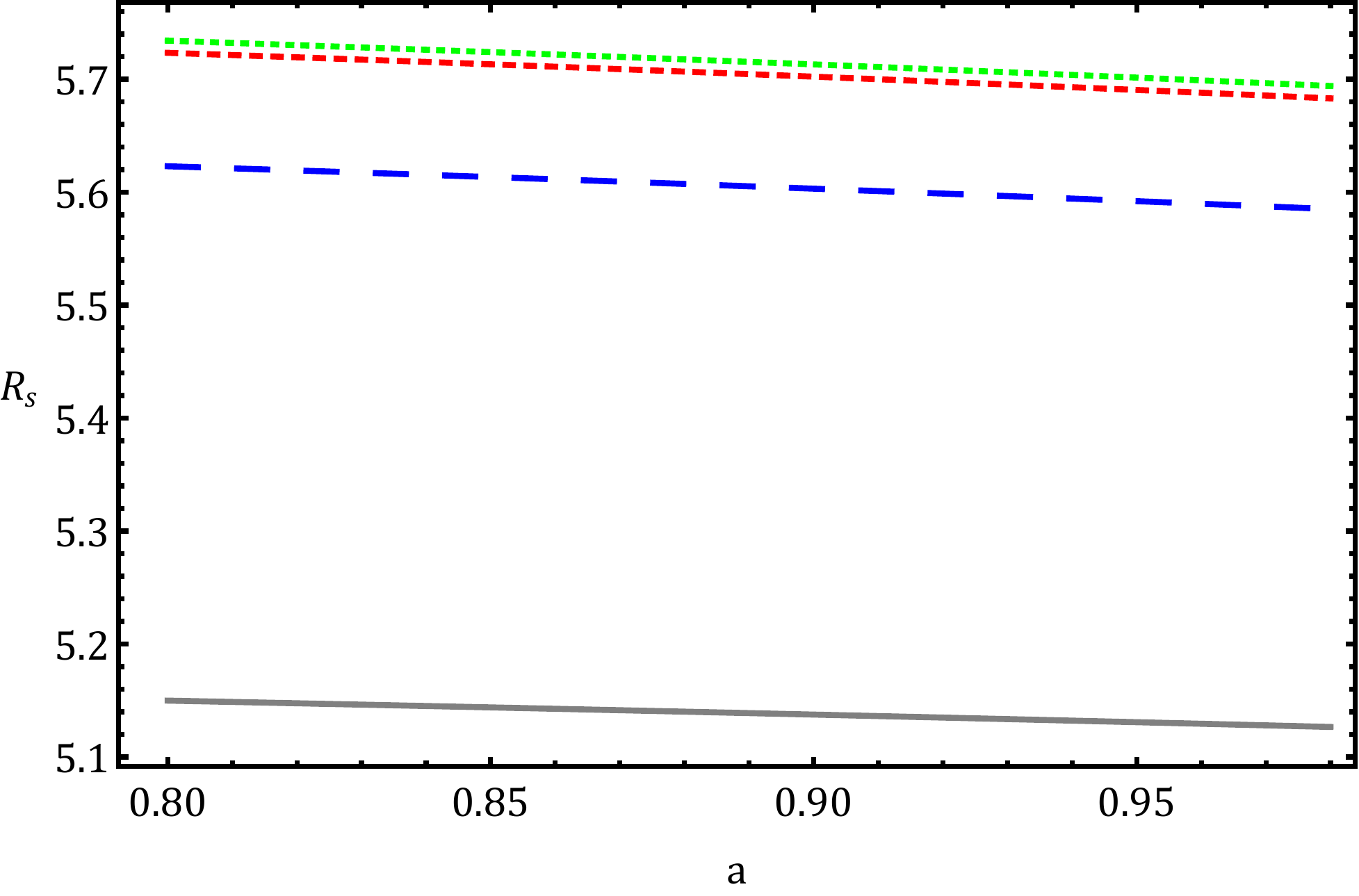}
	\includegraphics[scale=0.33]{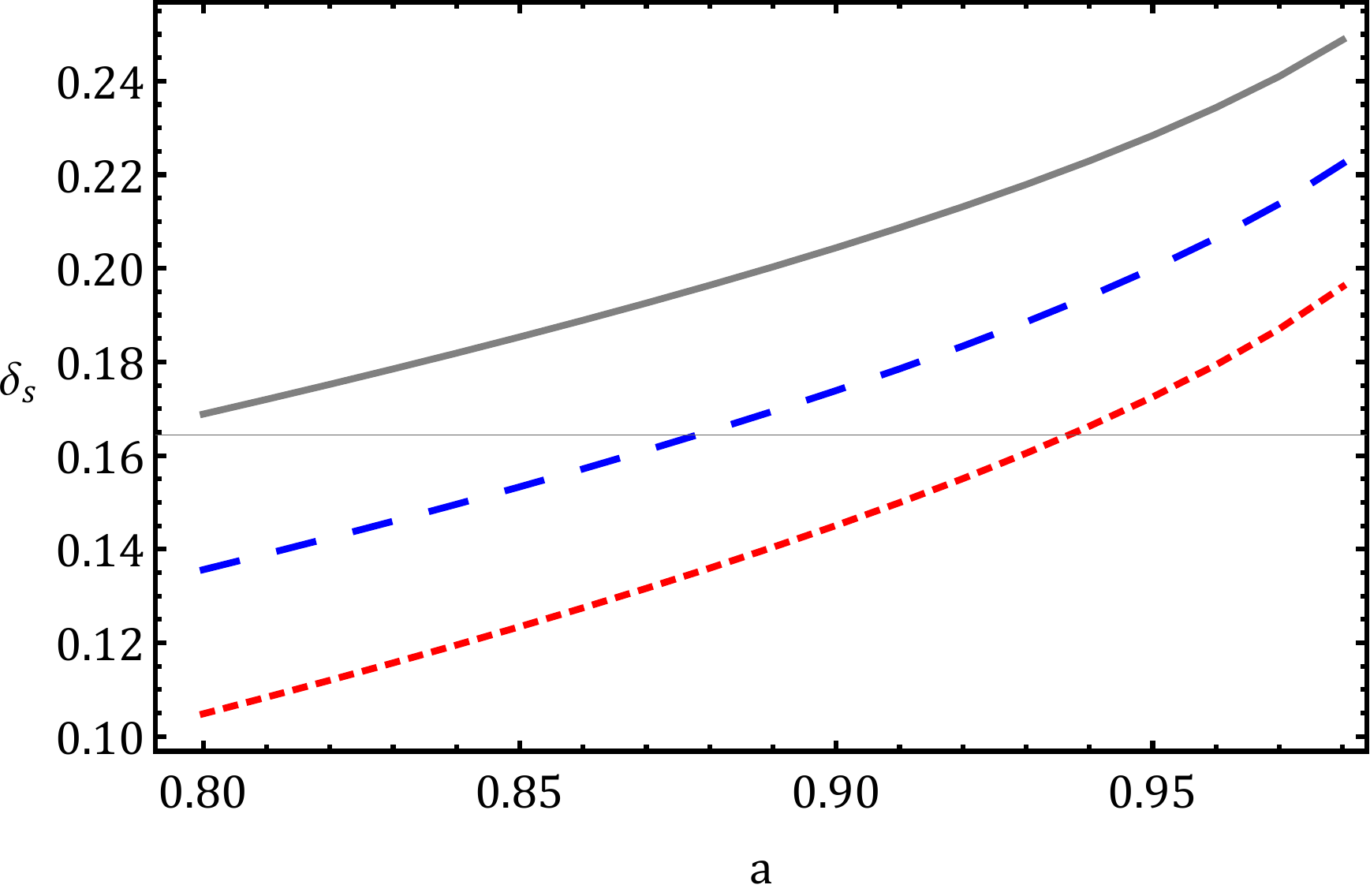}~~~
	\includegraphics[scale=0.33]{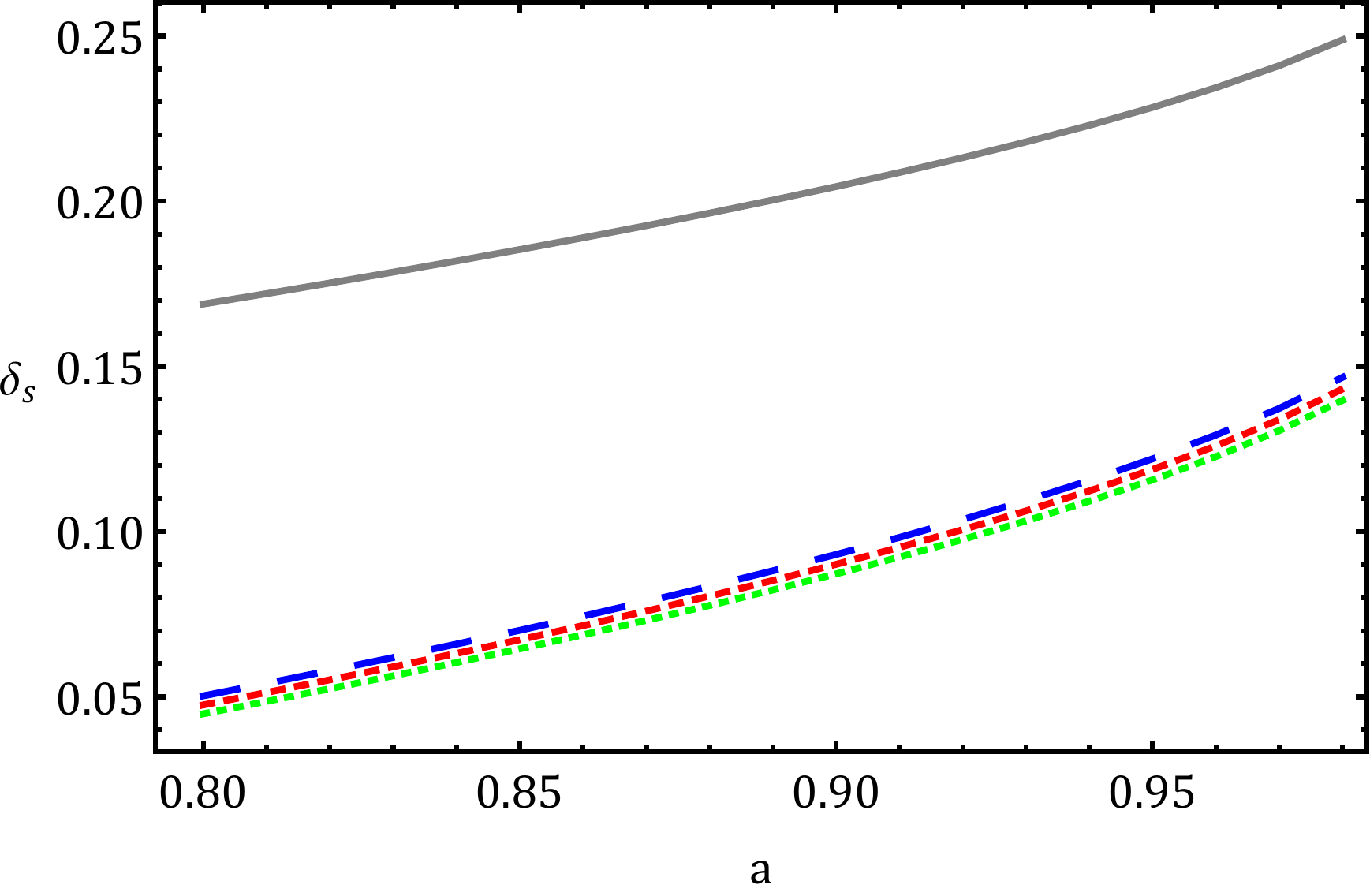}~~~
	\includegraphics[scale=0.33]{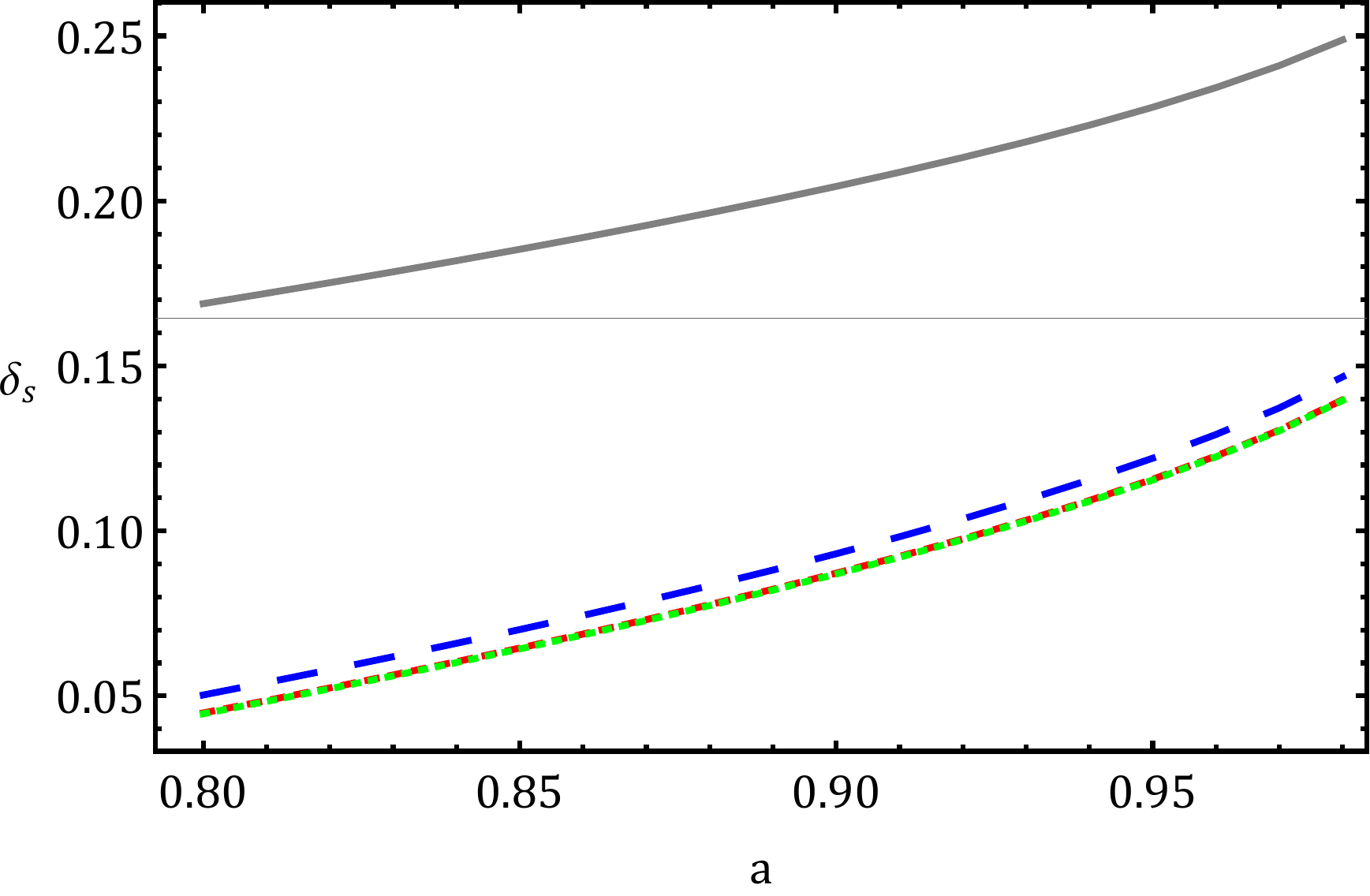}
	\caption{\textit{The plots $R_s-a$ (\textbf{Top row}) and $\delta_s-a$ (\textbf{bottom row}) for BH surrounded by a plasma with homogeneous power-law density distribution  ($h=0$) with different sets of $\{\chi_k,\chi_{ap},\chi_\varphi\}$. \textbf{Top left panel:} $\{0,0,0\}$ (gray-solid), $\{0.3,0,0\}$ (blue-long dashed ), $\{0.5,0,0\}$ (red-dashed). \textbf{Top middle panel:} $\{0,0,0\}$ (gray-solid), $\{0.5,0,0\}$ (blue-long dashed), $\{0.5,0.3,0.1\}$ (red-dashed), $\{0.5,0.5,0.1\}$ (green-dotted). \textbf{Top right panel:} $\{0,0,0\}$ (gray-solid), $\{0.5,0,0\}$ (blue-long dashed), $\{0.5,0.5,0.2\}$ (red-dashed), $\{0.5,0.5,0.4\}$ (green-dotted). \textbf{Bottom left panel:} $\{0,0,0\}$ (gray-solid), $\{0.15,0,0\}$ (blue-long dashed), $\{0.3,0,0\}$ (red-dashed). \textbf{Bottom middle panel:} $\{0,0,0\}$ (gray-solid), $\{0.5,0,0\}$ (blue-long dashed), $\{0.5,0.1,0.1\}$ (red-dashed), $\{0.5,0.15,0.1\}$ (green-dotted). \textbf{Bottom right panel:} $\{0,0,0\}$ (gray-solid), $\{0.5,0,0\}$ (blue-long dashed), $\{0.5,0.15,0.15\}$ (red-dashed), $\{0.5,0.15,0.3\}$ (green-dotted). We have set $\theta_0=\pi/3$ and $M=1$.}}
	\label{RD1}
\end{figure}
\begin{figure}[ht]
	\includegraphics[scale=0.3]{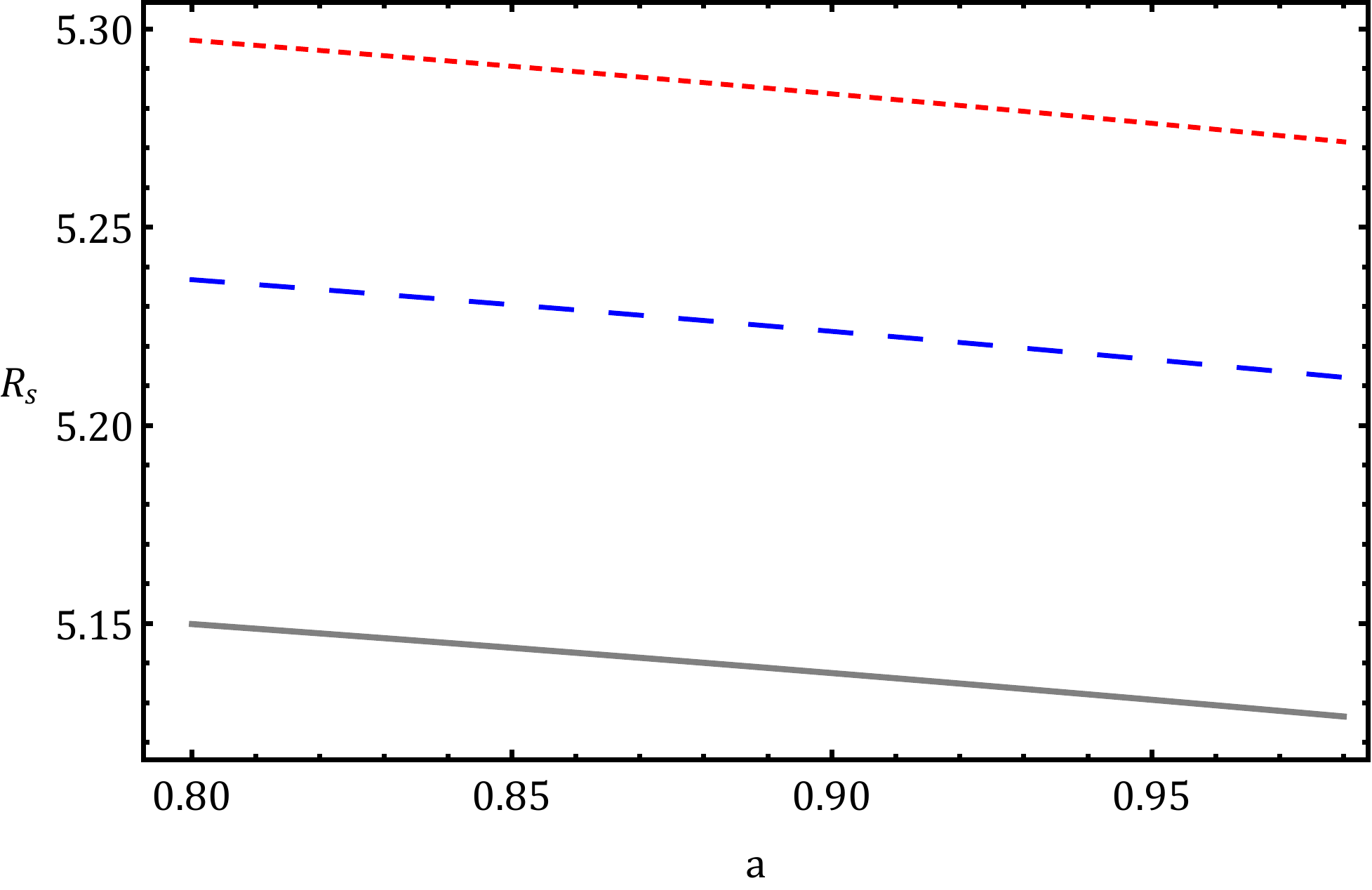}~~~
	\includegraphics[scale=0.3]{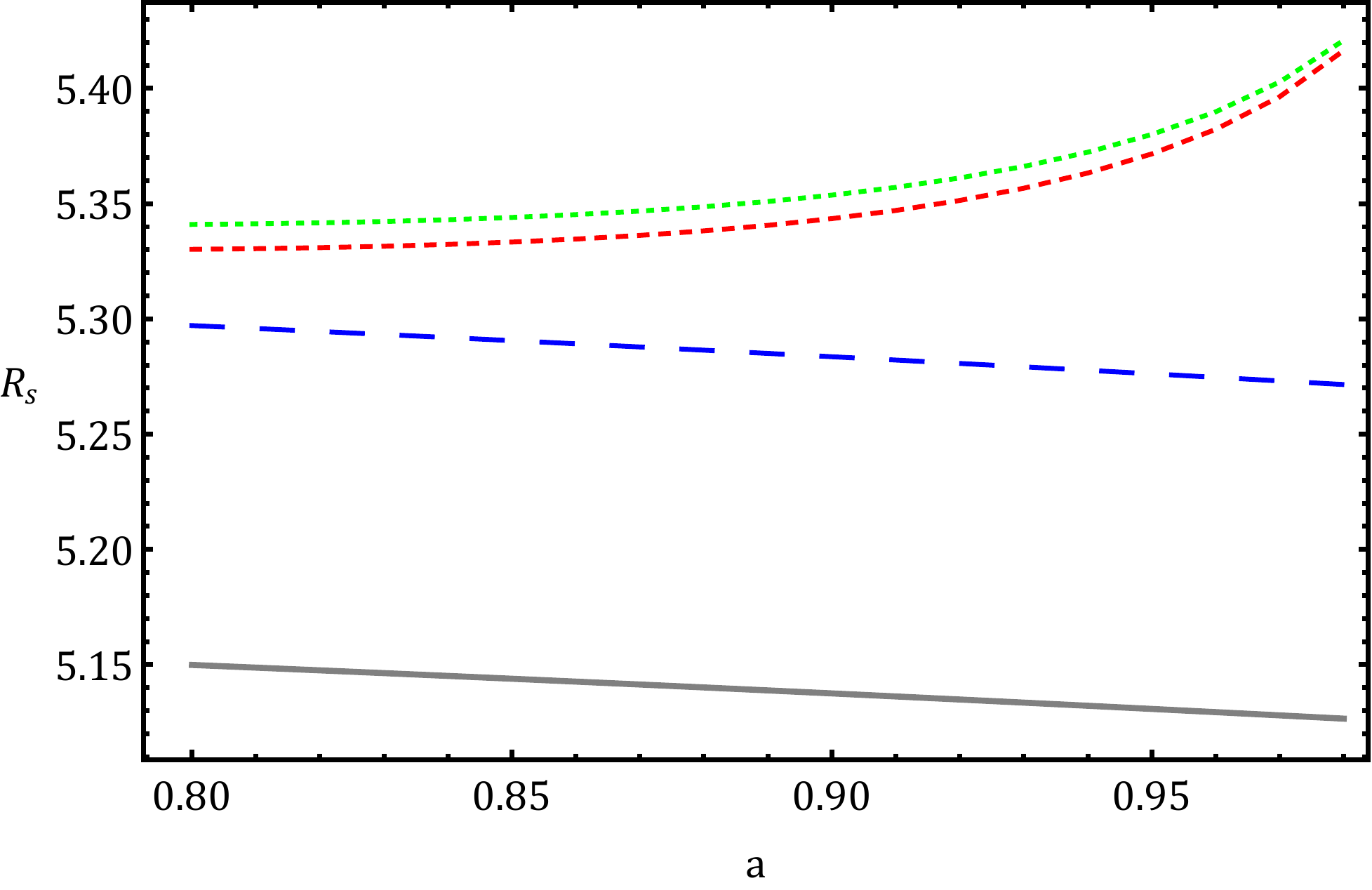}~~~
	\includegraphics[scale=0.3]{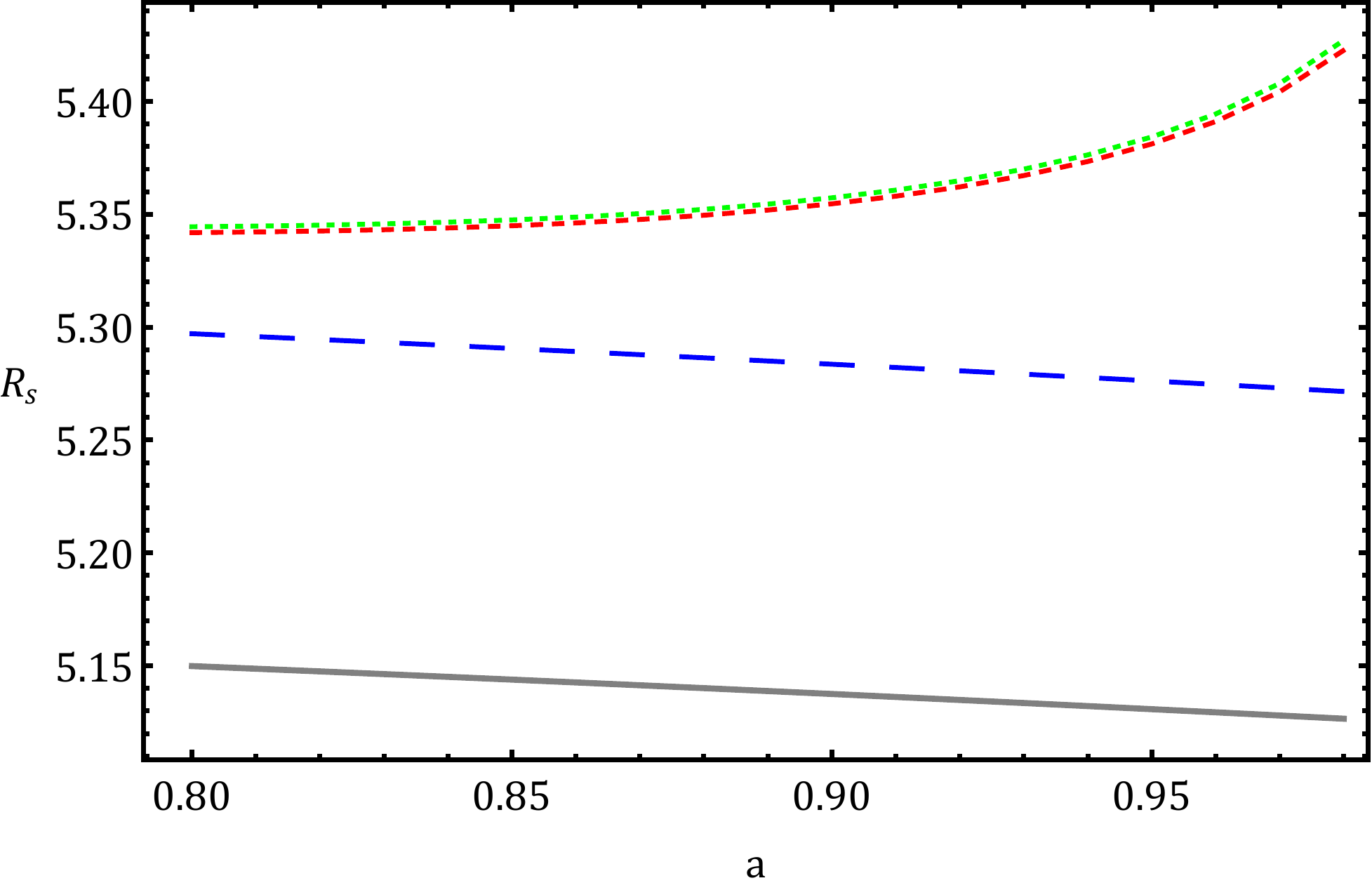}
	\includegraphics[scale=0.33]{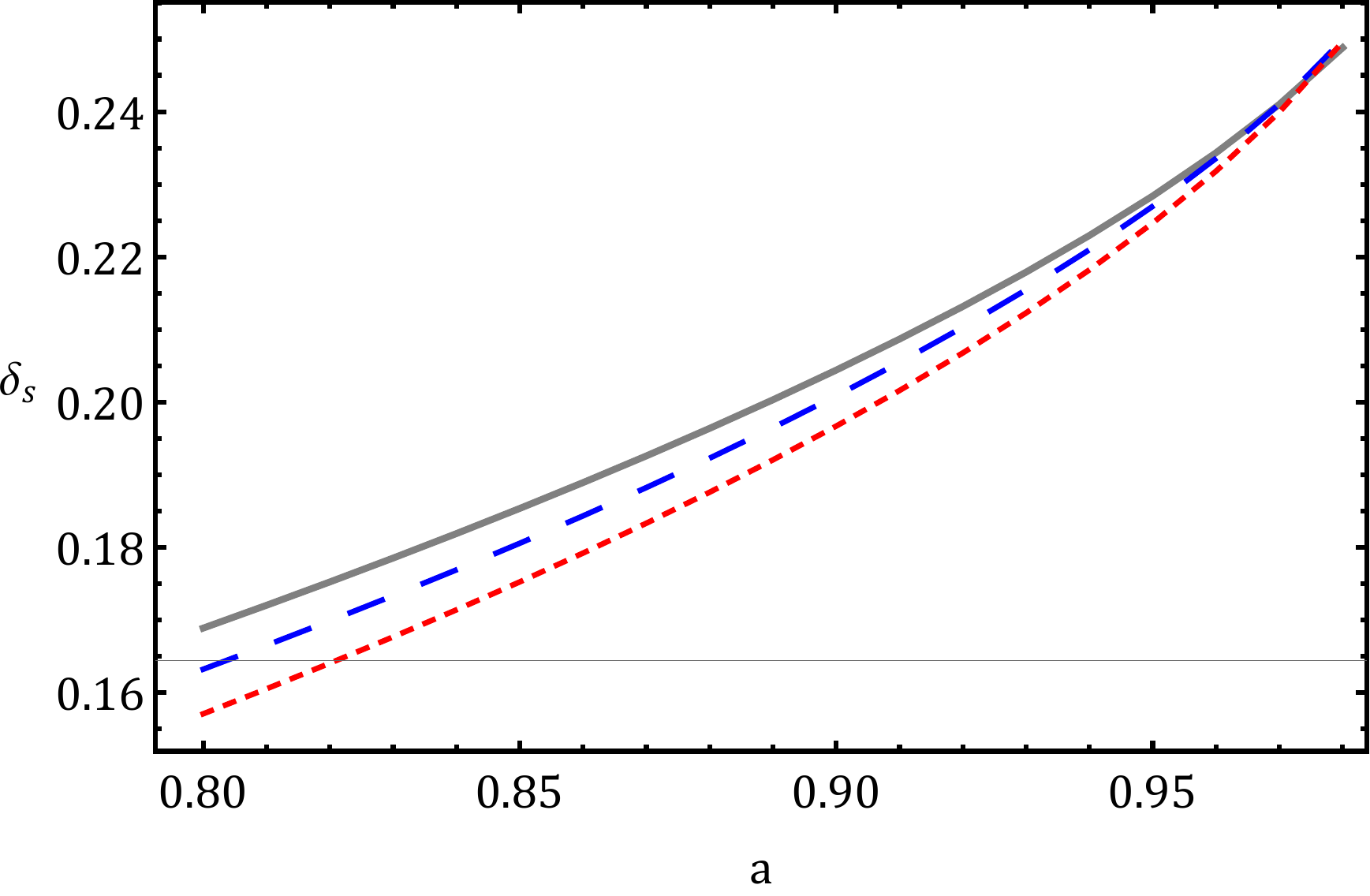}~~~
	\includegraphics[scale=0.33]{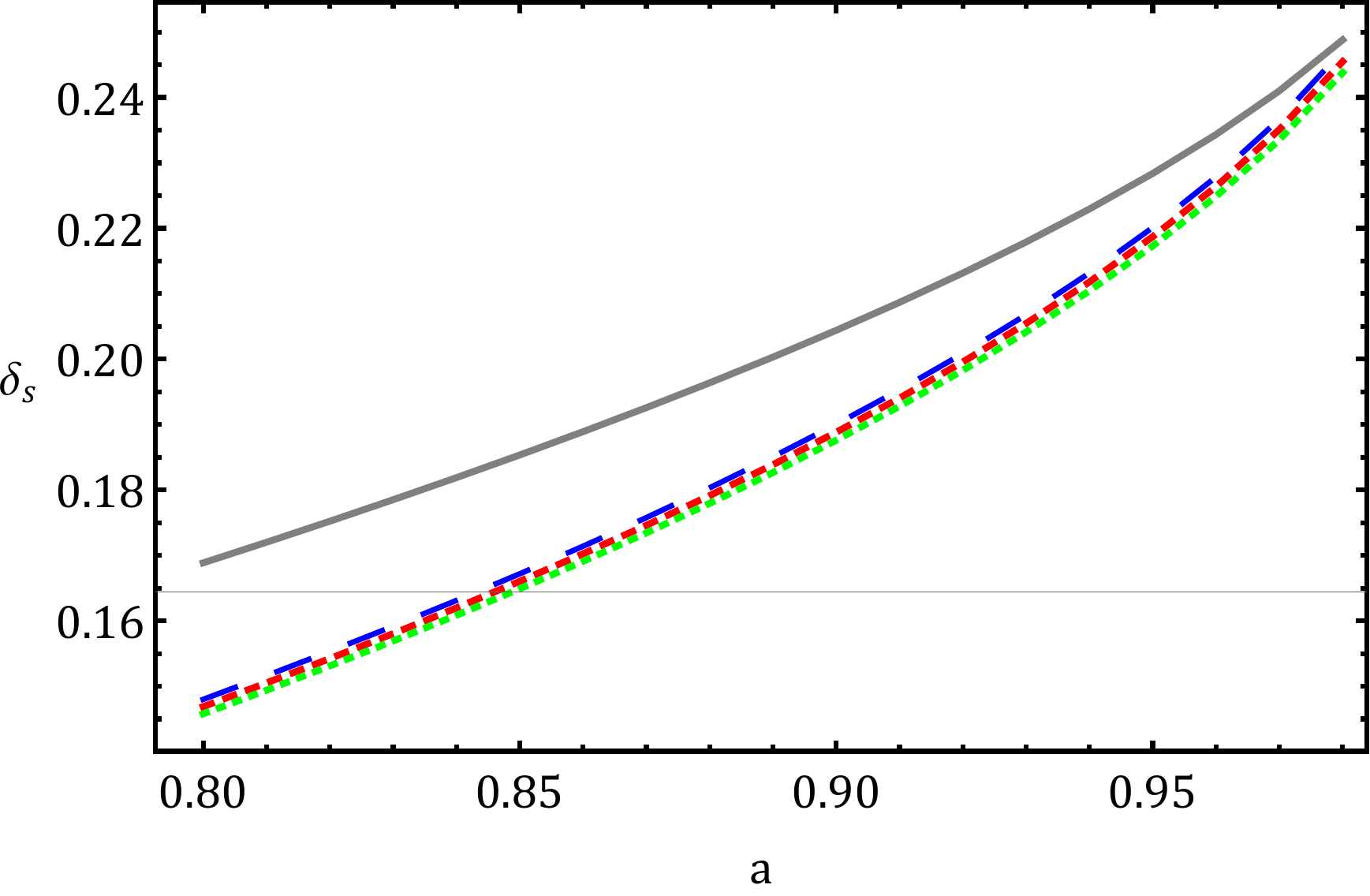}~~~
	\includegraphics[scale=0.33]{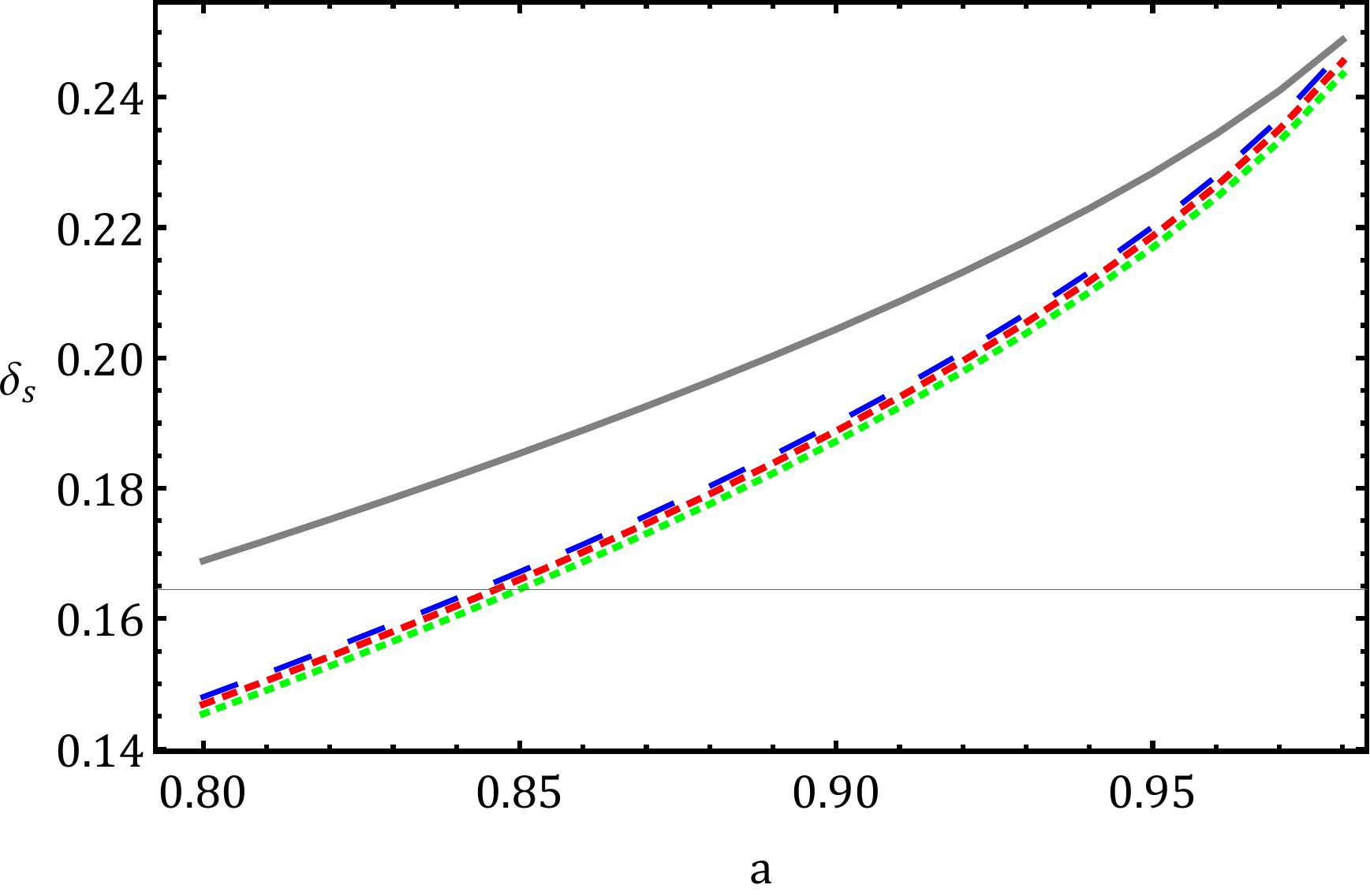}
	\caption{\textit{Same as Fig. \ref{RD1} but for a plasma with in-homogeneous power-law density distribution  ($h=1$).}}
	\label{RD2}
\end{figure}
\begin{figure}[ht]
	\includegraphics[scale=0.3]{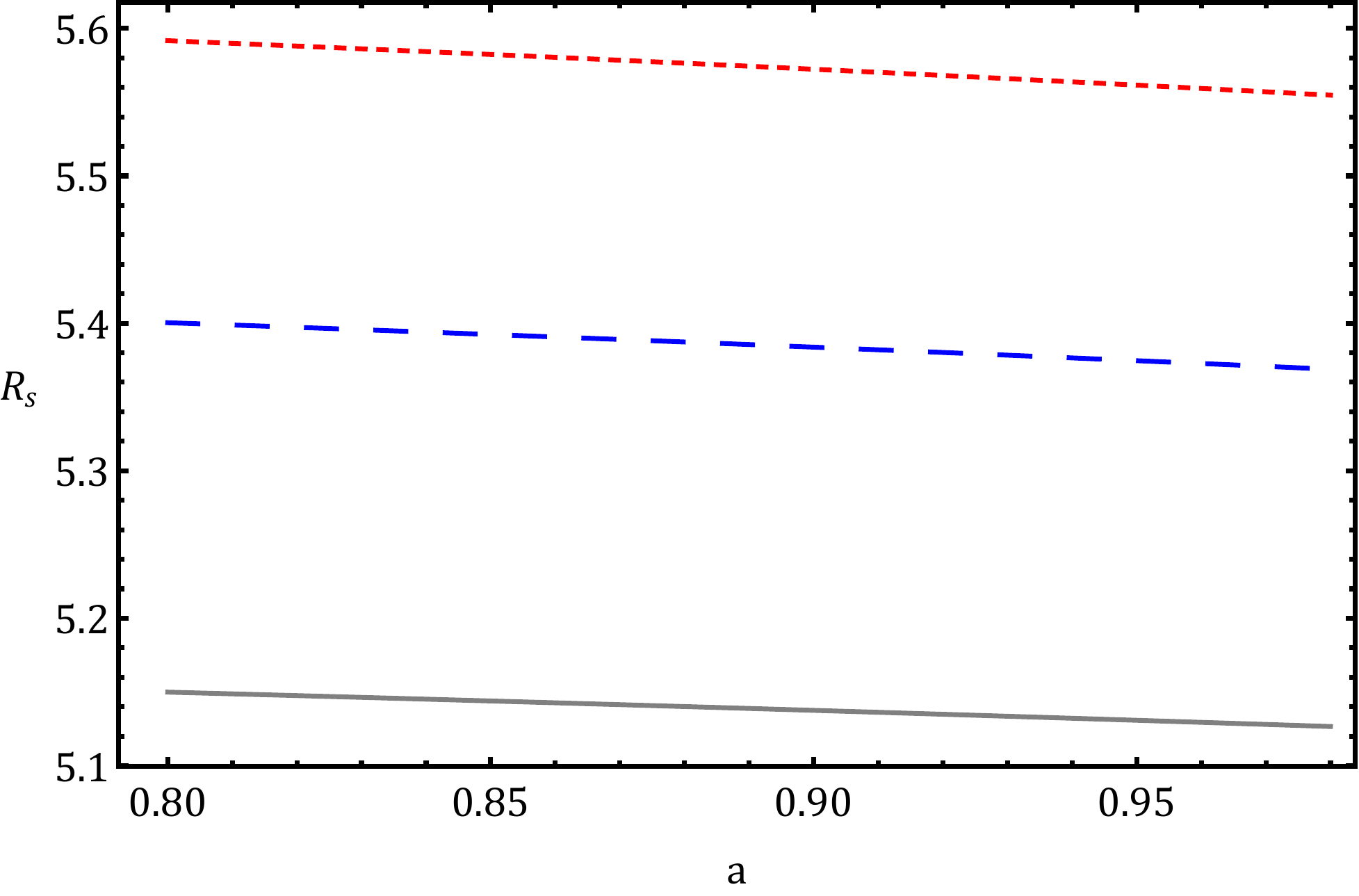}~~~
	\includegraphics[scale=0.3]{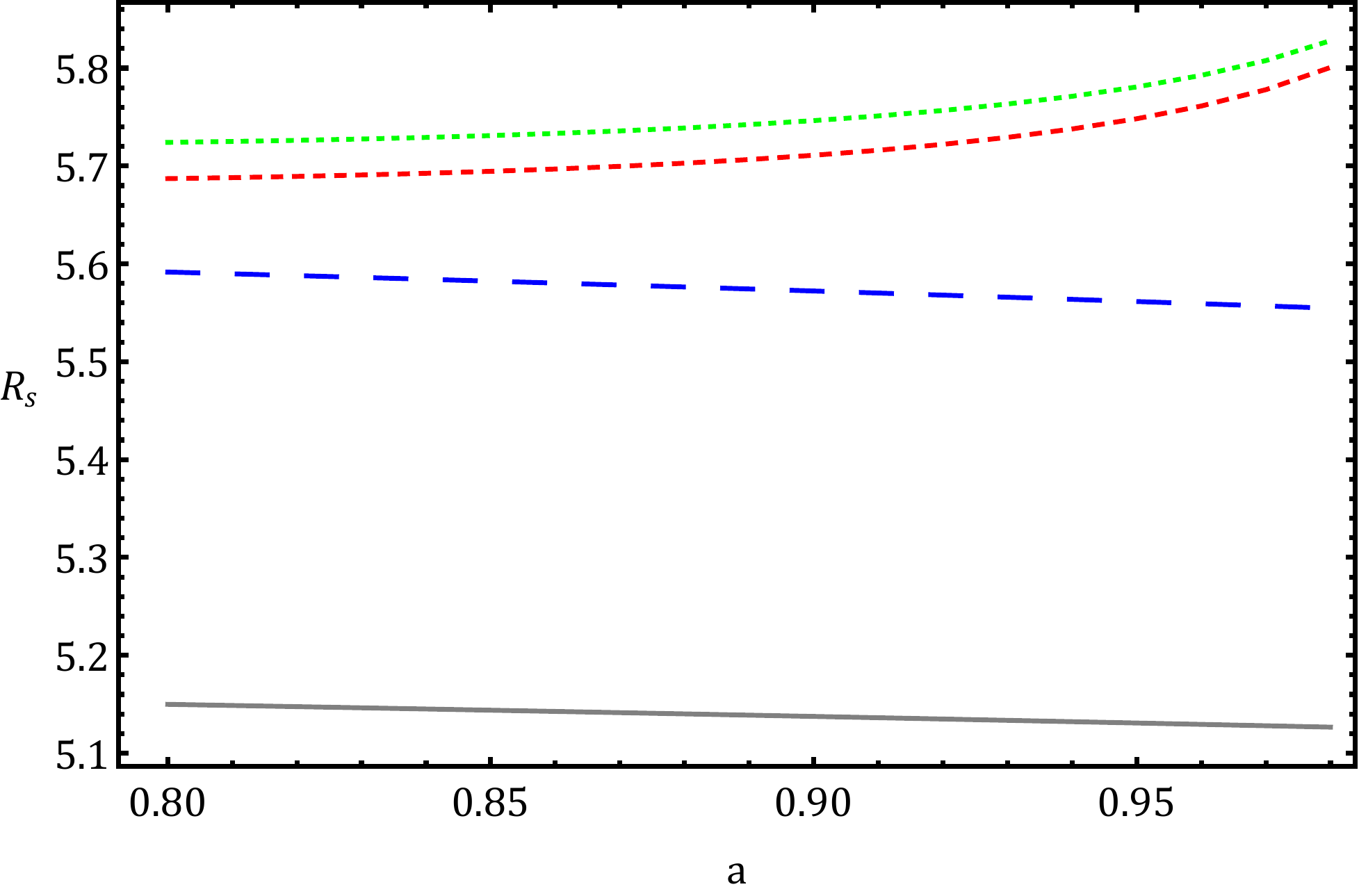}~~~
	\includegraphics[scale=0.3]{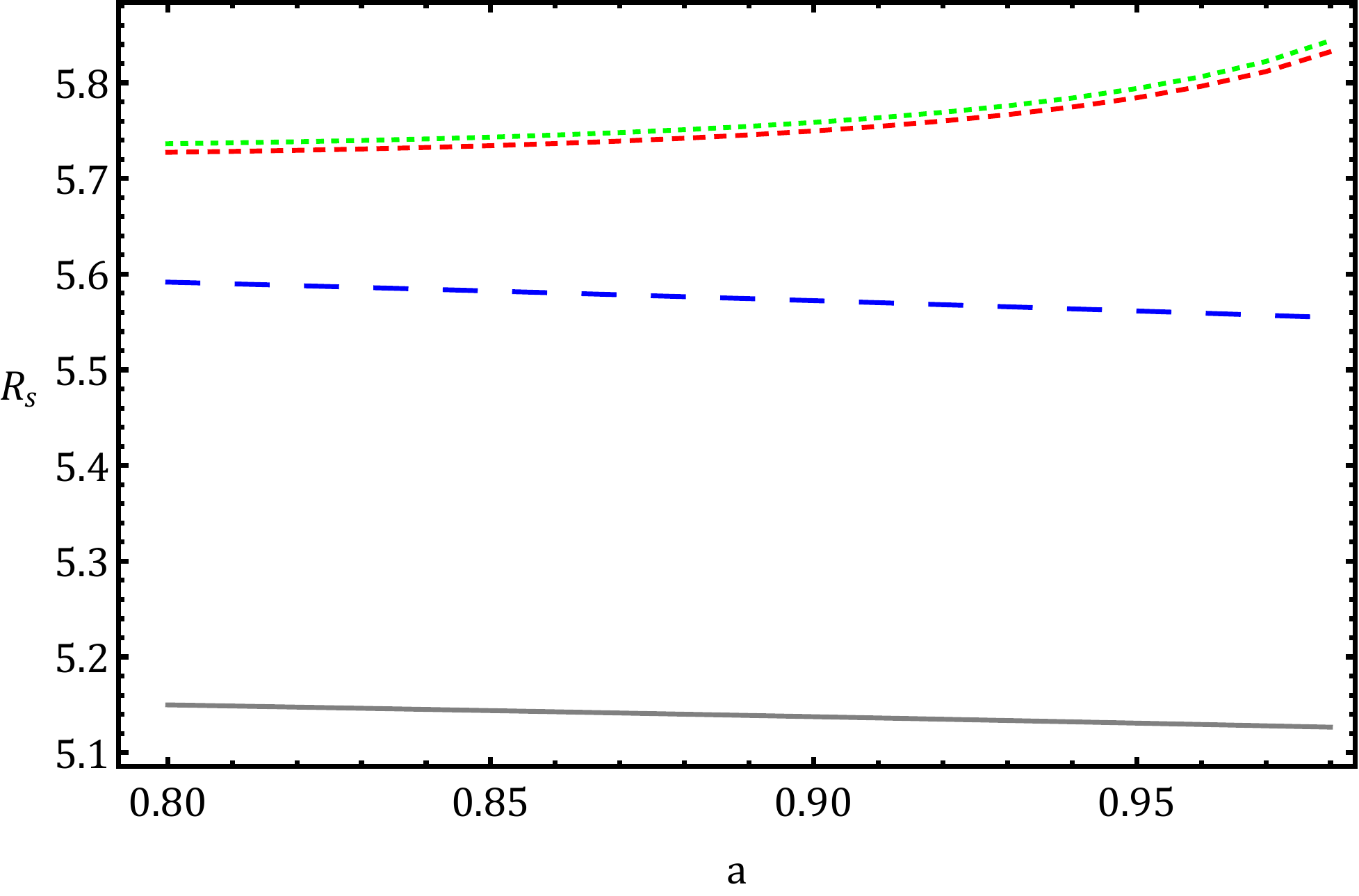}
	\includegraphics[scale=0.33]{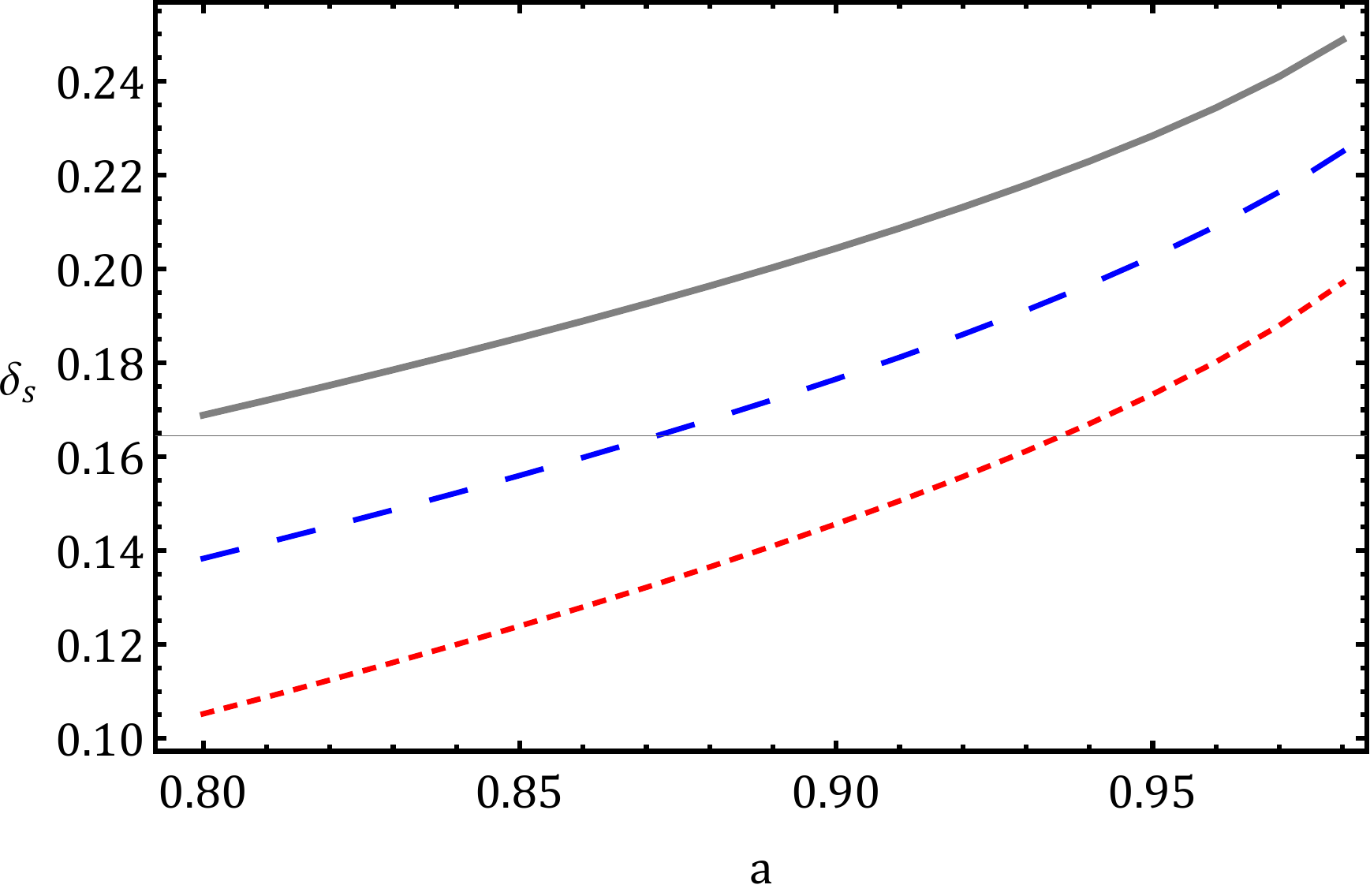}~~~
	\includegraphics[scale=0.33]{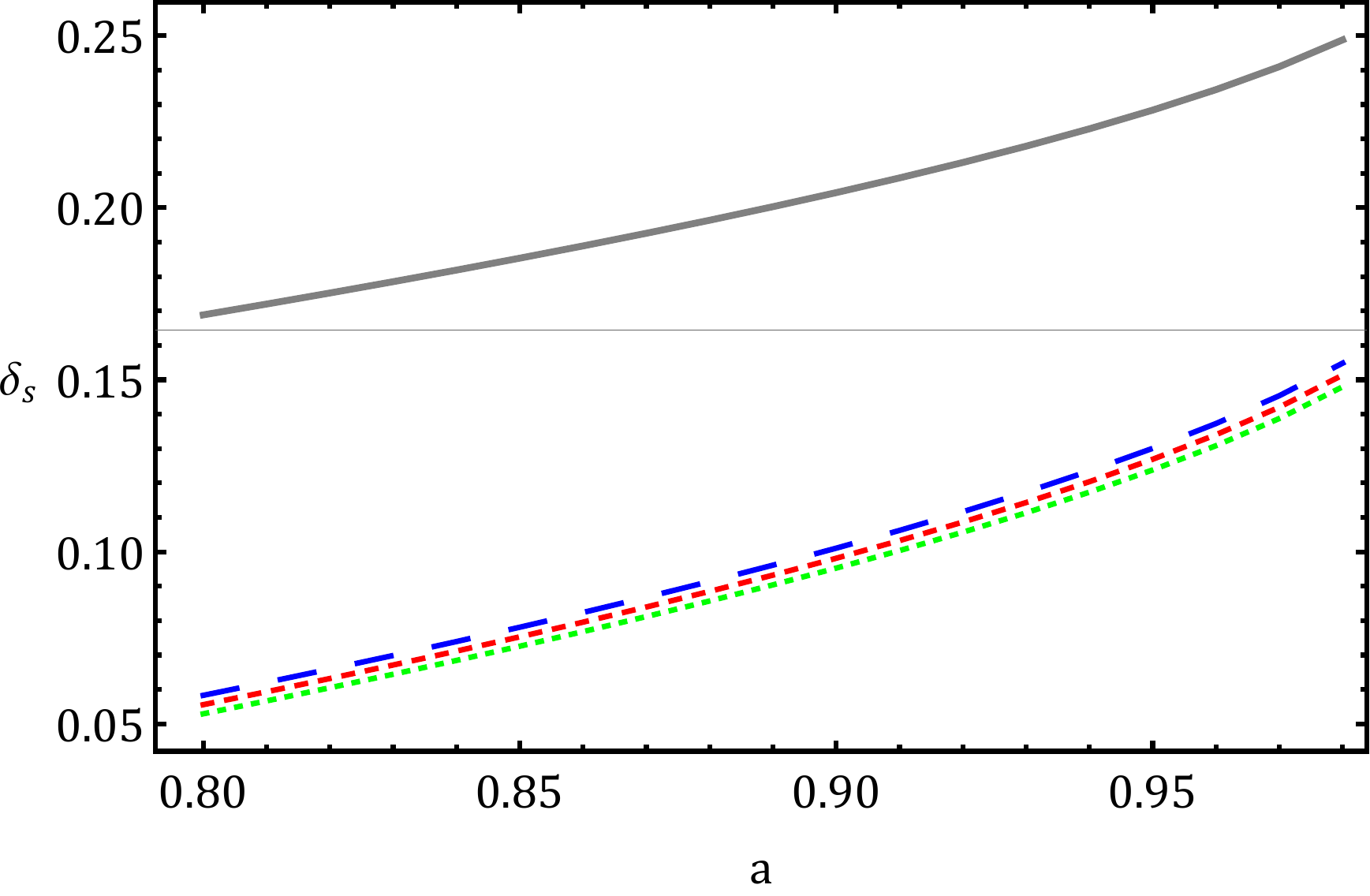}~~~
	\includegraphics[scale=0.33]{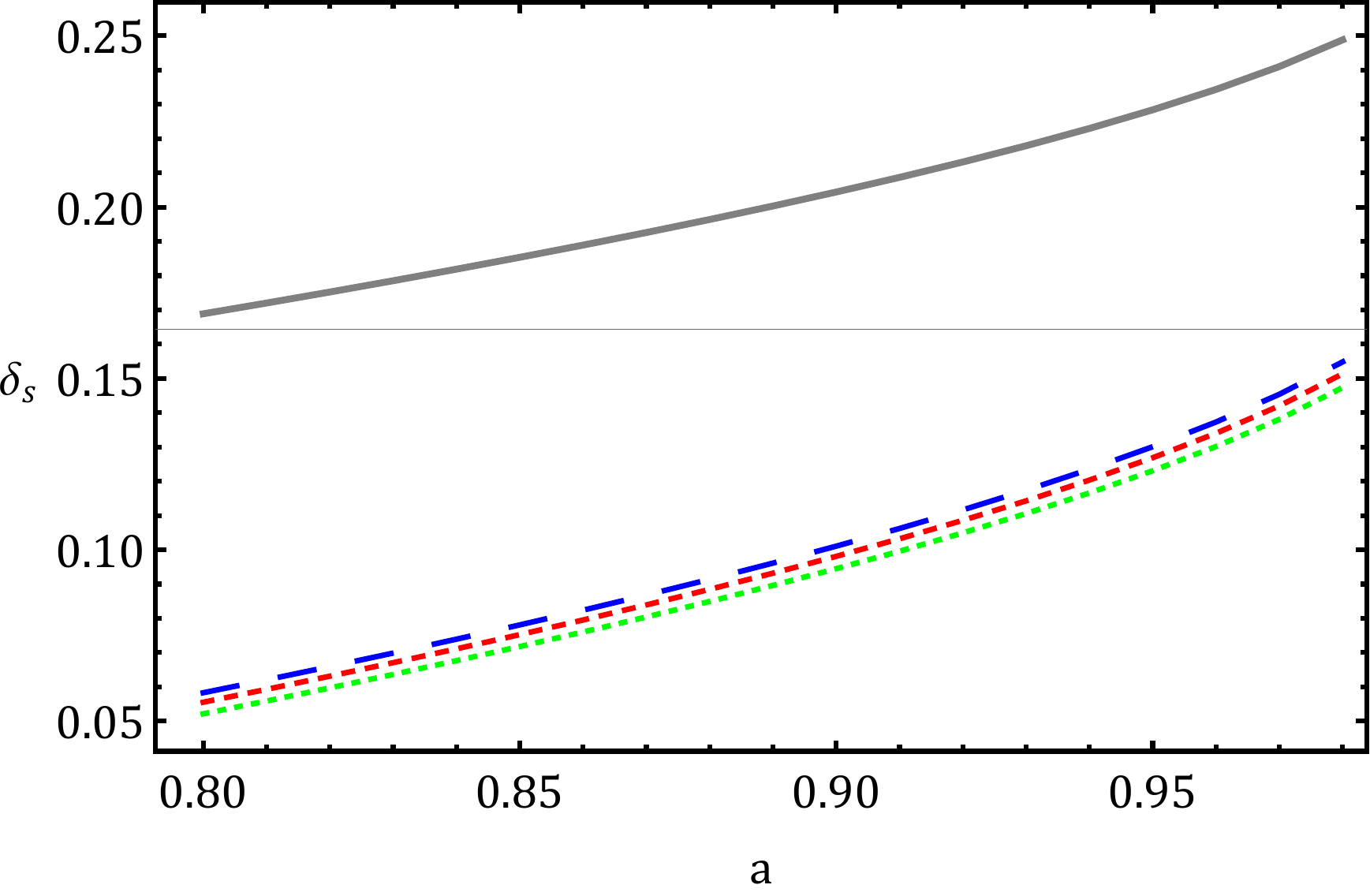}
	\caption{\textit{Same as Fig. \ref{RD1} but for a plasma with exponentially density distribution.}}
	\label{RD3}
\end{figure}
\begin{figure}[ht]
	\includegraphics[scale=0.33]{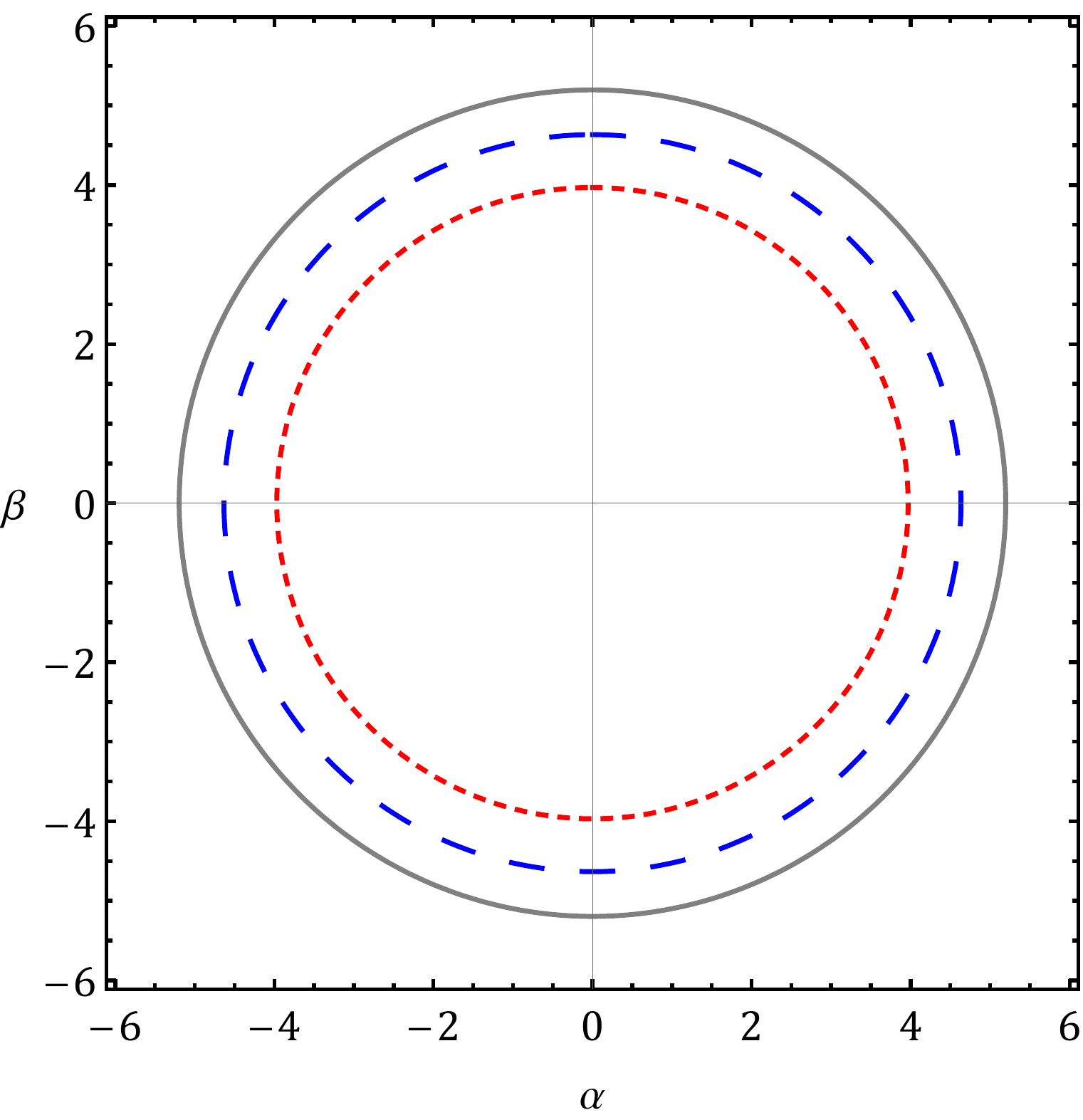}~~~
	\includegraphics[scale=0.33]{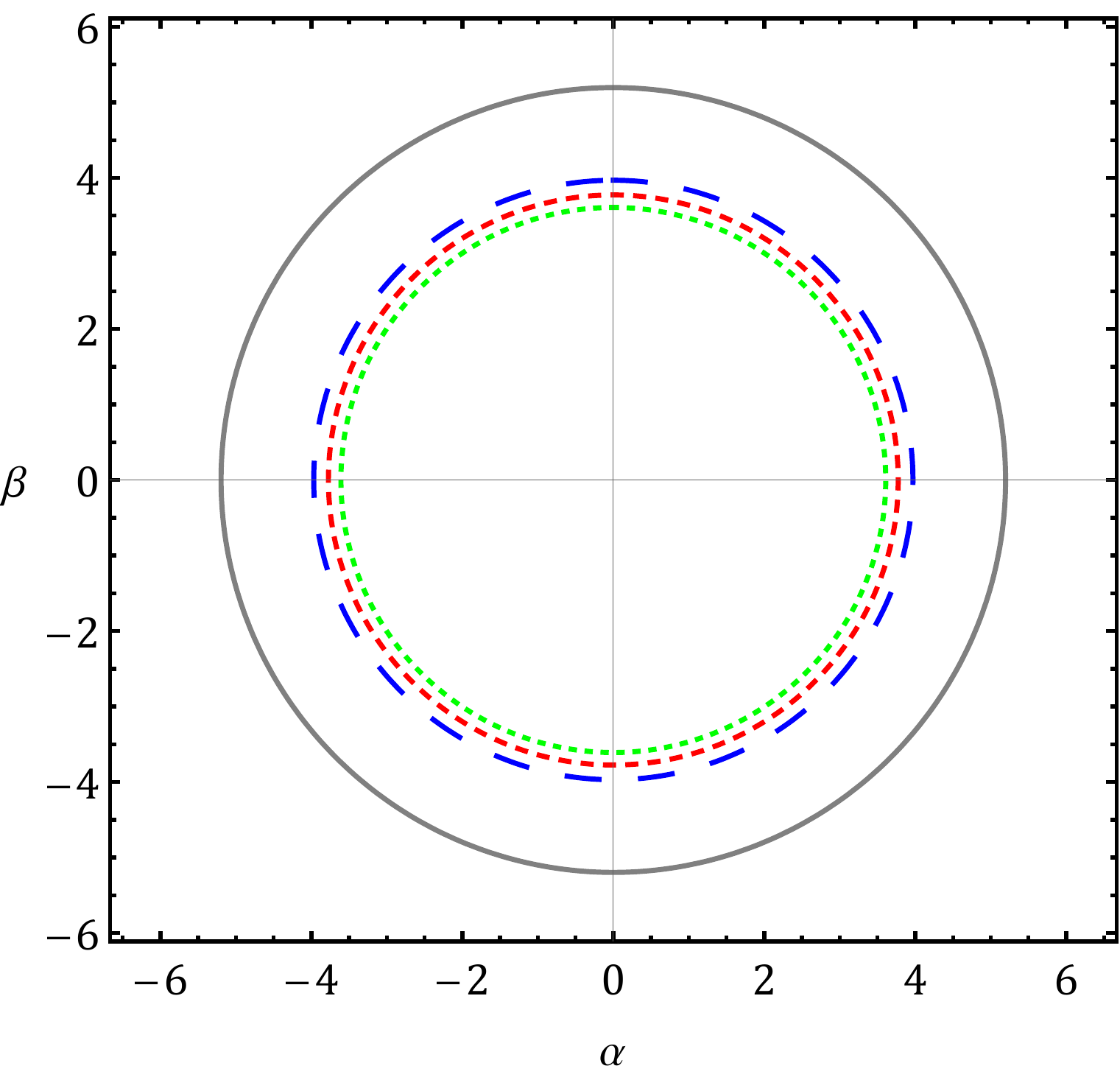}~~~
	\includegraphics[scale=0.33]{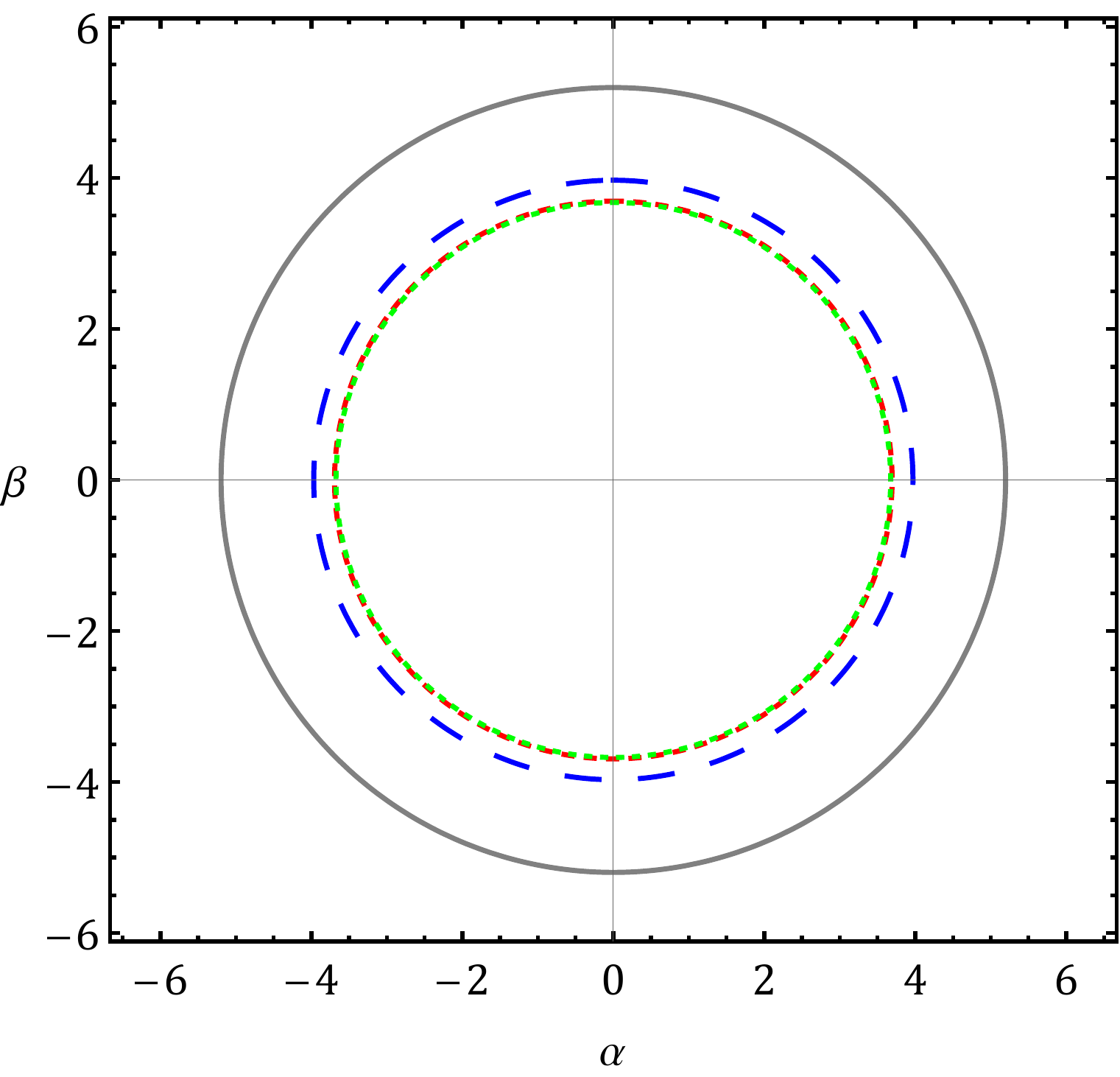}
	\caption{\textit{The shadow of Schwarzschild BH surrounded by a plasma with homogeneous power-law density distribution  ($h=0$) for different sets of $\{\chi_k,\chi_{ap},\chi_\varphi\}$. \textbf{Left:} $\{0,0,0\}$ (gray-solid), $\{0.25,0,0\}$ (blue-long dashed ), $\{0.5,0,0\}$ (red-dashed). \textbf{Middle:} $\{0,0,0\}$ (gray-solid), $\{0.5,0,0\}$ (blue-long dashed), $\{0.5,0.35,0.1\}$ (red-dashed), $\{0.5,0.7,0.1\}$ (green-dotted). \textbf{Right:} $\{0,0,0\}$ (gray-solid), $\{0.5,0,0\}$ (blue-long dashed), $\{0.5,0.5,0.25\}$ (red-dashed), $\{0.5,0.5,0.5\}$ (green-dotted).}}
	\label{Shadow4}
\end{figure}
\begin{figure}[ht]
	\includegraphics[scale=0.33]{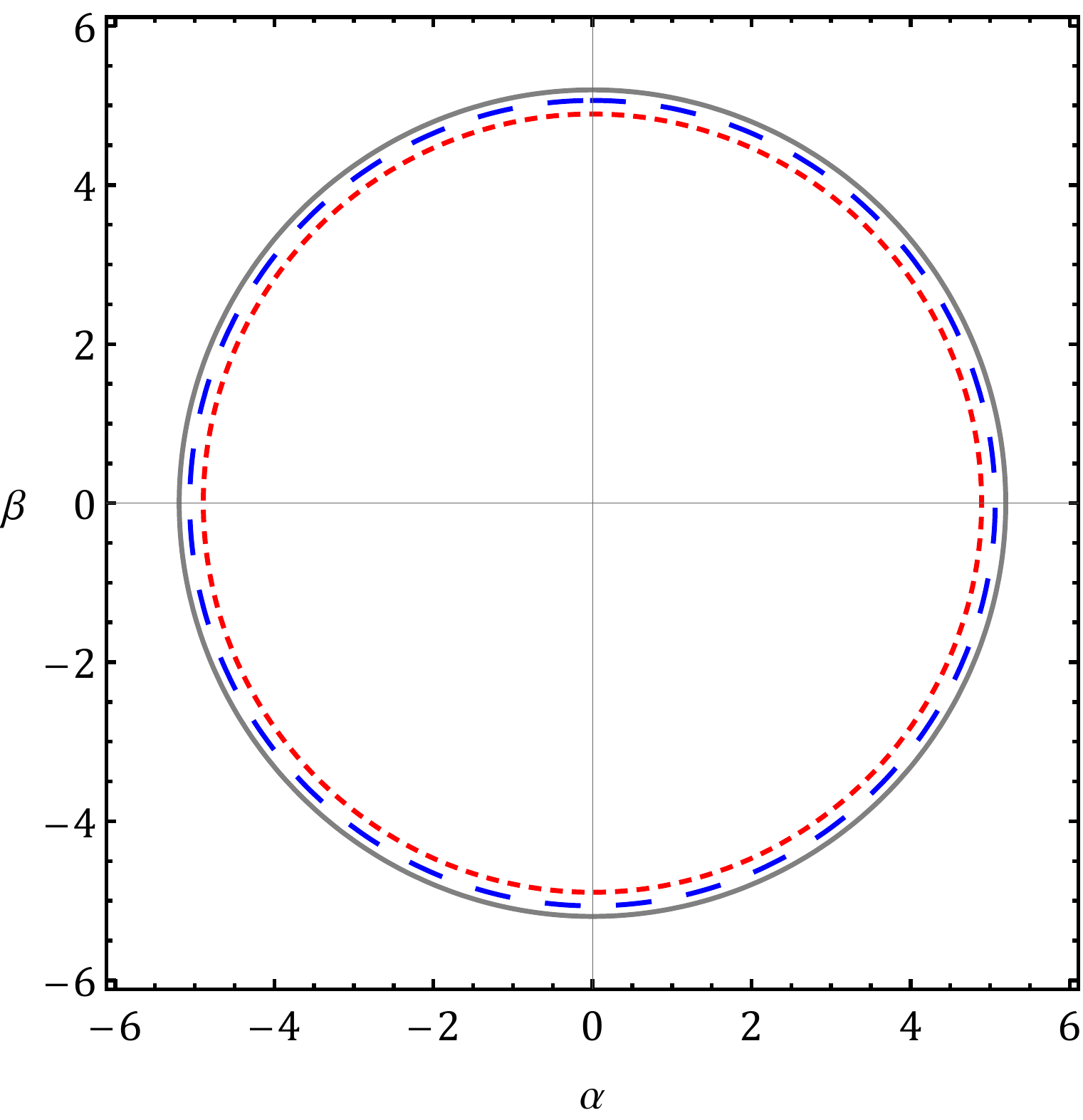}~~~
	\includegraphics[scale=0.33]{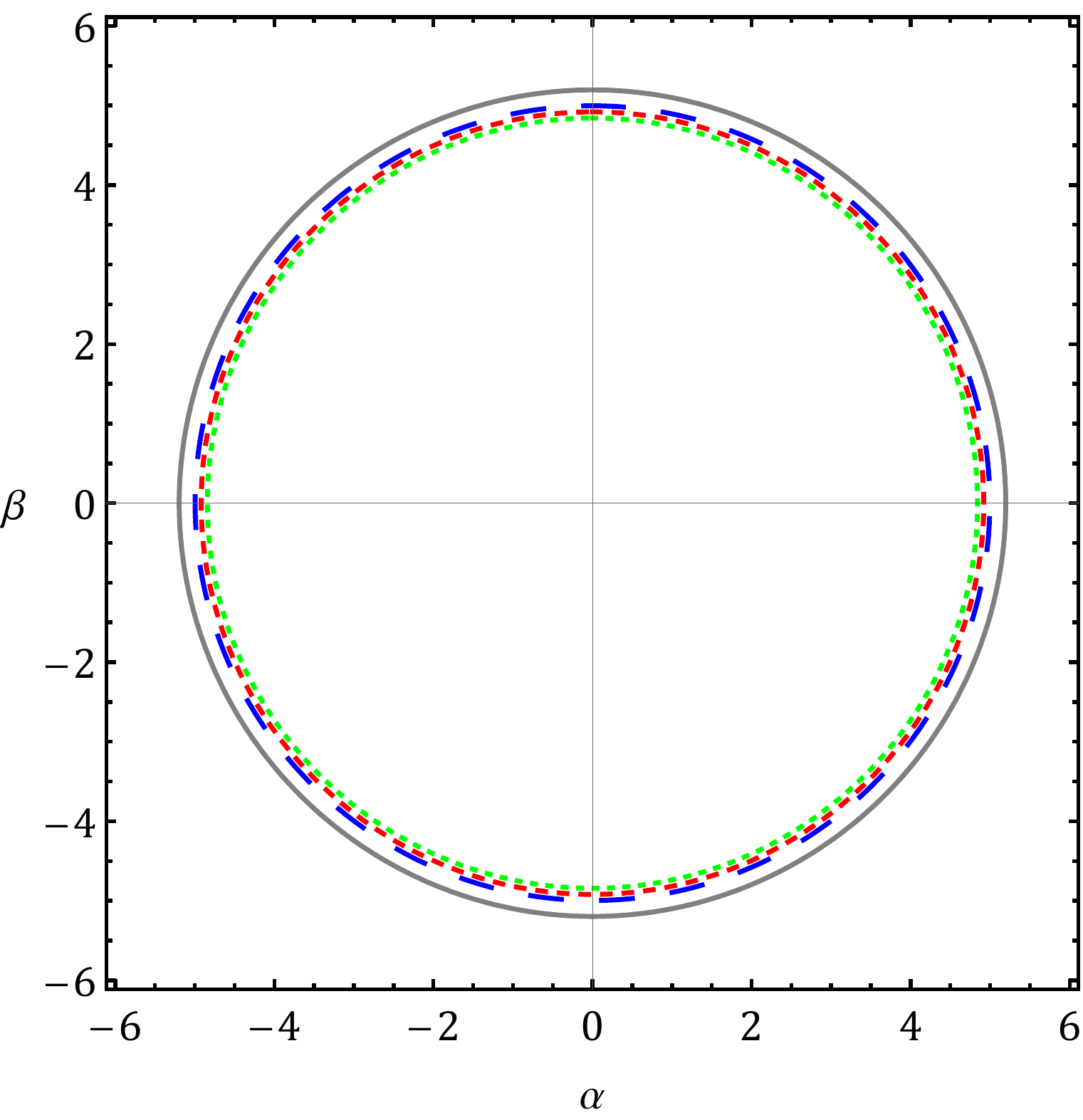}~~~
	\includegraphics[scale=0.33]{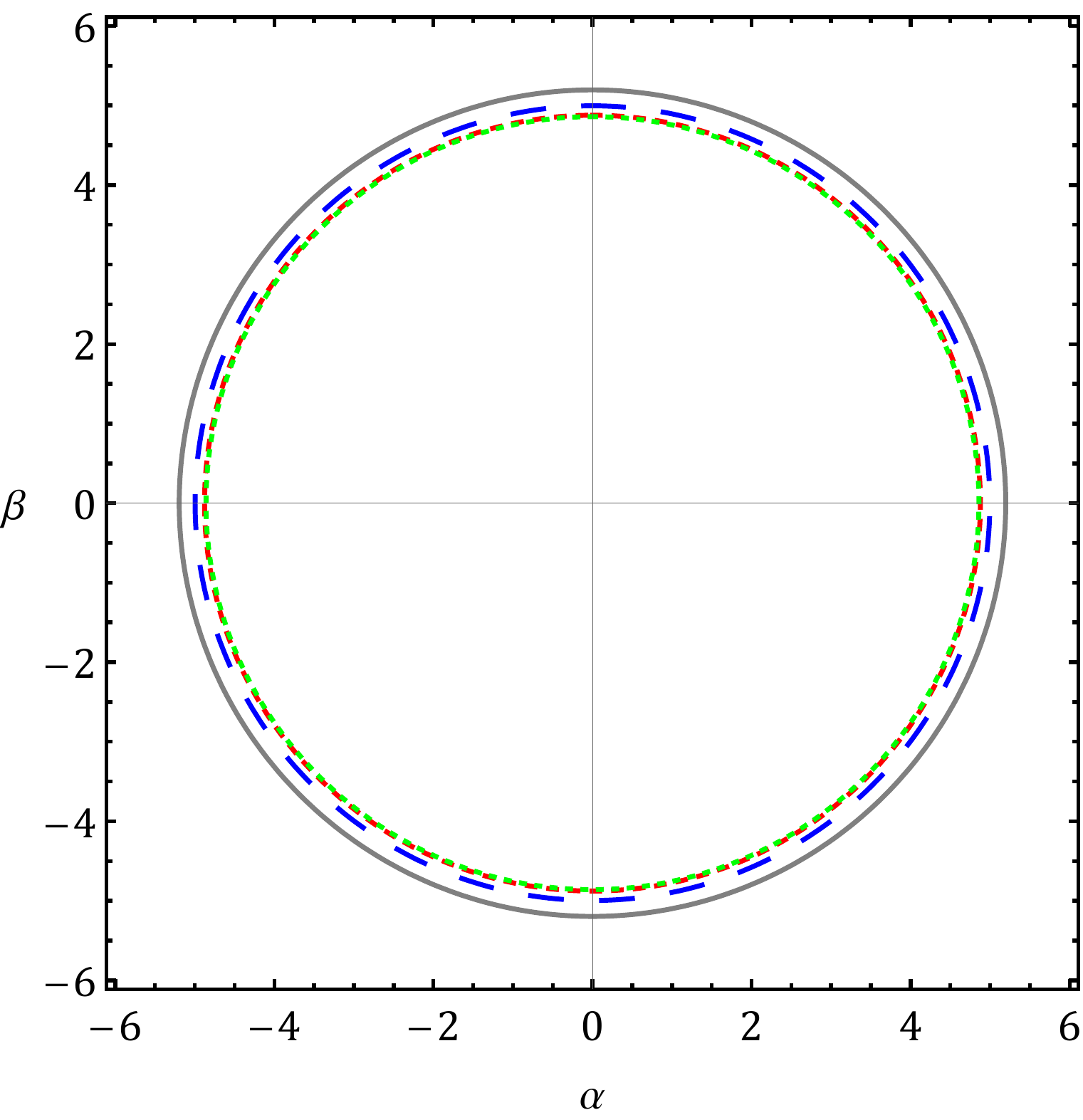}
	\caption{\textit{The shadow of Schwarzschild BH surrounded by a plasma with inhomogeneous power-law density distribution  ($h=1$) for different sets of $\{\chi_k,\chi_{ap},\chi_\varphi\}$. \textbf{Left:} $\{0,0,0\}$ (gray-solid), $\{0.35,0,0\}$ (blue-long dashed ), $\{0.7,0,0\}$ (red-dashed). \textbf{Middle:} $\{0,0,0\}$ (gray-solid), $\{0.5,0,0\}$ (blue-long dashed), $\{0.5,0.35,0.1\}$ (red-dashed), $\{0.5,0.7,0.1\}$ (green-dotted). \textbf{Right:} $\{0,0,0\}$ (gray-solid), $\{0.5,0,0\}$ (blue-long dashed), $\{0.5,0.5,0.25\}$ (red-dashed), $\{0.5,0.5,0.5\}$ (green-dotted).} }
	\label{Shadow5}
\end{figure}
By seeing the shadow curves displayed in Figs~\ref{Shadow1}-\ref{Shadow3}, it seems that due to their close behaviors the curves to be not goodly distinguishable. Besides, they also do not give us information on the role of the changes of rotation parameters. To fix this, here we employ two astronomical observables: the shadow radius $R_s$ and the distortion parameter $\delta_s$, which are widely used in the literature relevant to the study of BH shadow.
The former is the radius of the reference circle passing via three points: the top one $B(\alpha_t, \beta_t)$, the bottom one  $D(\alpha_b, \beta_b)$ and the most right one $A(\alpha_r, 0)$.
The letter denotes the distance between the most left position (C) of the shadow and the most right position (E) of the reference circle, as one can see in Fig. \ref{shadow_ref}. More exactly, the points $C(\alpha_p, 0)$ and $E(\bar{\alpha}_p, 0)$ respectively denote where the circle of the shadow and the reference circle cross the horizontal axis at the opposite side of $A(\alpha_r, 0)$. In general, $R_s$ approximately reveals the shadow size of BH, while $\delta_s$ measures its deformation with respect to the reference circle.
Now with this assumption that the geometry of the rotating BH shadow describe schematically by Fig. \ref{shadow_ref}, then these two mentioned astronomical observables obtain via the following relations \cite{Hioki:2009na}
\begin{equation}\label{RRp}
R_s = \dfrac{(\alpha_t-\alpha_r)^2 + \beta_t^2}{2|\alpha_r-\alpha_t|},~~~~~~~
\delta_s=\frac{\bar{\alpha}_p-\alpha_p}{R_s}.
\end{equation}
With a carefully look at Fig. \ref{shadow_ref}, one will find that in the absence of rotation (axial-symmetry), the deformation parameter to be zero. The plots of $R_{s}-a$ and $\delta_s-a$ in terms of the free parameters involved in any two types plasma distributions functions at hand can be seen from Figs. \ref{RD1}-\ref{RD3}\footnote{While drawing these figures, we found that due to taking the rotation parameter $a$ as a variable so any three figures addressing the behavior of $\delta_s-a$, for values $\chi_{ap}>0.15$, will face challenges. Hence in all of plots $\delta_s-a$, we fixed values $\chi_{ap}\leq0.15$.}. In any three figures, one can be seen that the shadow radius in the presence of magnetized plasma is bigger than the non-magnetized plasma and vacuum solutions. The shadow radius for the case of the magnetized plasma with homogeneous power-law density distribution decreases as the rotation parameter increases (similar to non-magnetized plasma and vacuum solutions) while, for in-homogeneous power-law and exponentially density distributions, it increases. As one can see, the deformation parameter $\delta_s$ in the presence of the magnetized plasma is smaller relative to two other counterparts. In any three plasma distributions, $\delta_s$ related to the magnetized plasma solution like those two grows with the increase of the spin parameter.  Due to the relation between axion frequency and axion mass (as can be seen in (\ref{mass})), the top/bottom right panels in Figs. \ref{RD1}-\ref{RD3} give this message to us that the heavier the axion induced of the magnetized plasma, it may leave some imprints, on the shadow of BH that are subtly distinguishable from non-magnetized plasma. 
\begin{figure}[ht]
	\includegraphics[scale=0.33]{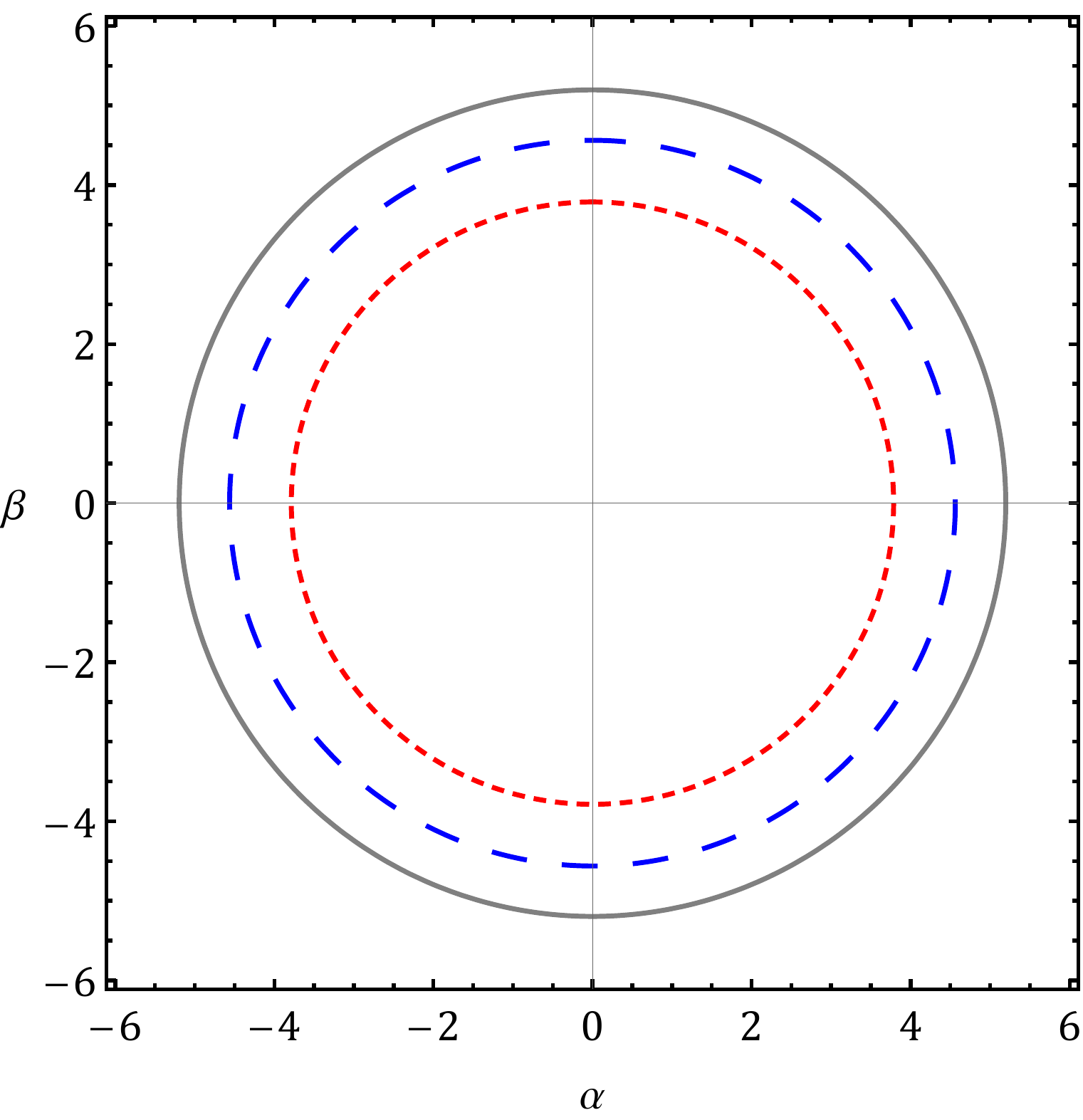}~~~
	\includegraphics[scale=0.33]{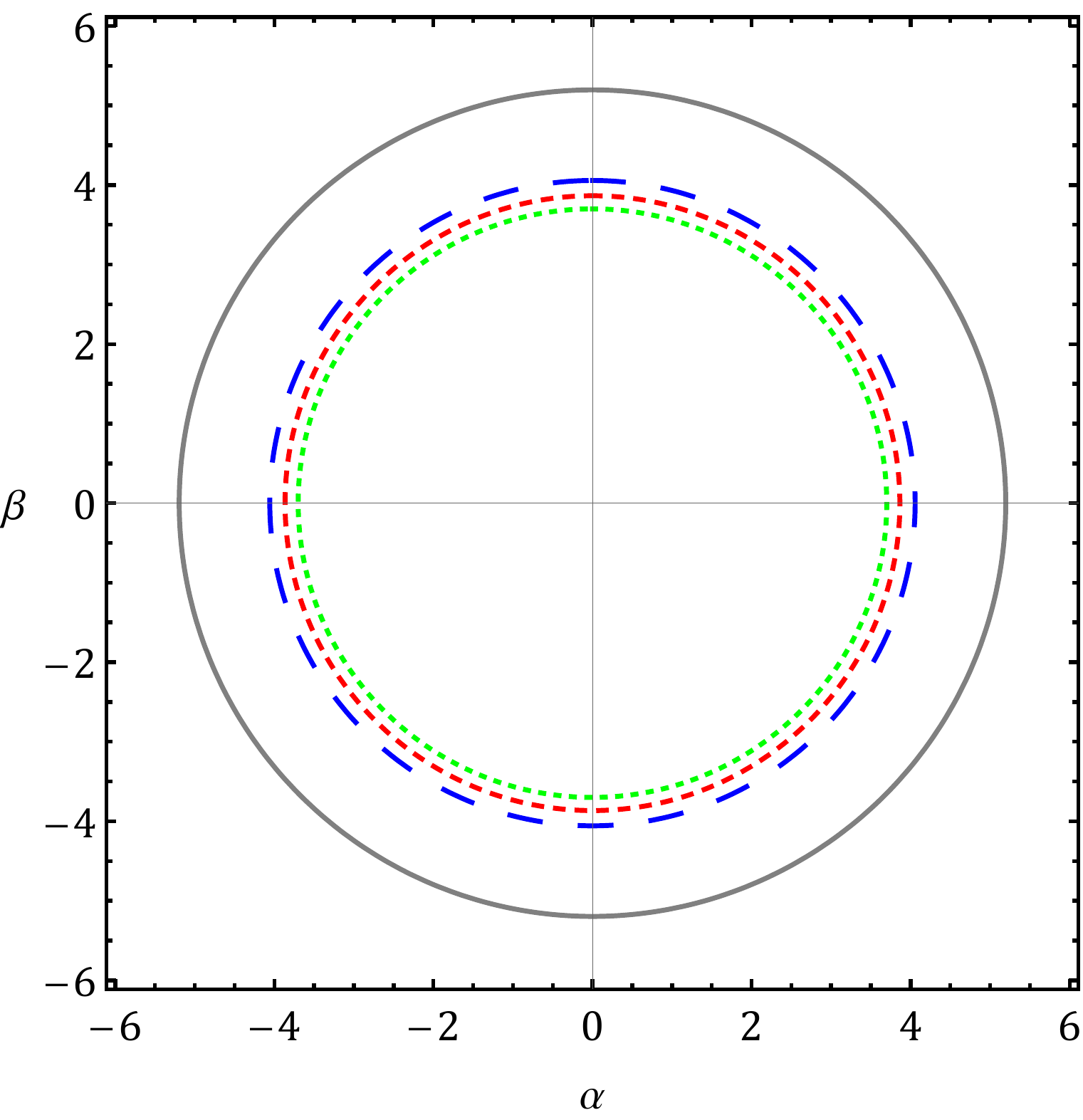}~~~
	\includegraphics[scale=0.33]{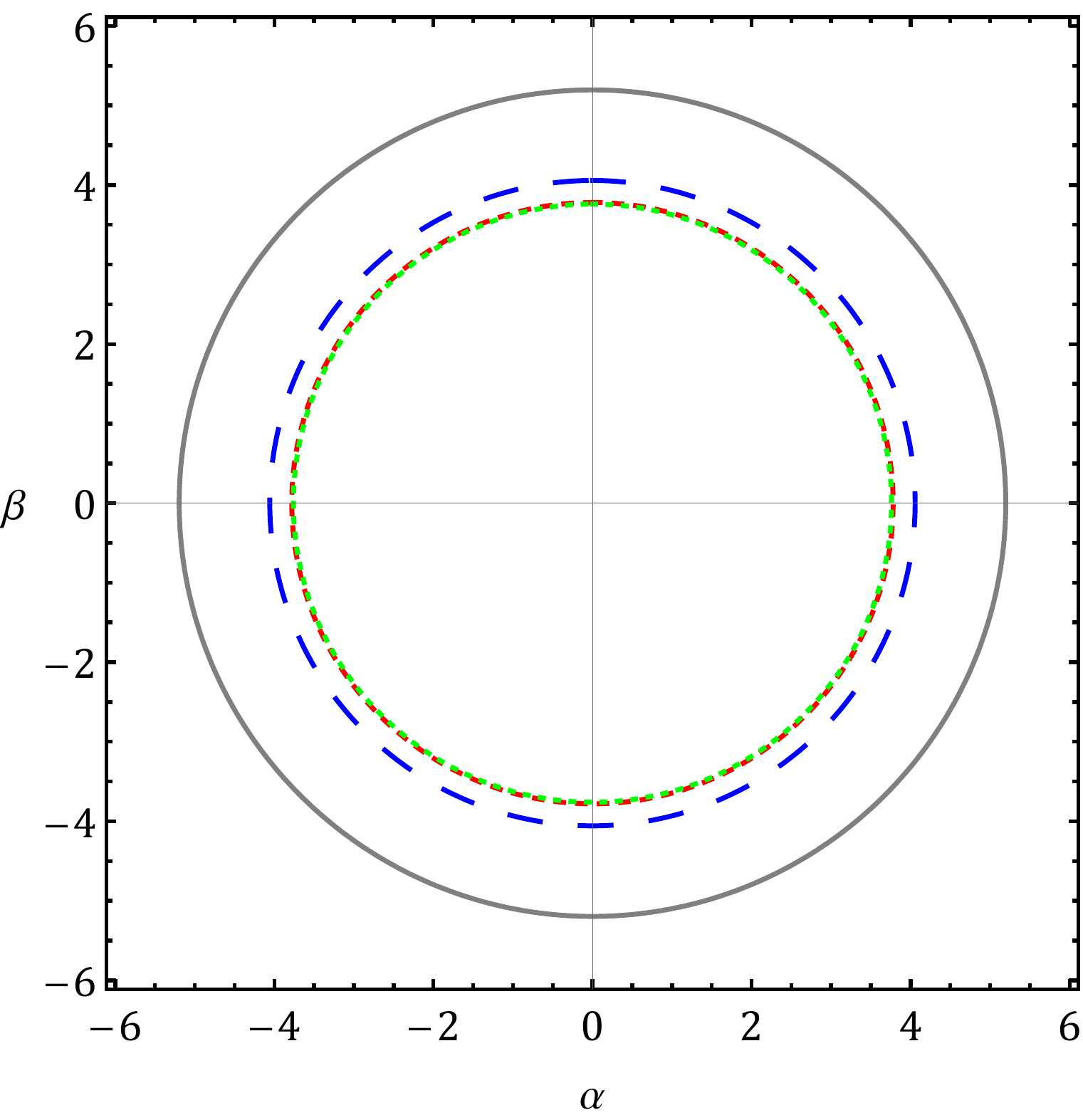}
	\caption{\textit{The shadow of Schwarzschild BH surrounded by a plasma with in-homogeneous exponentially density distribution for different sets of $\{\chi_\kappa,\chi_{ap},\chi_\varphi\}$. \textbf{Left:} $\{0,0,0\}$ (gray-solid), $\{0.35,0,0\}$ (blue-long dashed), $\{0.7,0,0\}$ (red-dashed). \textbf{Middle:} $\{0,0,0\}$ (gray-solid), $\{0.5,0,0\}$ (blue-long dashed), $\{0.5,0.35,0.1\}$ (red-dashed), $\{0.5,0.7,0.1\}$ (green-dotted). \textbf{Right:} $\{0,0,0\}$ (gray-solid), $\{0.5,0,0\}$ (blue-long dashed), $\{0.5,0.5,0.25\}$ (red-dashed), $\{0.5,0.5,0.5\}$ (green-dotted). We have set $M=1$ and $l_0=50$.} }
	\label{Shadow6}
\end{figure}
\begin{figure}[ht]
	\includegraphics[scale=0.3]{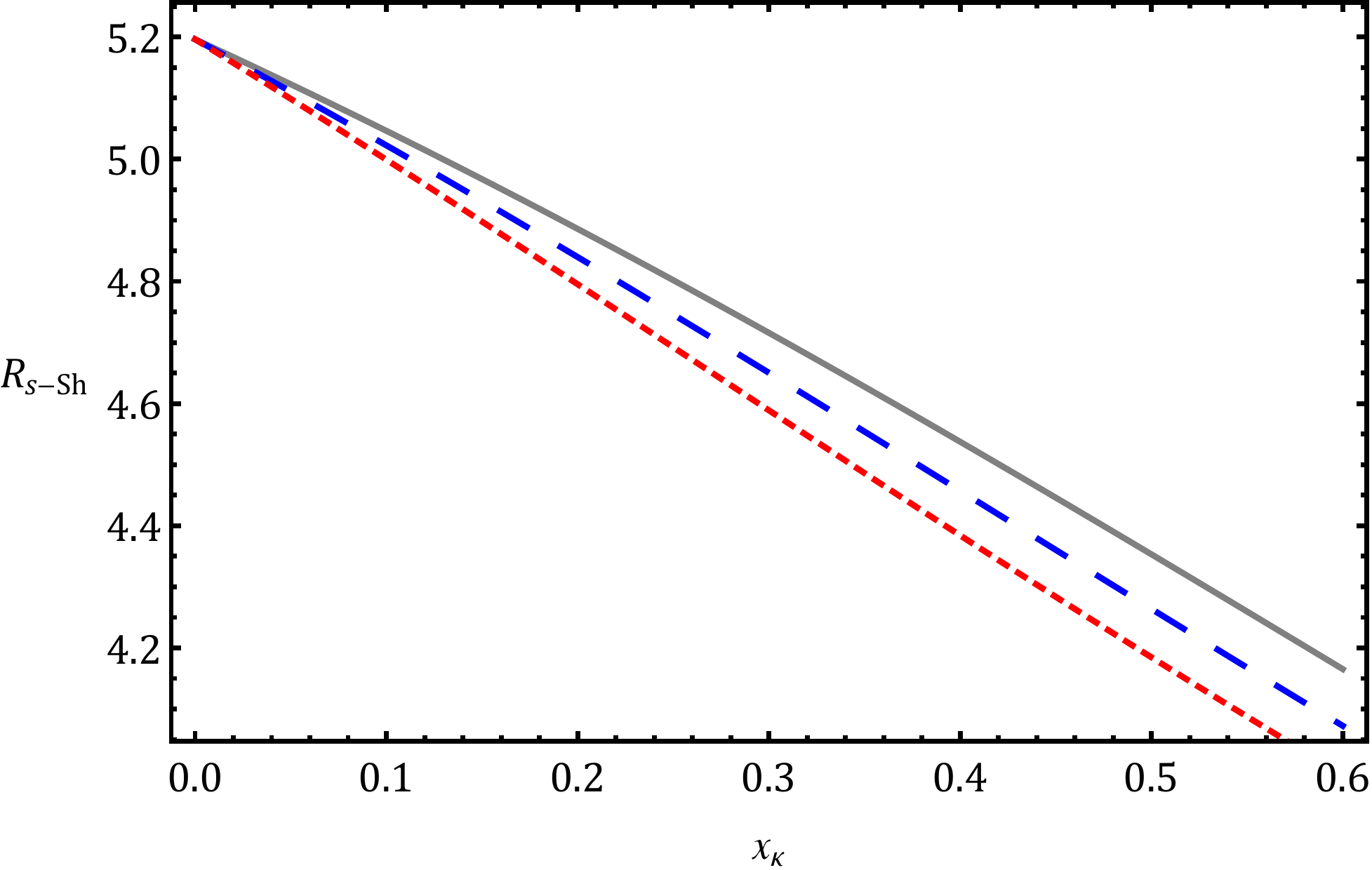}~~~
	\includegraphics[scale=0.3]{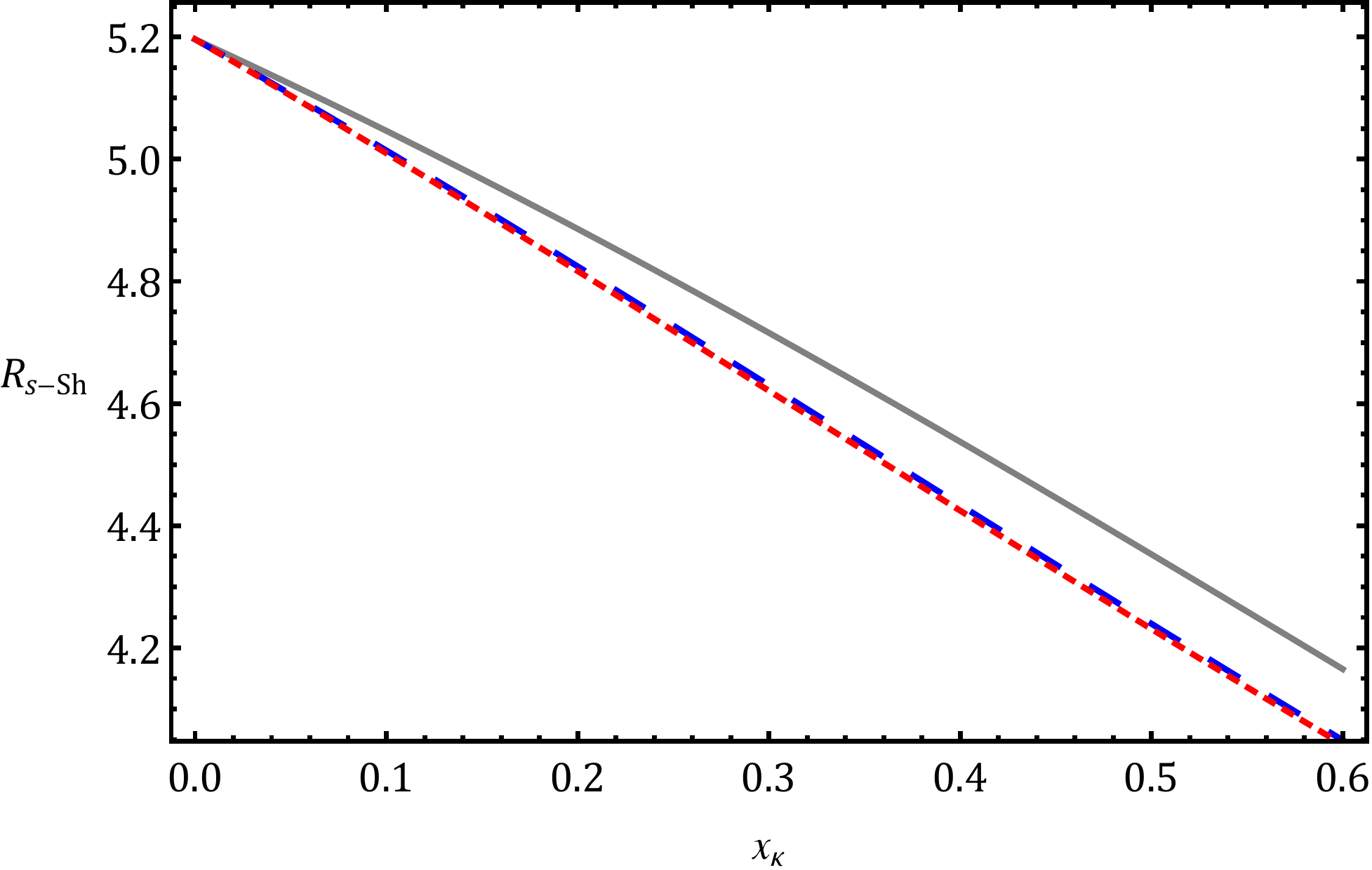}\\
	\includegraphics[scale=0.3]{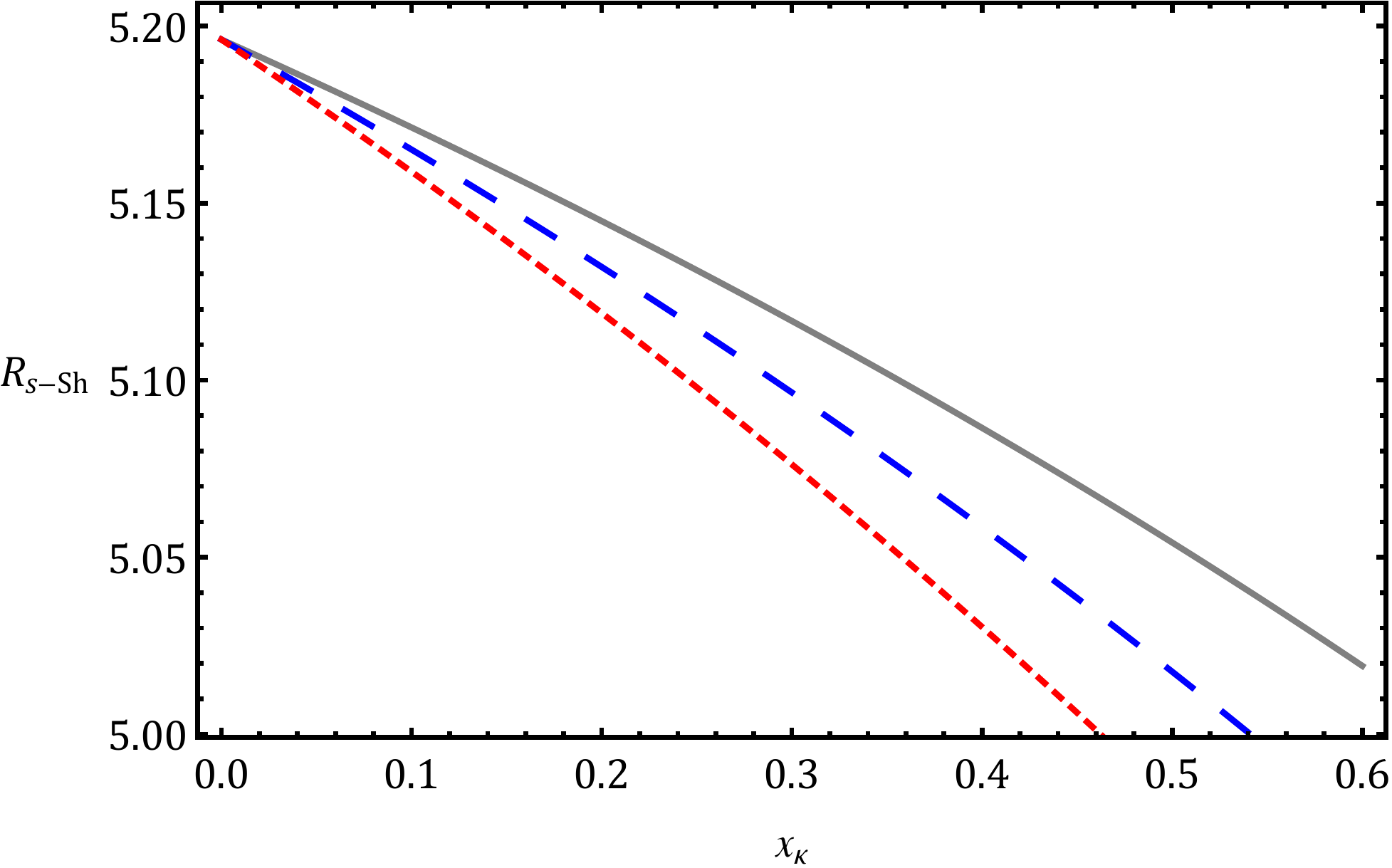}~~~
	\includegraphics[scale=0.3]{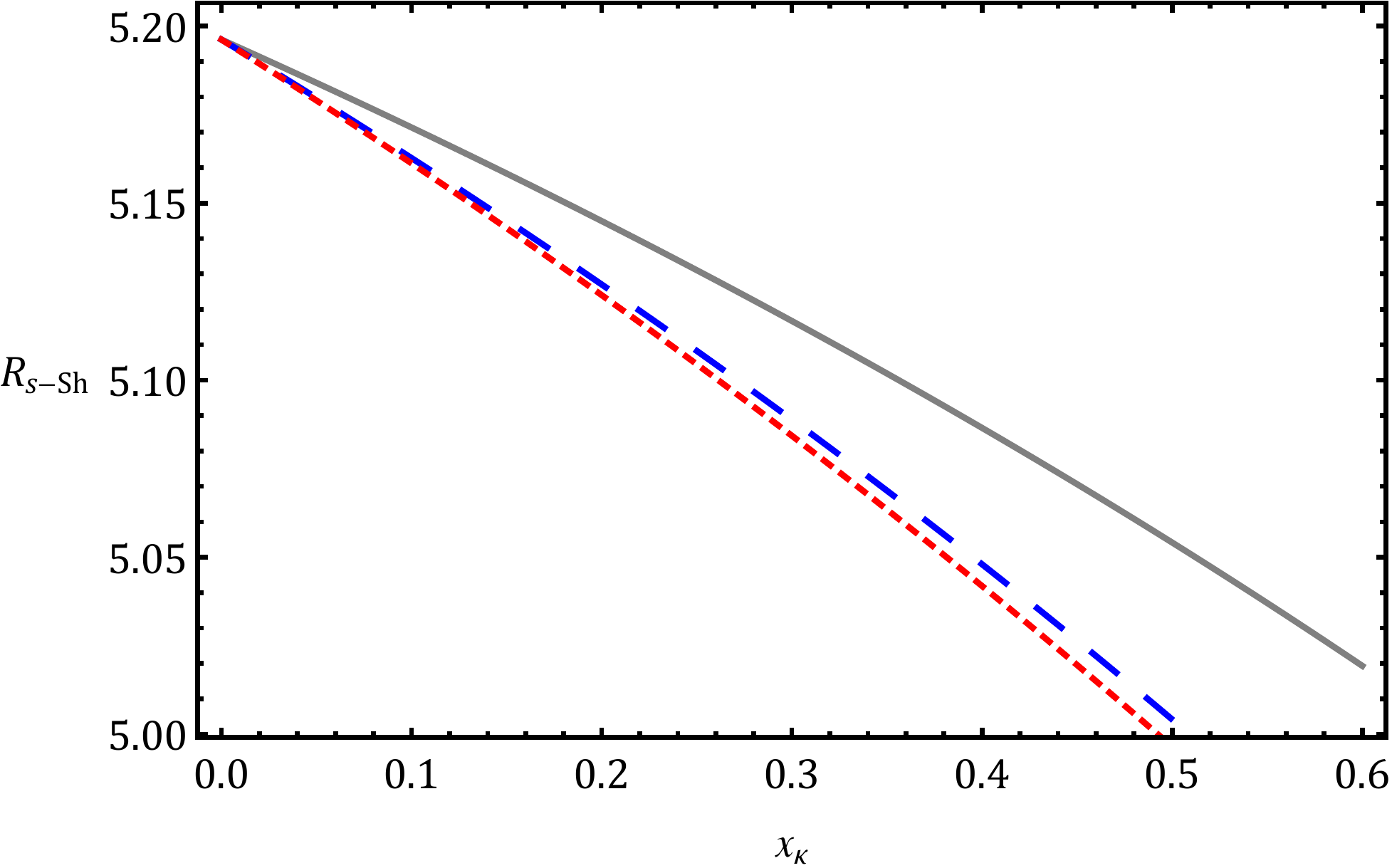}\\
	\includegraphics[scale=0.3]{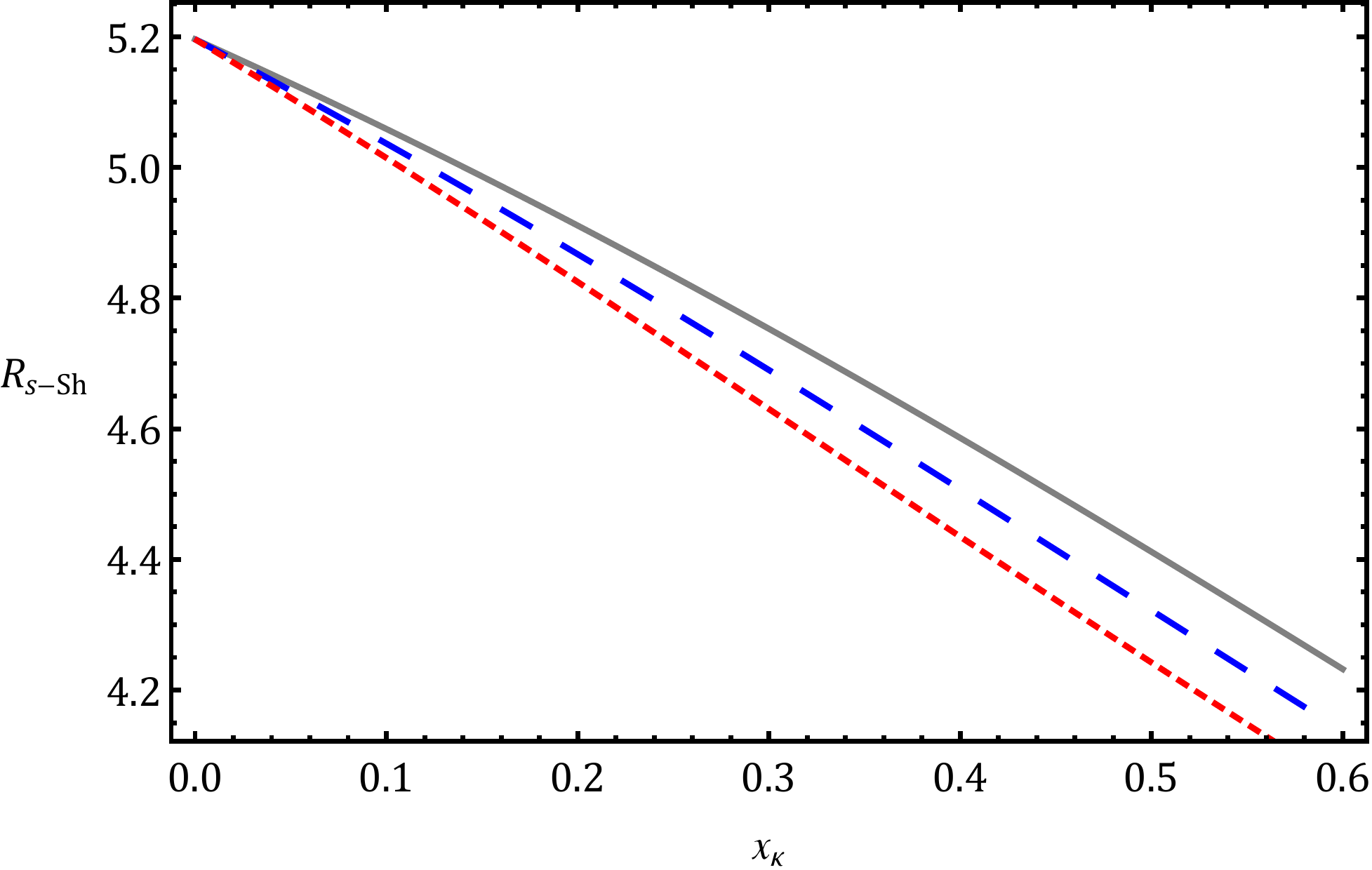}~~~
	\includegraphics[scale=0.3]{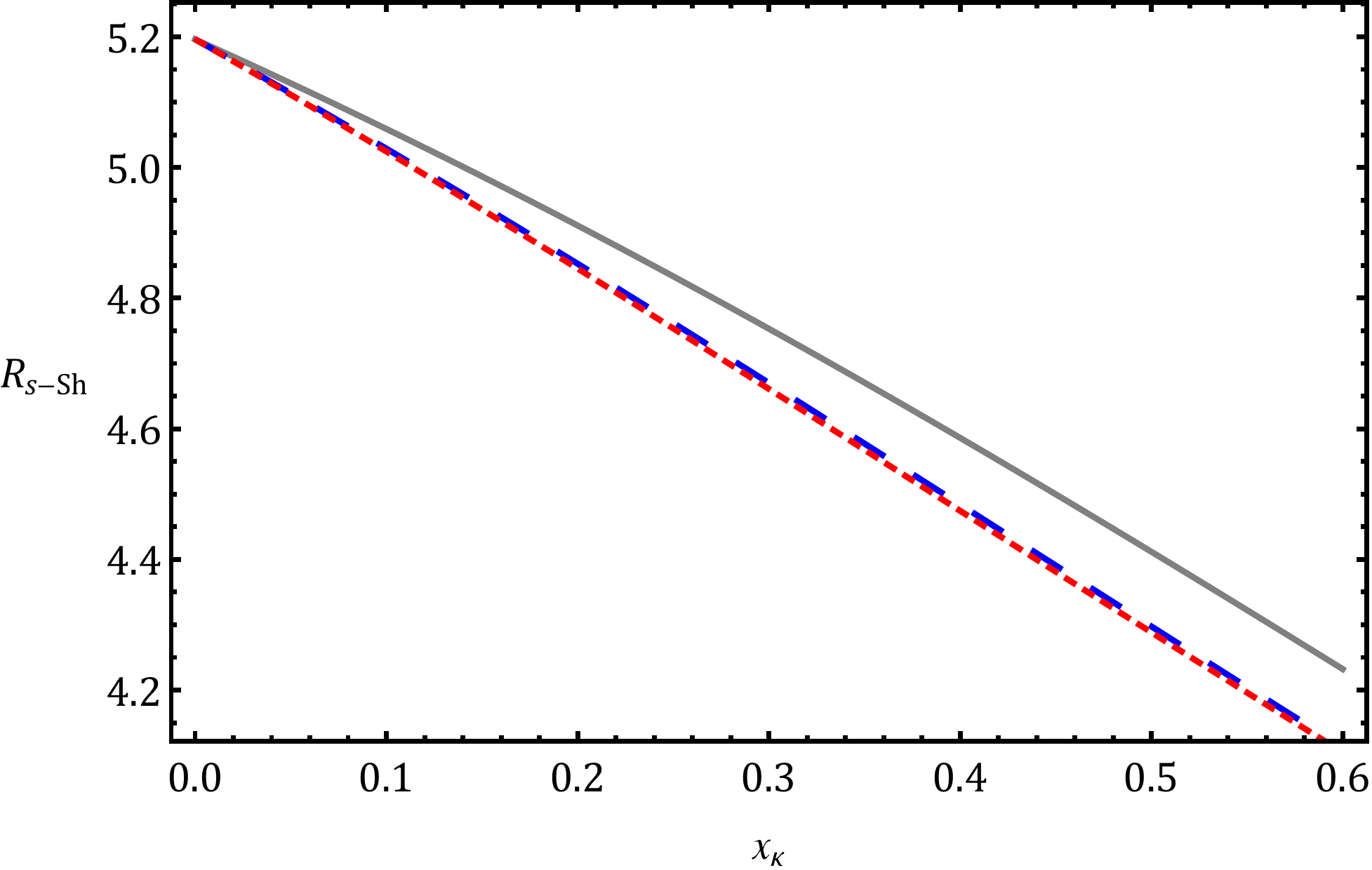}
	\caption{\textit{The plot of $R_{s-Sh}-\chi_k$ in terms of different sets of $\{\chi_{ap},\chi_\varphi\}$ for Schwarzschild BH surrounded by the plasma with homogeneous power-law (\textbf{up row}), in-homogeneous power-law (\textbf{middle row}), and in-homogeneous exponentially (\textbf{bottom row}) density distributions. In the left panel of any row, we have set values $\{0,0\}$ (gray-solid), $\{0.25,0\}$ (blue-long dashed), and $\{0.5,0\}$(red-dashed), while for the right panel of any row, we have set values $\{0,0\}$ (gray-solid), $\{0.3,0.25\}$ (blue-long dashed), and $\{0.3,0.5\}$(red-dashed).} }
	\label{RSH1}
\end{figure}
\begin{figure}[ht]
	\includegraphics[scale=0.27]{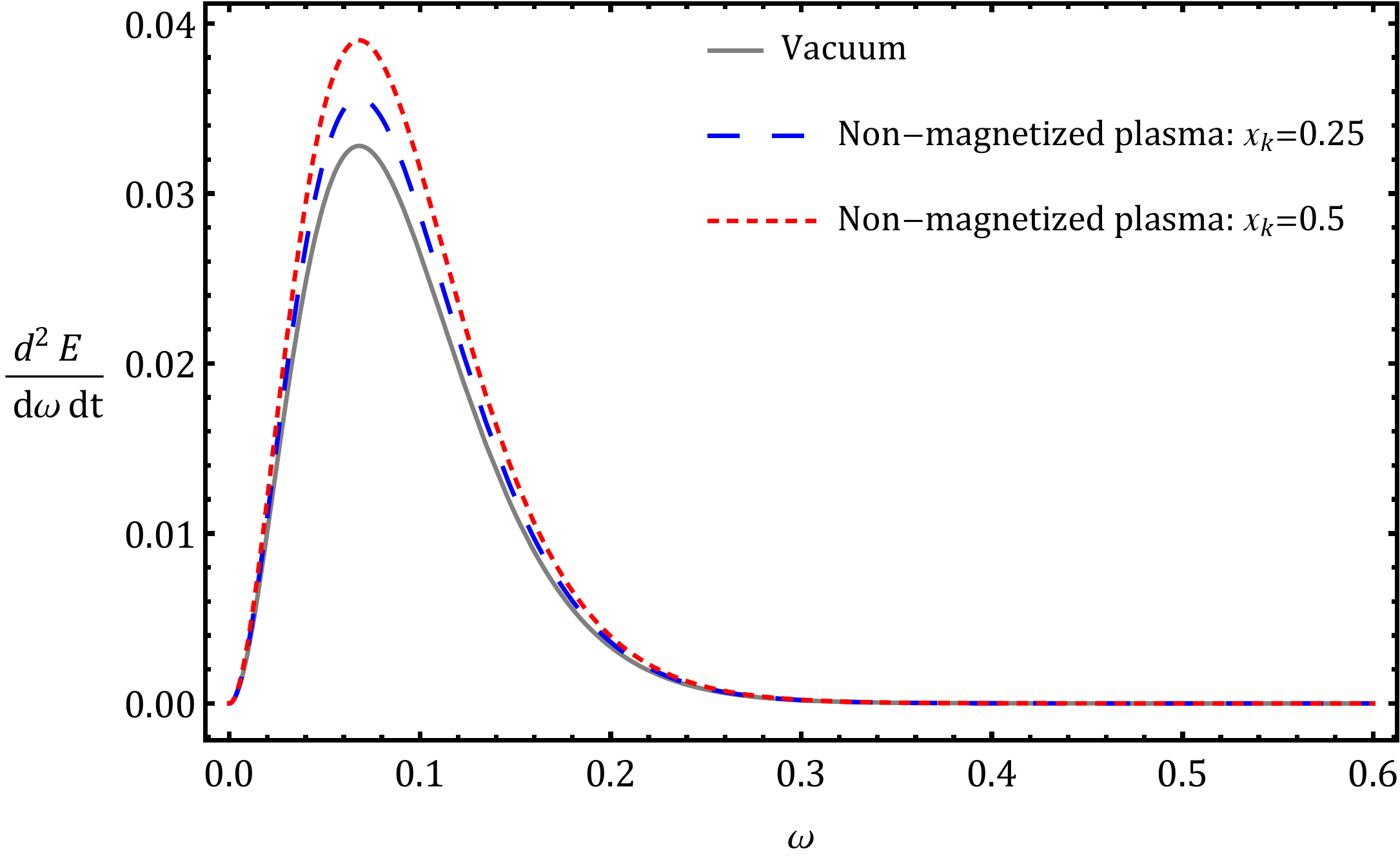}~~~
	\includegraphics[scale=0.25]{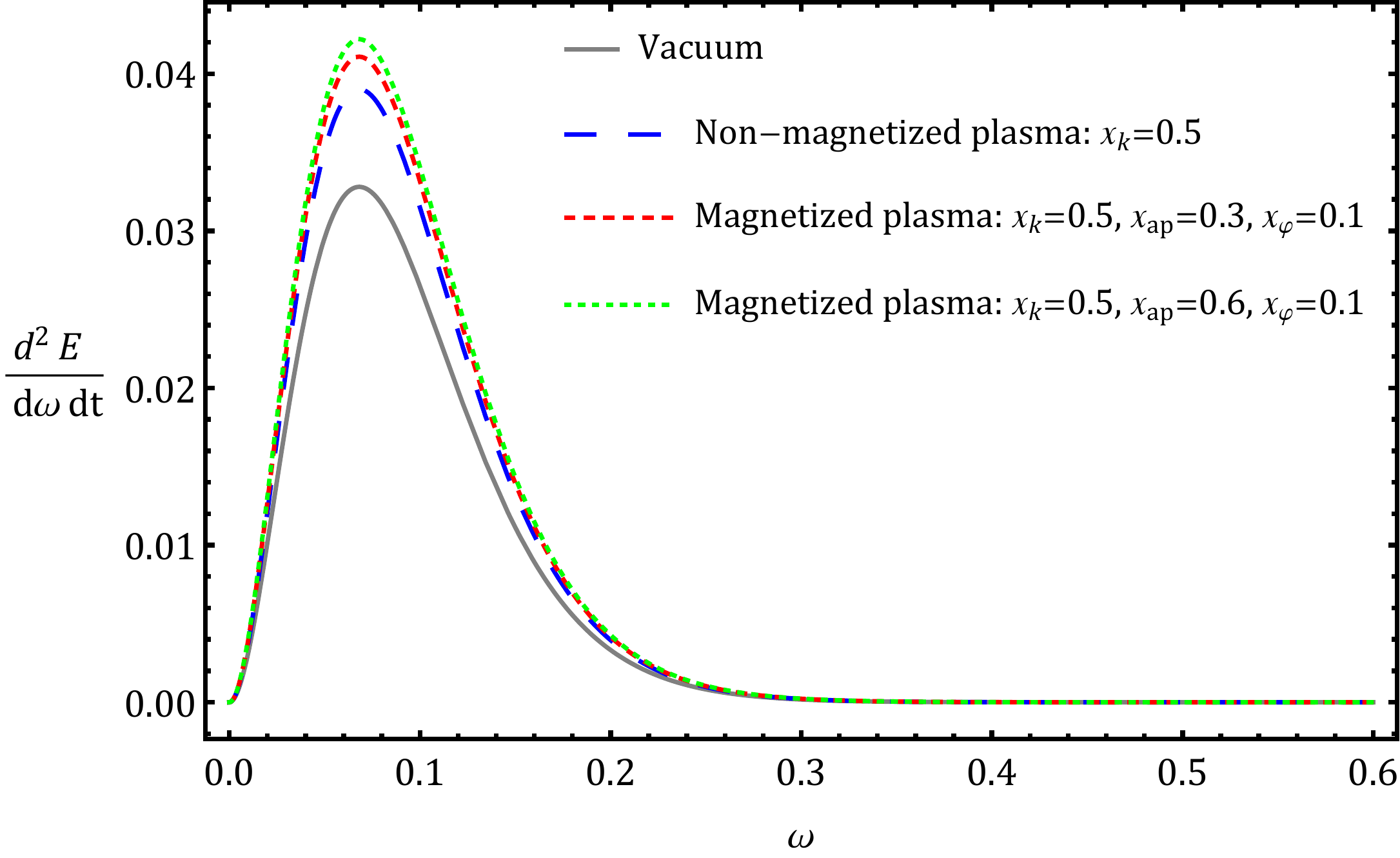}~~~
	\includegraphics[scale=0.25]{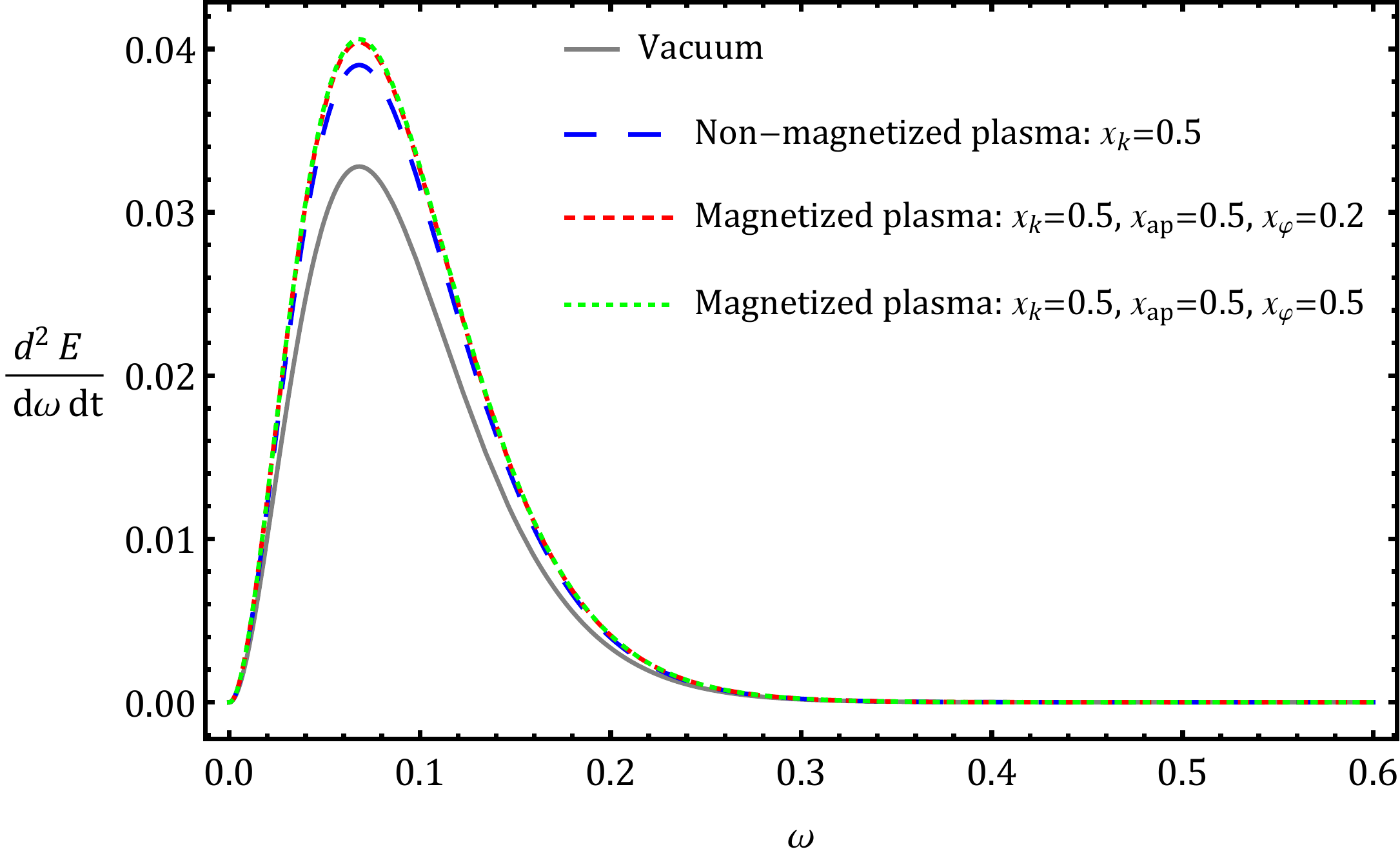}
	\includegraphics[scale=0.27]{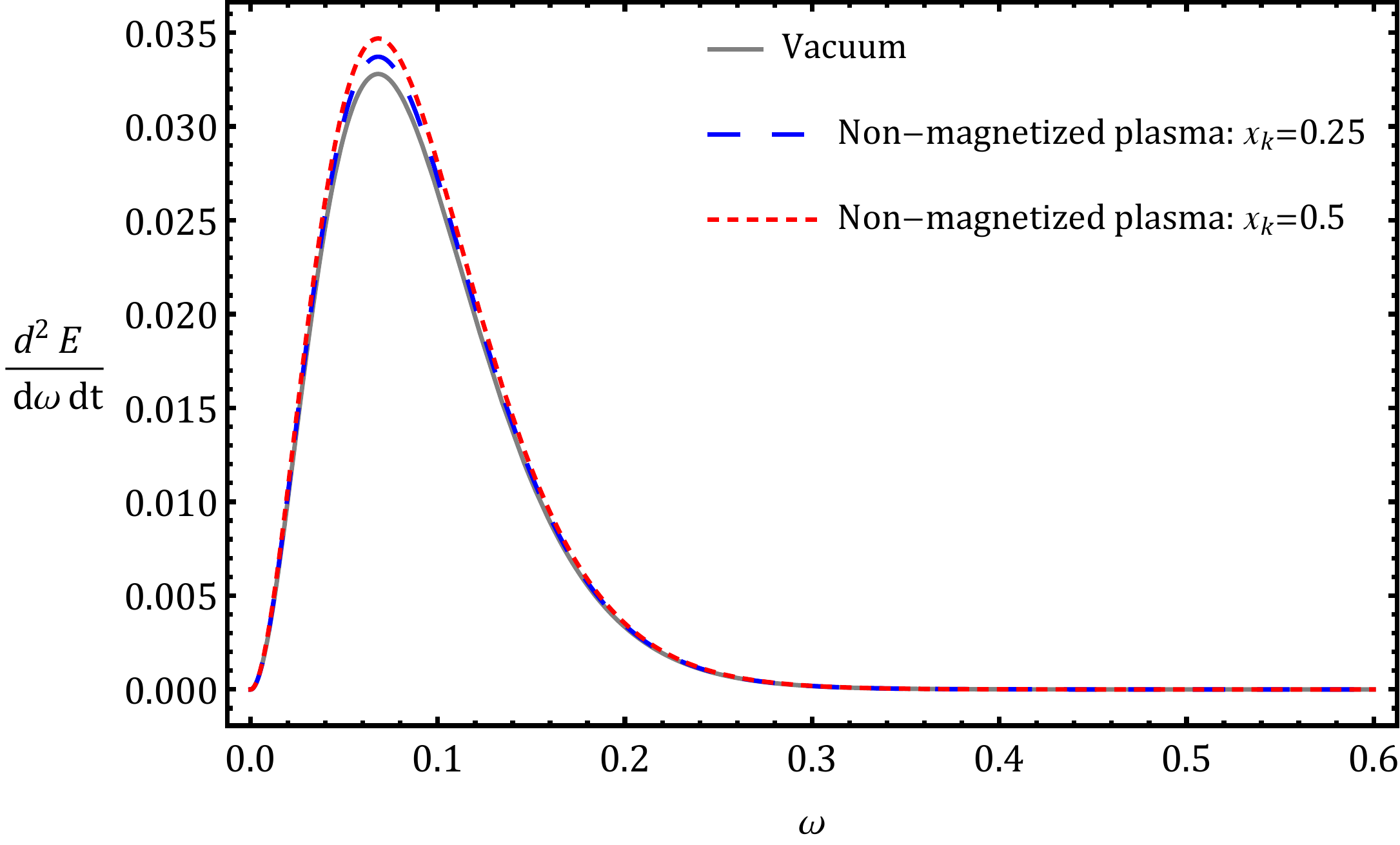}
	\includegraphics[scale=0.25]{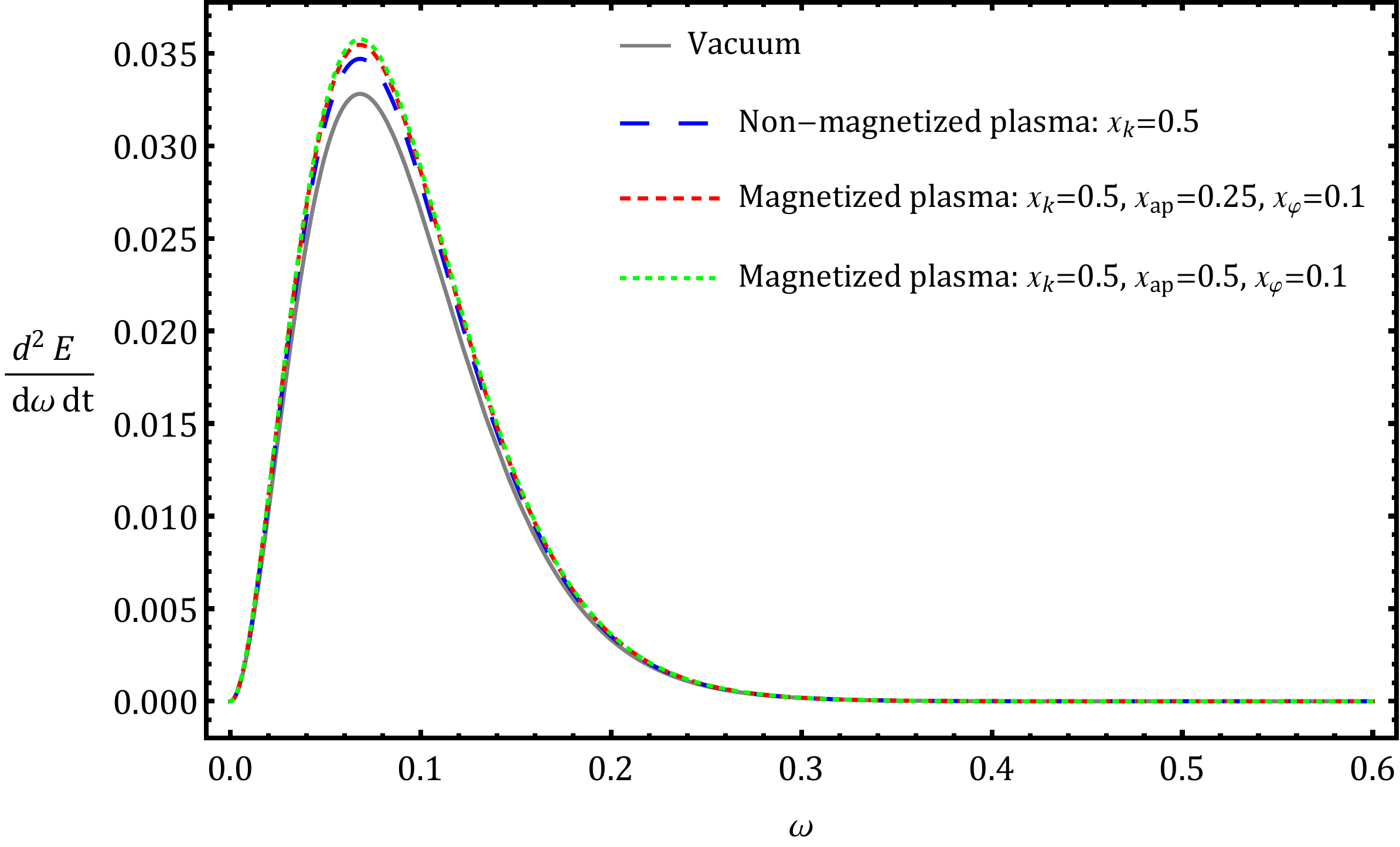}
	\includegraphics[scale=0.25]{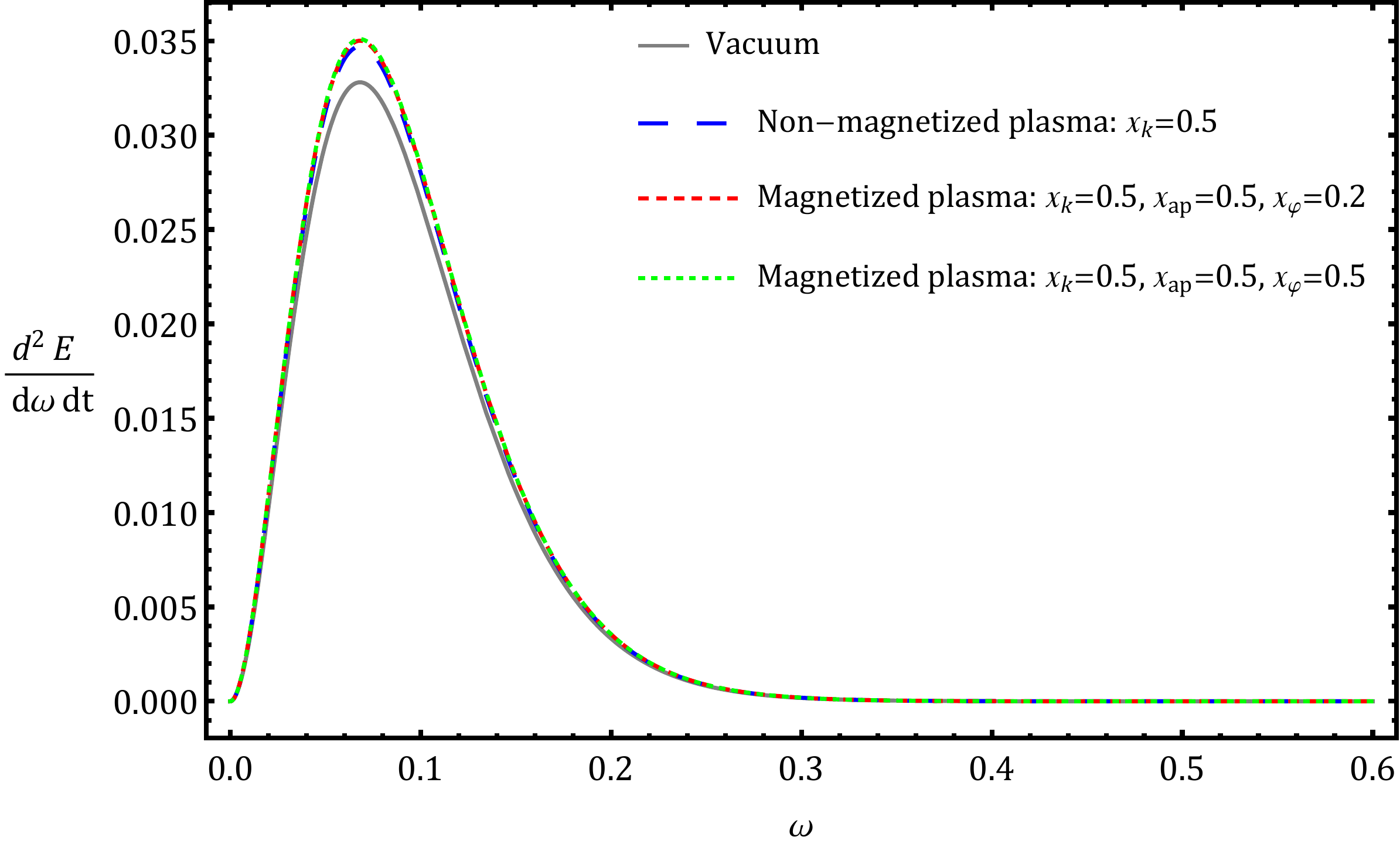}
	\caption{\textit{The EER of rotating BH $(a=0.9)$ surrounded by a plasma with homogeneous power-law density distribution  ($h=0$) (\textbf{up row}) and in-homogeneous power-law density distribution  ($h=1$) \textbf{(bottom row}) for different sets of $\{\chi_k,\chi_{ap},\chi_\varphi\}$. We have set $\theta_0=\pi/3,~M=1$.}}
	\label{Emissionr1}
\end{figure}
\begin{figure}[ht]
	\includegraphics[scale=0.27]{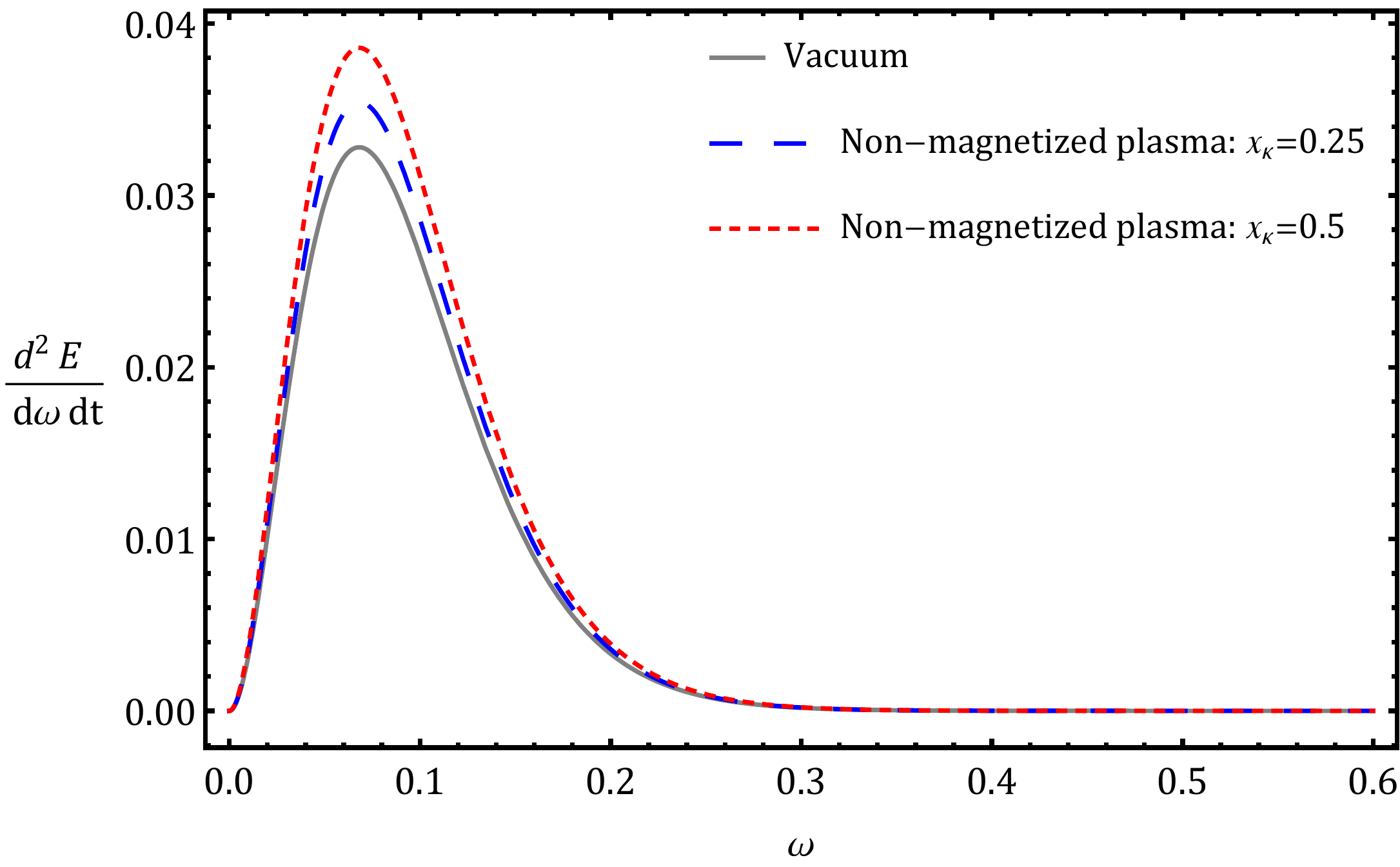}~~~
	\includegraphics[scale=0.25]{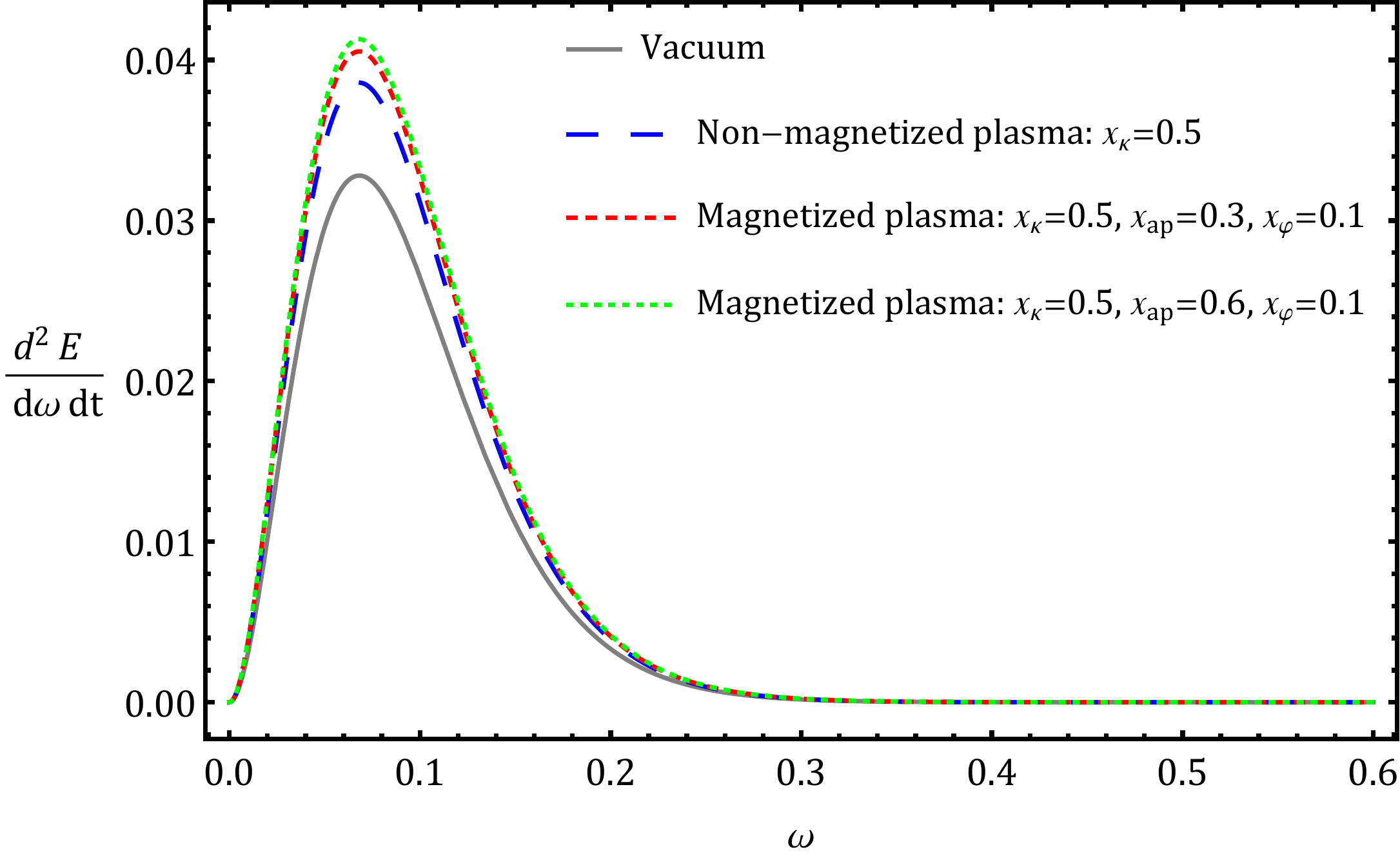}~~~
	\includegraphics[scale=0.25]{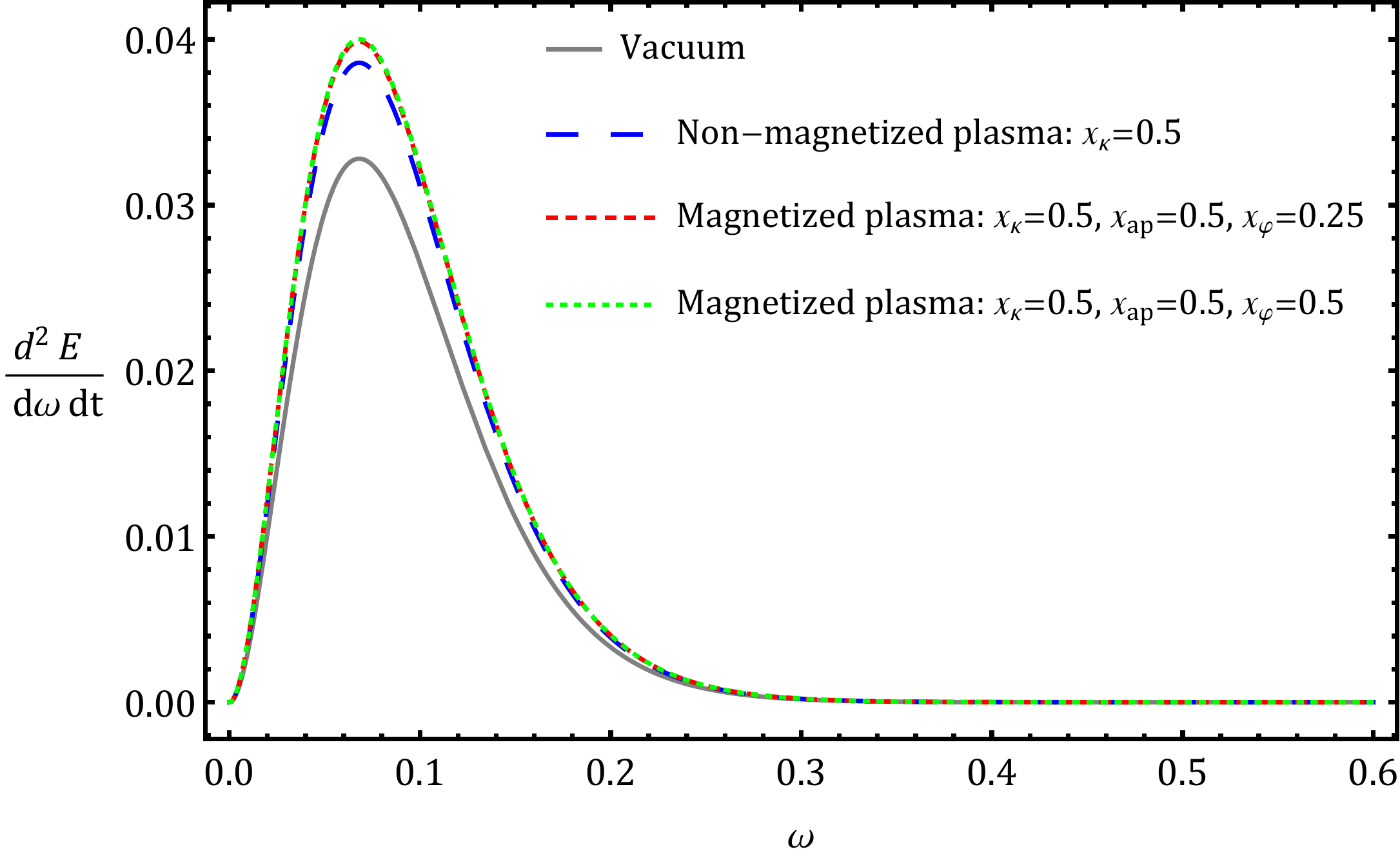}
	\caption{\textit{The EER of rotating BH $(a=0.9)$ surrounded by a plasma with in-homogeneous exponentially density distribution for different sets of $\{\chi_k,\chi_{ap},\chi_\varphi\}$. We have set $\theta_0=\pi/3,~M=1$ and $l_0=50$.}}
	\label{Emissionr2}
\end{figure}
\begin{figure}[ht]
	\includegraphics[scale=0.27]{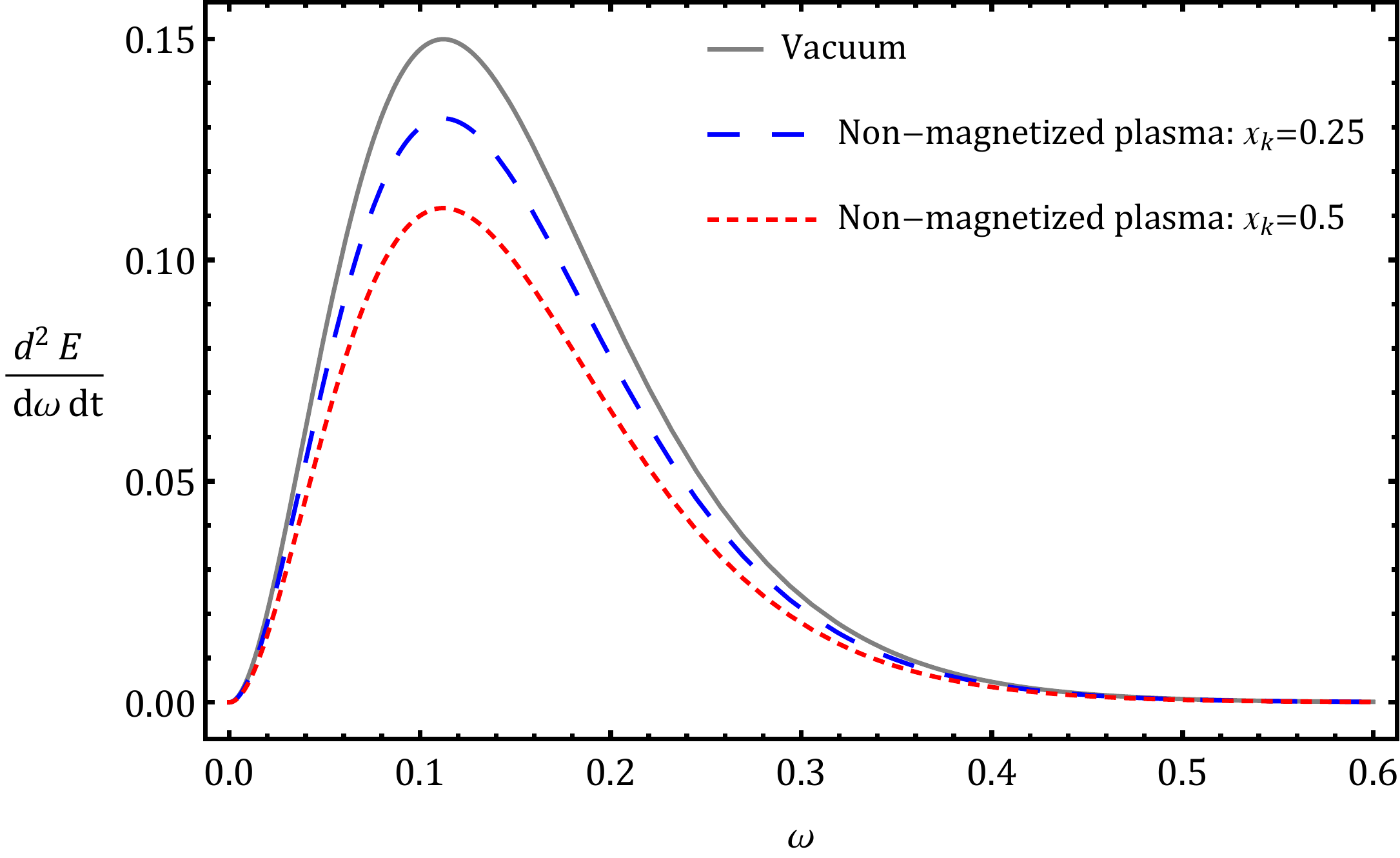}~~~
	\includegraphics[scale=0.25]{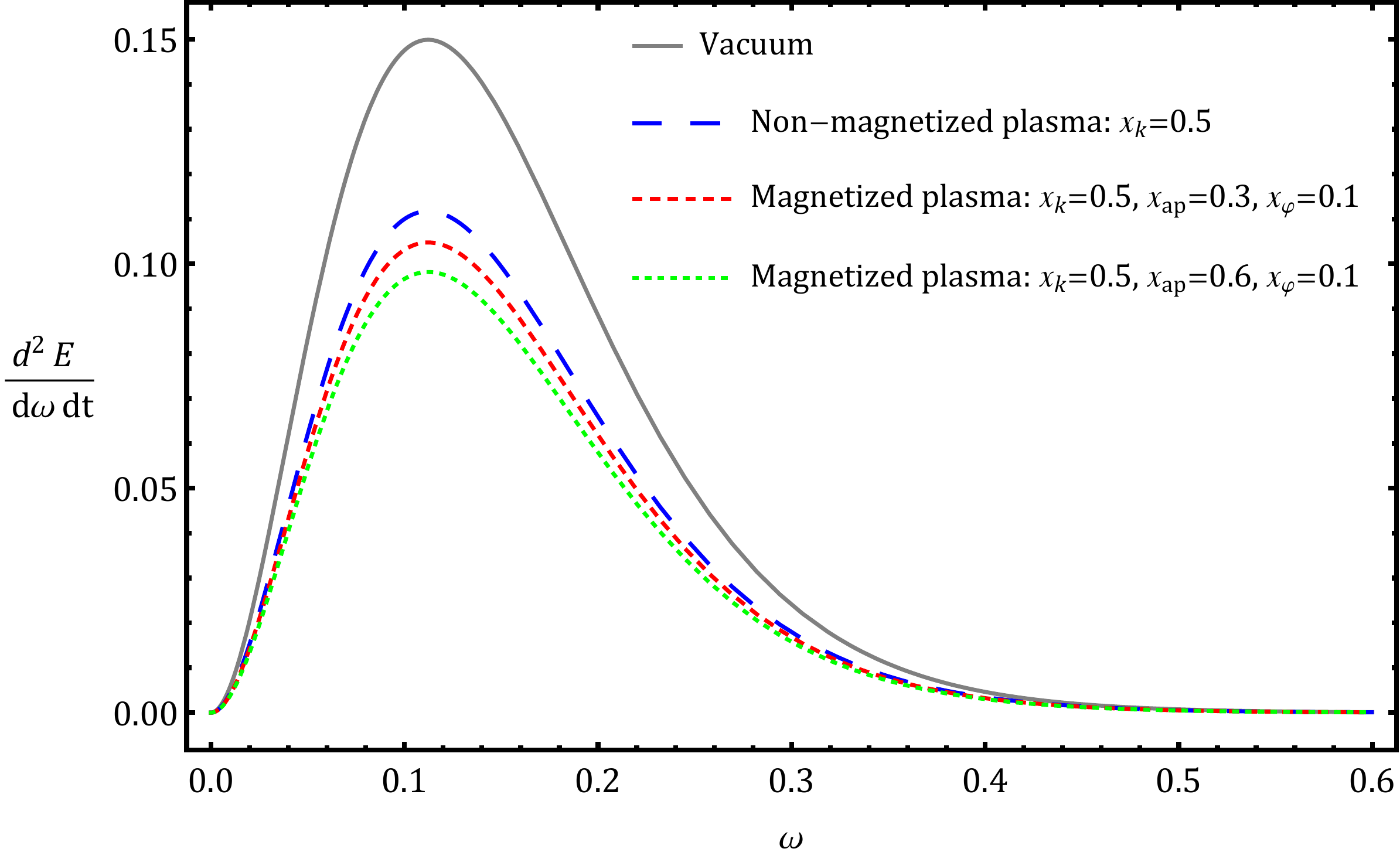}~~~
	\includegraphics[scale=0.25]{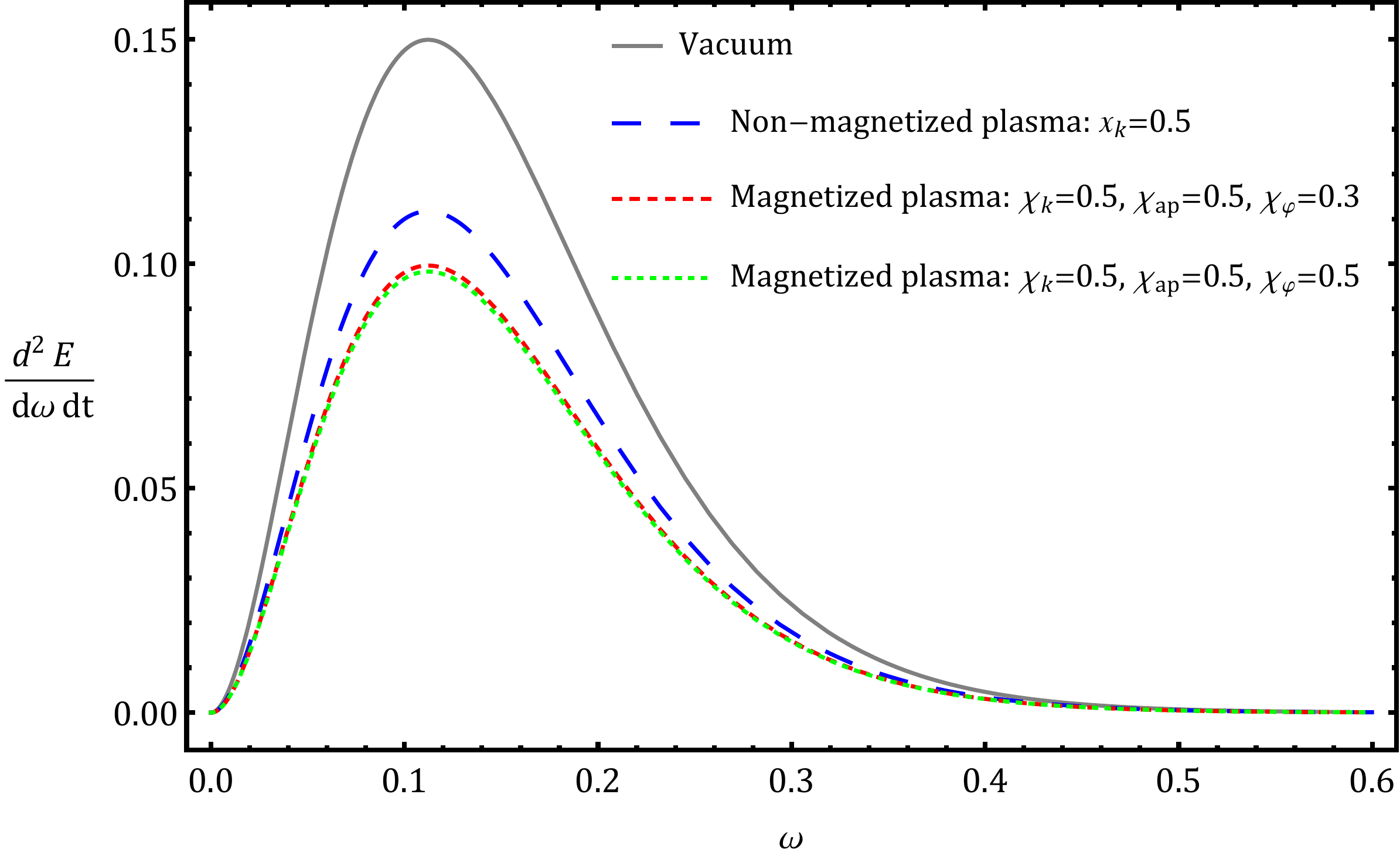}
	\includegraphics[scale=0.27]{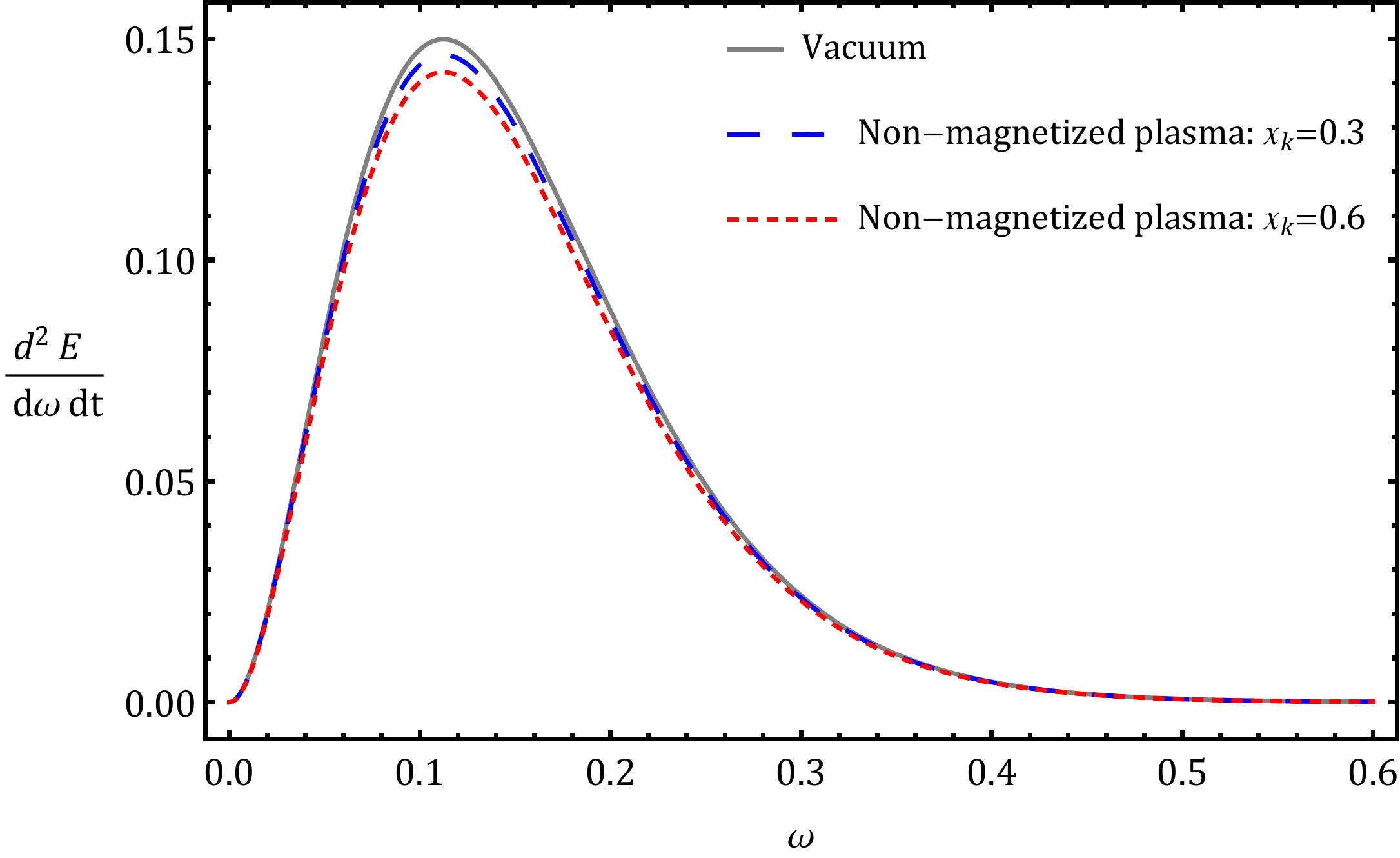}
	\includegraphics[scale=0.25]{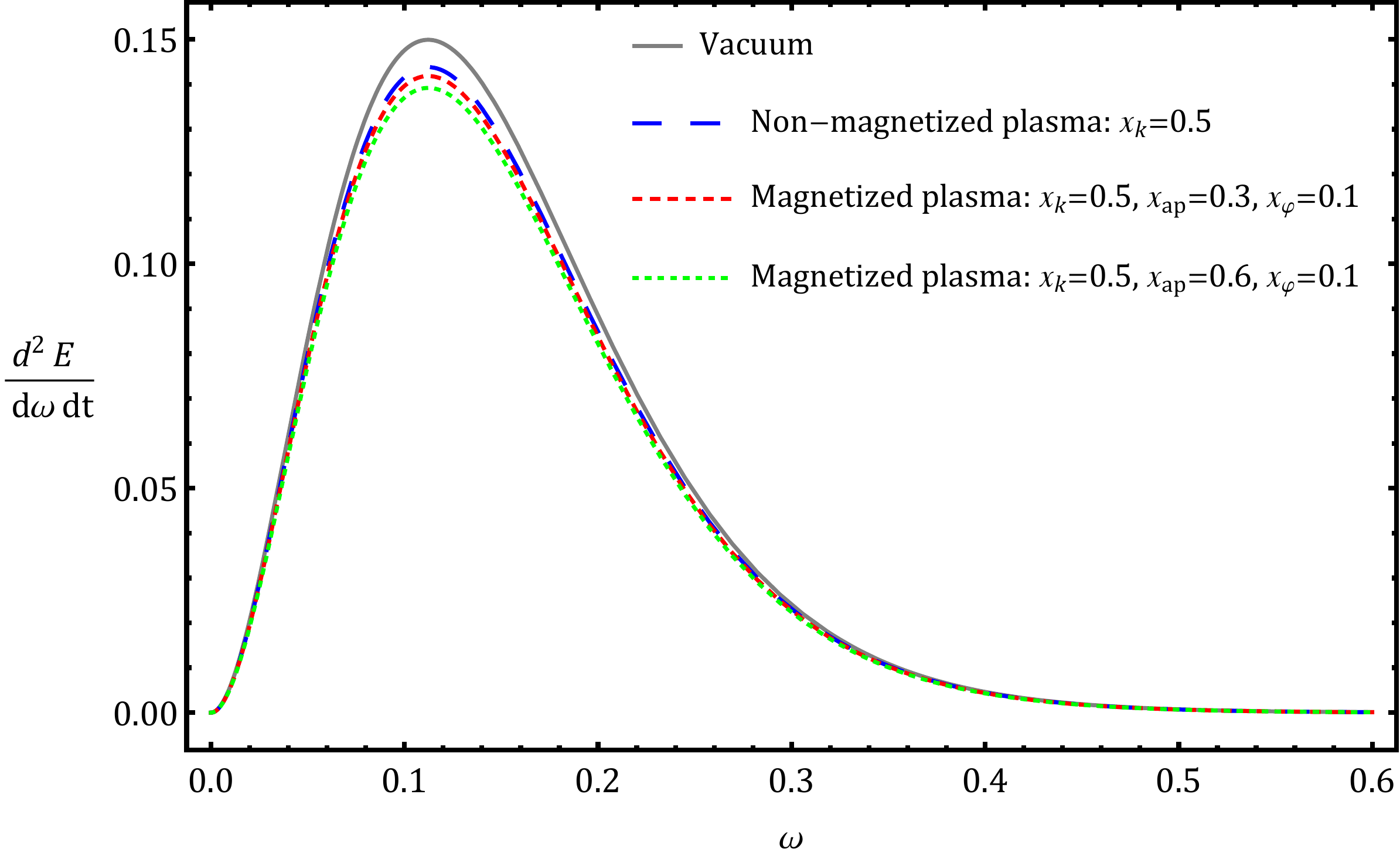}
	\includegraphics[scale=0.25]{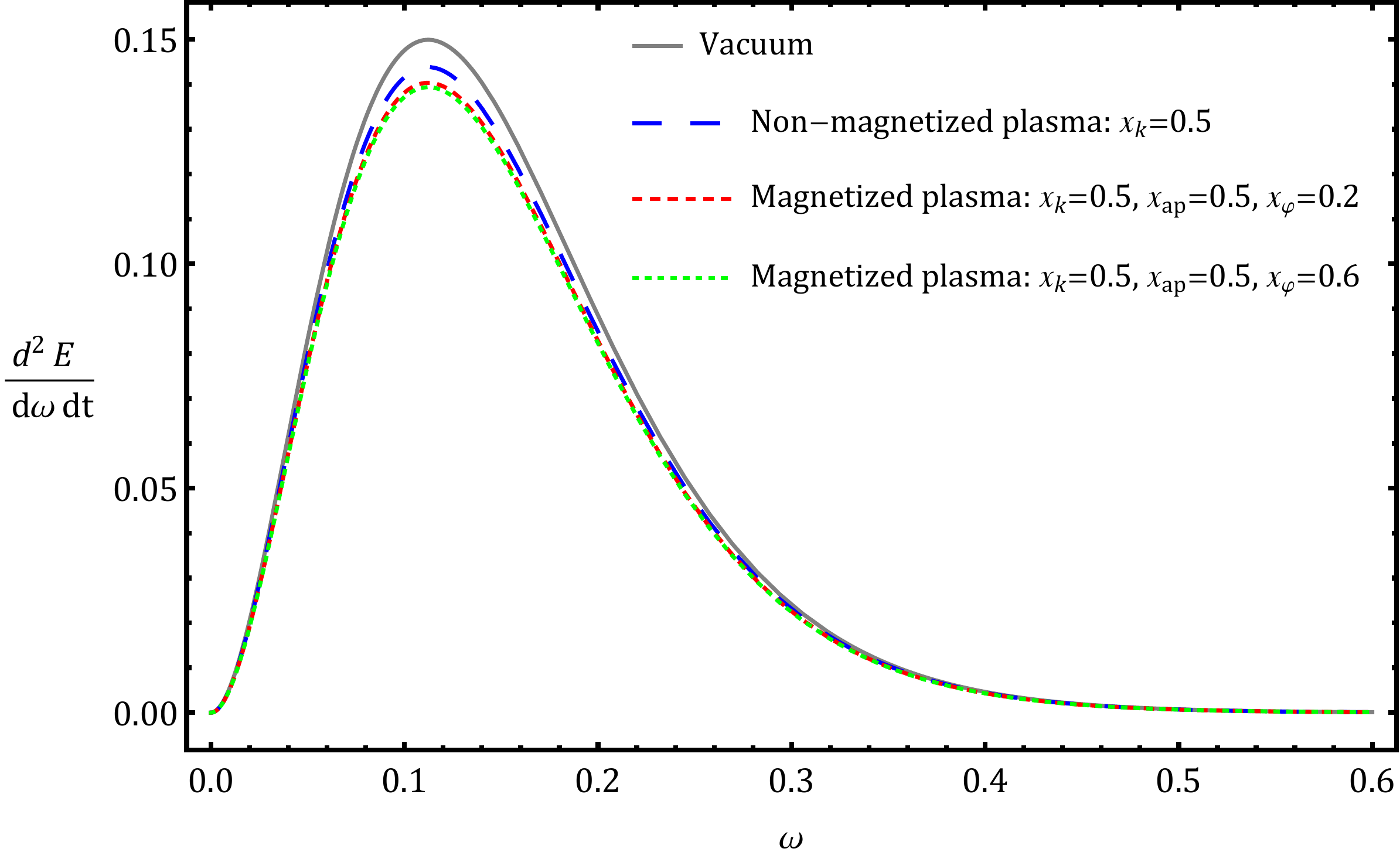}
	\caption{\textit{The EER of Schwarzschild BH surrounded by a plasma with homogeneous power-law density distribution  ($h=0$) (\textbf{up row}) and inhomogeneous power-law density distribution  ($h=1$) \textbf{(bottom row}) for different sets of $\{\chi_k,\chi_{ap},\chi_\varphi\}$.} }
	\label{Emissions1}
\end{figure}
\begin{figure}[ht]
	\includegraphics[scale=0.27]{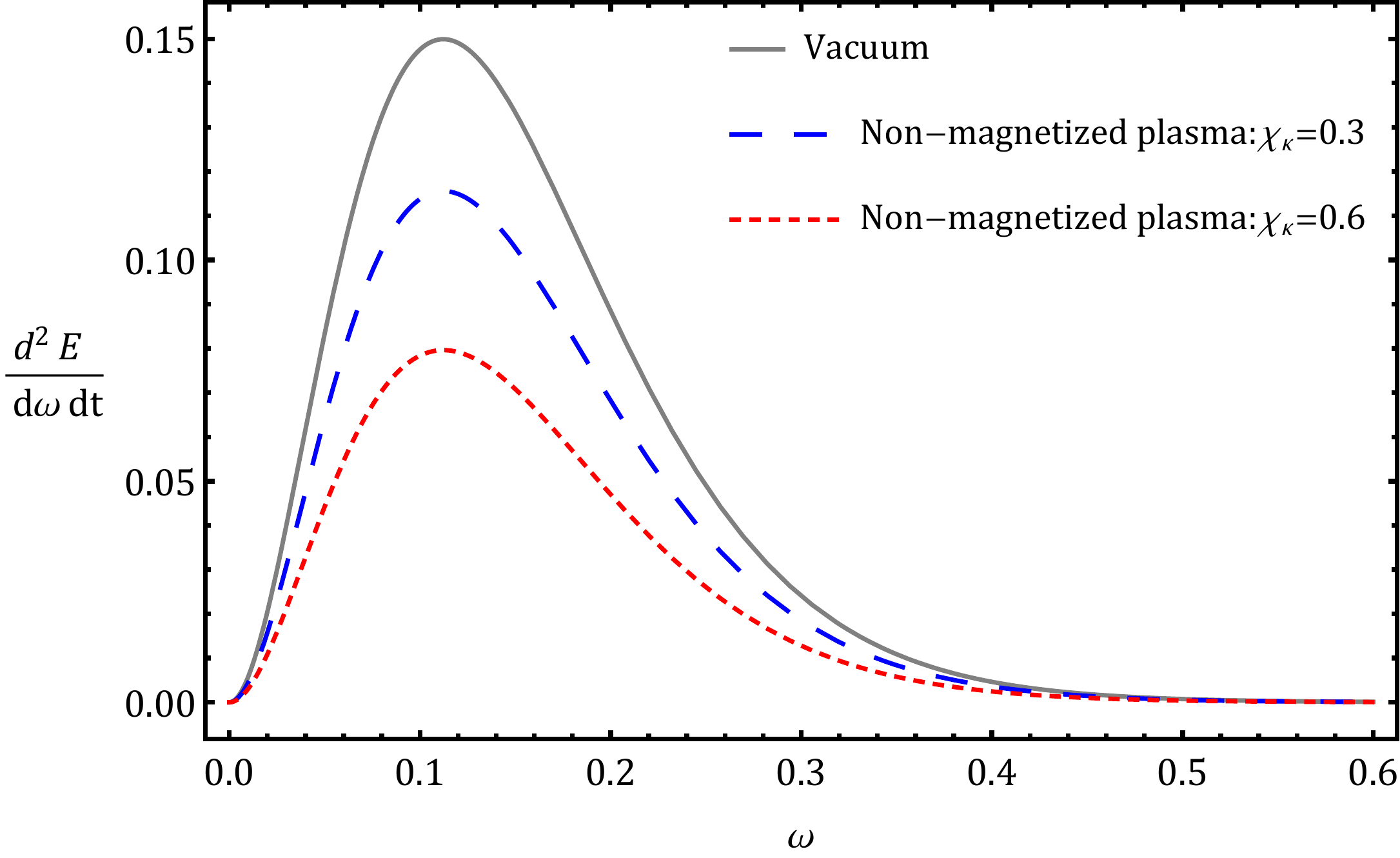}~~~
	\includegraphics[scale=0.25]{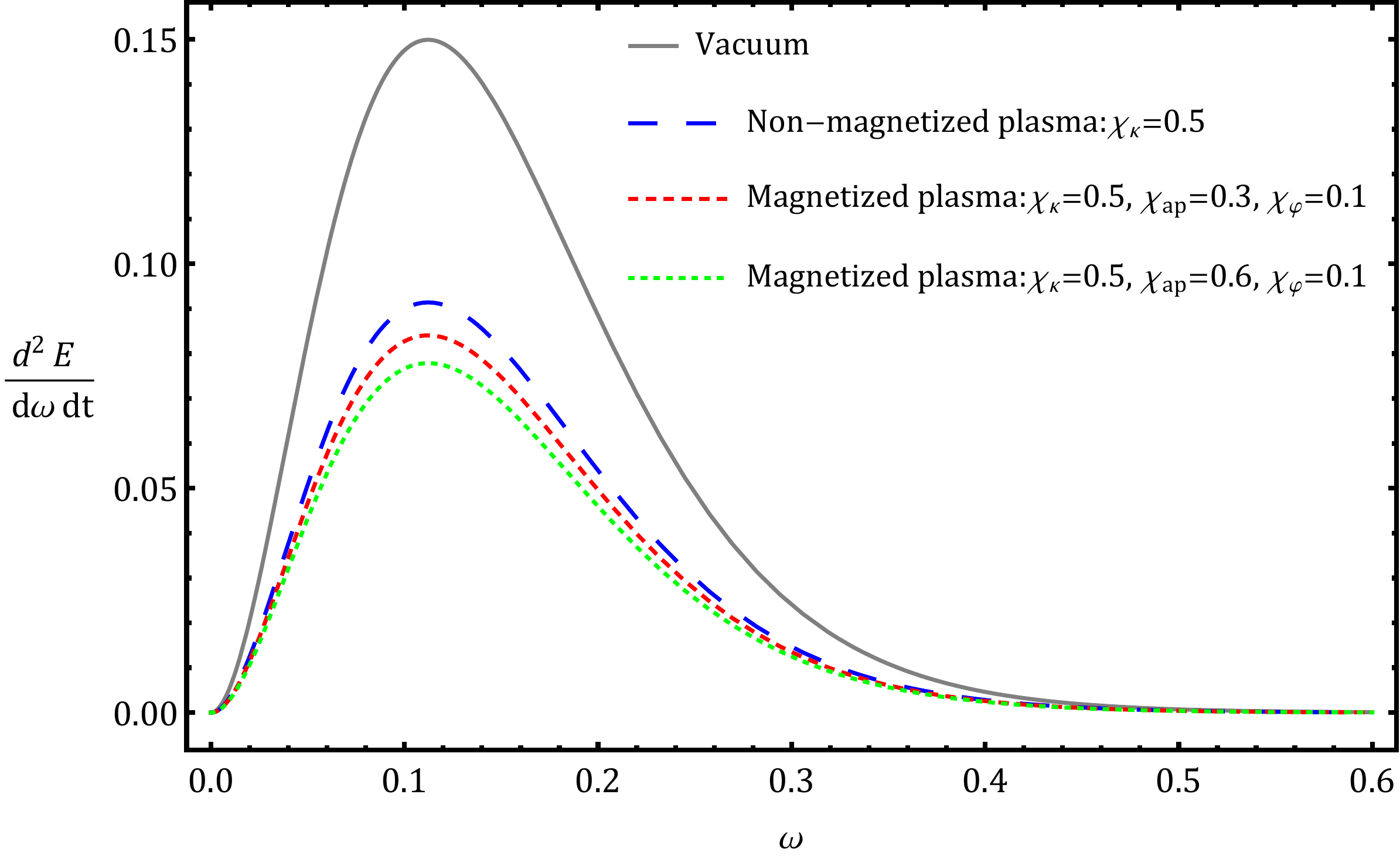}~~~
	\includegraphics[scale=0.25]{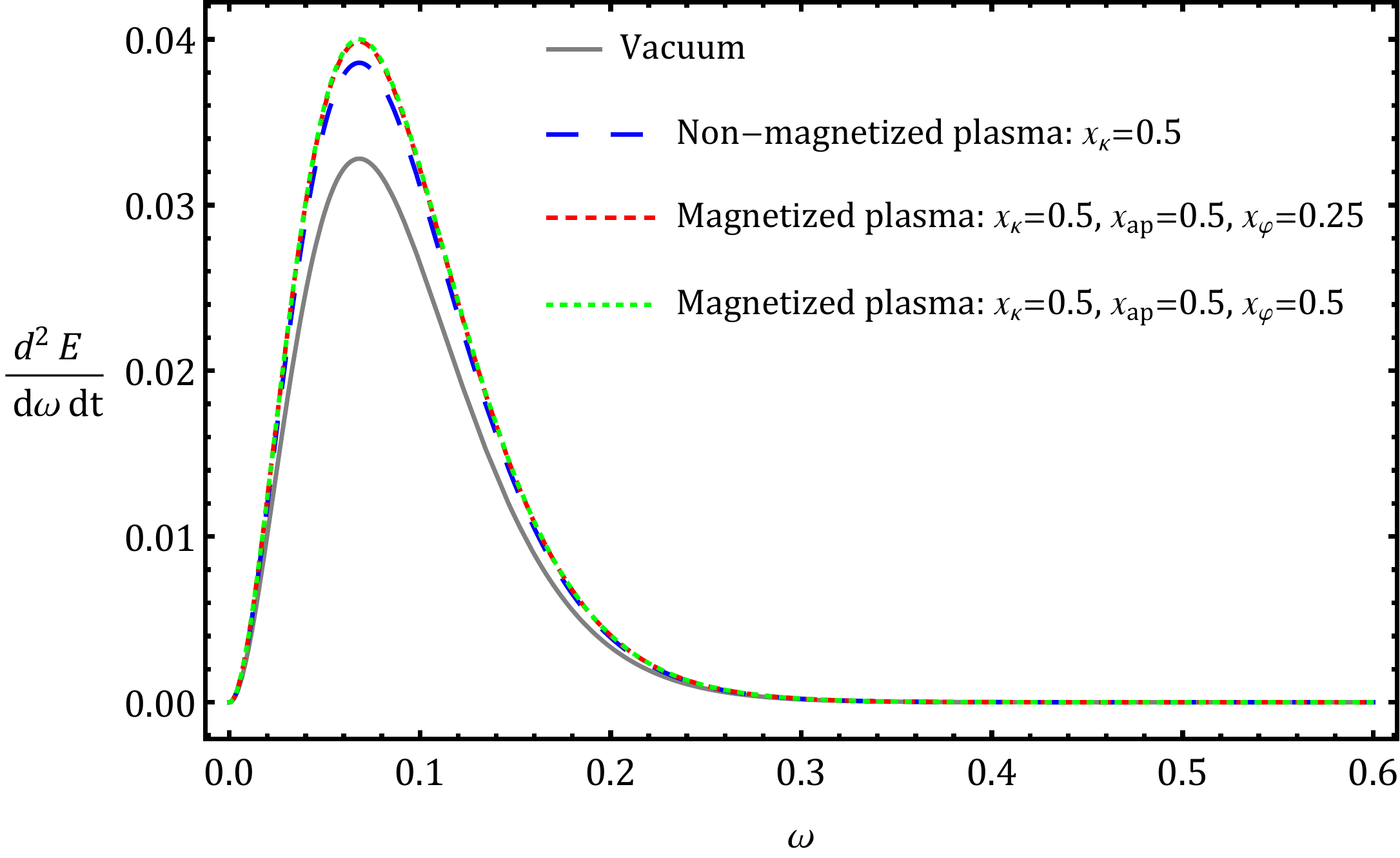}
	\caption{\textit{The EER of Schwarzschild BH surrounded by a plasma with in-homogeneous exponentially density distribution for different sets of $\{\chi_k,\chi_{ap},\chi_\varphi\}$.}}
	\label{Emissions2}
\end{figure}
\section{Non-rotating BH shadow}
\label{secs.4}
Although nature does not seem to agree with the existence of a static or Schwarzschild BH, one can be considered it as a theoretical well-approximation of reality. Now by discarding the axisymmetry property for the BH spacetime, we are plan to do the above analysis for a non-rotating spherical symmetry BH. In other words, we are interested in see in the presence of the magnetized plasma at hand, how will change the size and shape of shadow related to a static BH. Although such a subject has recently been studied directly through confronting the Schwarzschild BH with magnetized plasma in \cite{Atamurotov:2021cgh}, here we will do it as a non-rotating limit of the above analysis. By taking the limit $a\longrightarrow0$ in expressions of the celestial coordinates $(\alpha,\beta)$ released in  (\ref{alpha}) one easily arrive at the shadow radius of static BH surrounded by plasma in the form
\begin{eqnarray}
R_{s-Sh}&=& \frac{1}{n(r_p-M)}\Bigg(2 r_p^3 (r_p-M) n^2   +r_p^4 nn'(r_p-M)-2 r_p^2 M^2   + 2M r_p^2  \bigg(n r^2 (n+r_p n') - (4 n +3r_p n') n Mr_p+\nonumber\\&&
M^2 (1+3 n^2+2r_p nn')\bigg)^{1/2}\Bigg)^{1/2}\ ,\label{shadrad}
\end{eqnarray}
where $n'=(\partial {n}(r)/\partial r)_{r=r_p}$ so that $r_p$ is the unstable circular orbits of photons defined by $dr/d\sigma = 0=\partial {\cal R}(r)/\partial r$
\begin{eqnarray}
2M n' r_p^2-(n'+2n)r_p+6Mn=0~. \label{un}
\end{eqnarray}
Note that for the reader to be able to easily distinguish between
the shadow radius of Schwarzschild BH from its rotating counterpart, we here use the subscript ``Sh''. In the case of vacuum $(n=1, n'=0)$, one has the standard value of the photon sphere radius as $r_p=3M$ that by inserting into (\ref{shadrad}) obtain well known value $R_{s-Sh}=3\sqrt{3}$ for shadow radius of Schwarzschild BH ~\cite{chandra98}. Now by having  (\ref{shadrad}) and  (\ref{un}), one can for the plasma distribution models at hand display the relevant Schwarzschild BH shadow in terms of different values of involved parameters, see Figs. \ref{Shadow4},  \ref{Shadow5} and \ref{Shadow6}.
These figures openly show that, unlike the rotating case, the size of Schwarzschild BH shadow radius due to being surrounded by the magnetic and non-magnetic plasma, turns smaller relative to the vacuum solution. For both plasma distributions, there is a possibility of resolution of the magnetized plasma shadow from its non-magnetized counterpart. Of course, in the case of the in-homogeneous power-law model, this is poorer since the changes of its shadow radius are insignificant. As a complementary work, it would be interesting if we display the plot of $R_{s-Sh}$ in terms of $\chi_k$ for different values of $\chi_{ap}$ and $\chi_\phi$. Note that due to the spherical symmetry of Schwarzschild BH shadow curves, so $\delta_{s-Sh}$, not applicable here any longer. The plots released of $R_{s-Sh}-\chi_k$ in Fig. \ref{RSH1}, explicitly confirm the changing trend of shadow size in Figs. \ref{Shadow4}-\ref{Shadow6}, which by going from vacuum solution toward non-magnetized and magnetized plasma solutions, turns smaller.
As can be seen, unlike the rotating case here just for in-homogeneous power-law density distribution, there is a possibility for leaving a separable imprint of heavy axion mass on the Schwarzschild BH shadow.
\section{The effect of Axion-plasmon on emission energy rate (EER) of BH}
\label{emission}
The idea of absorption and emission of particles close to the strong gravity-dominated regions has got attention to study the different phenomena arising from the interactions of fields and BHs, for example see some seminal papers \cite{Mashhoon:1973zz}-\cite{Das:1996we}. It is well-known at high energy physics that the absorption cross-section is utilized as a measure for the probability of an absorption process and is of great importance to investigate the absorption and scattering of different types of fields around the BH (see, e.g., \cite{Jung:2004yh,Doran:2005vm,Crispino:2007qw}). 
Besides, the photon sphere, indeed is a hypersurface of unstable null circular geodesics, so that can be imagined it as a capture cross-section of the BH. As a result, from wive of a observer located at infinity the BH shadow can address the high energy absorption cross section of BH.  This is because the photon sphere is located at the maximum of the effective potential and at the critical impact parameter related to coming null rays from the infinity that reach the photon sphere by circling. In \cite{Sanchez:1977si} was showed that the absorption cross section oscillates around the limiting constant geometric-optics value for a BH which is linked with the radius of the photon sphere, i.e, $\sigma_{lim} \approx \pi R_s^2$. Note that for cases of Schwarzschild and rotating BHs, we should take into account $R_{s-Sh}$ and $R_{s}$, respectively. By taking the $\sigma_{lim}$ in the account of the EER of BH which has a dependency on the limiting constant value of the absorption cross-section \cite{Wei:2013kza}, we have 
\begin{equation}\label{EM}
\dfrac{d^2E(\omega)}{d\omega dt} = \dfrac{2\pi^3R_s^2 }{e^{\omega/T}-1}\omega^3~,
\end{equation}
where $\omega$ is the frequency of photon and $T$ is the Hawking temperature for the outer event horizon $r_+$ and is defined by
\begin{equation}
T = \lim_{\theta=0,  r \to r_+} \dfrac{\partial_r \sqrt{g_{tt}}}{2\pi \sqrt{g_{rr}}},
\end{equation}
where after some straightforward calculation, it takes the following form 
\begin{equation}\label{TK}
T_{Kerr} = \dfrac{r_+^2-a^2 }{ 4\pi r_+ (r_+^2+a^2) }~.
\end{equation} 
By relaxing the rotation parameter $a=0$,  the Hawking temperature related to event horizon of the Schwarzschild BH reads as
\begin{equation}\label{Tsh}
T_{Sh} = \dfrac{1}{ 8\pi M}~.
\end{equation}
Now, by taking (\ref{RRp}),  and (\ref{TK}) into (\ref{EM}), we depict the EER related to a rotating BH surrounded by both plasma distribution functions at hand, see Figs. \ref{Emissionr1},  and \ref{Emissionr2}. As one can see, by going from the case of vacuum solution to Kerr BH surrounded by the magnetized plasma, the peak of EER increases while, the frequencies corresponding to peaks are identical. Generally, by choosing some values of free parameters, one can find peaks that separate, albeit subtly, the magnetized plasma from its non-magnetized and Kerr vacuum counterparts. 

At the end, by discarding the rotation parameter and mixing Eqs. (\ref{shadrad}),  (\ref{un}), and (\ref{Tsh}) into (\ref{EM}), we depict the EER related to these plasma models in Figs. \ref{Emissions1}, and \ref{Emissions2}.  Unlike the rotating case, here by going from the case of vacuum solution to Kerr BH surrounded by the magnetized plasma, the peak of EER decreases. As before, the peaks address separably the maximum EER related to cases: vacuum solution, the non-magnetized plasma, and magnetized one, without a shift in their frequencies.

A common point in Figs. \ref{Emissionr1}-\ref{Emissions2} is that the change of axion mass (addressed by $\chi_\varphi$) cannot leave  separable imprint on the EER. It is because EER has a direct relation with the squared shadow size from one side, and the changes of $R_s$ in terms of $\chi_\varphi$ are very tiny from another side. 
 
\section{Conclusion}
\label{con}
By exploiting BHs and their environment as extreme places for tracing axion via its interaction with the photon, we in this paper have taken into account a laboratory-based axion-producing model of the magnetized plasma \cite{Mendonca:2019eke, Tercas:2018gxv}, as an axion-plasmon cloud surrounding a rotating BH. This extension from laboratory to the environment of BH is safe because the first estimations released by EHT of the magnetic field strength and the average plasma density around supermassive M87* BH satisfies the required laboratory condition in the axion-plasma configuration, i.e., $\omega_e\gg gB_0$. The principal purpose of this paper has been dedicated to investigate the effect of magnetized plasma induced by an axion-plasmon cloud on the BH shadow geometry, in particular the shape and size of shadow.

To do so, we have employed two type distribution functions for the magnetized plasma: the radial power-law density $N(r)=N_0 r^{-h} (h\geq0)$ \cite{Rogers:2015dla} and the exponential density $N(r)=N_0 e^{-r/l_0}$ \cite{Er:2013efa}. Concerning the former, we have demonstrated that both the shape and size of rotating BH are affected by the magnetized plasma and results in deformation in the BH shape and an increase in its size relative to cases of non-magnetized plasma and vacuum solution. More precisely, we have found that by taking a homogeneous distribution ($h=0$), the axion-plasmon background leaves some separable imprints relative to non-magnetic plasma, as well as vacuum solution on the shadow, while for the case of in-homogeneous ($h=1$), its separability turns more difficult.  By taking the latter plasma distribution into account, also the shadow size increases, as before, of course with this difference that the value of free parameter $l_0$, i.e., the scale radius of the central region of the plasma, has an essential role for separability of imprints on the BH shadow. 

In this regard, to provide a more exact analysis, we also have used two observables: $R_s$ and $\delta_s$, which are usually defined to characterize the shape of the BH shadow. The former addresses the size of the shadow, while the latter the amount of deformation of the BH shape. The benefit of utilizing these two astronomical quantities is that lets us consider into play the role of change of rotation parameter $a$ in the interplay with other involved parameters. In agreement with the results discussed above, it is shown that the shadow size in the presence of magnetized plasma turns bigger than two other counterparts. However, the plots of $R_s-a$ indicate that the shadow size for the case of the magnetized plasma with homogeneous power-law density distribution decreases as the rotation parameter increases (similar to non-magnetized plasma and vacuum solutions) while, for in-homogeneous power-law and exponentially density distributions, it increases. Concerning the deformation of the shadow shape, we found that by taking the magnetized plasma into account, it becomes smaller relative to two other counterparts. The plots of $\delta_s-a$ indicate that in any three plasma distributions the deformation of shadow  $\delta_s$ related to the magnetized plasma solution like those two grows with the increase of the spin parameter. Another interesting result obtained from plots $R_s-a$ and $\delta_s-a$ is that the heavier the axion particles, thereby, the shadow curves show changes distinguishable from the other two solutions.

In the next step, by relaxing the rotation parameter ($a=0$), we investigated the shadow size of a Schwarzschild BH surrounded by the axion-plasmon cloud.  For the plasma distribution functions at hand, it is shown that the size of BH shadow in the absence of rotation compresses is compared to non-magnetized plasma and vacuum solutions. We confirmed this result by drawing the plot  $R_{s-Sh}$ in terms of $\chi_k$, for all plasma distribution functions at hand.

In the end, by studying the energy emission from the rotating and non-rotating BHs in magnetized plasma has shown that the maximum value of energy emission rate from the BH respectively increases and decreases compared to its two other counterparts. It happens without any change in the location of the peak of energy. This trend is natural since the magnetized plasma in rotating and no-rotation BHs increases and decreases shadow size, respectively.
\vspace{0.5cm}
\begin{center}
{\bf ACKNOWLEDGMENTS}
\end{center}
The author is grateful to Luca Visinelli for carefully reading the manuscript and enlightening discussions.

\end{document}